\documentclass[aps,pra,groupedaddress,secnumarabic,floatfix, longbibliography]{revtex4-1} 
\usepackage{amsmath, amssymb,amsbsy, graphicx, booktabs, array, natbib, bbold, color, datetime, bbm} 
\usepackage[bottom]{footmisc}
\usepackage[colorlinks=true, citecolor=red, linkcolor=blue,urlcolor=blue ]{hyperref}

\newcommand{\bv}[1]{\mathbf{#1}} 
\newcommand{\dd}{\; \mathrm{d}} 
\newcommand{\ul}{ \underline}
\newcommand{\bsym}[1]{ \boldsymbol{#1}}  
\newcommand{\Epsilontensor}{\ul{\ul{\ul{\boldsymbol{\epsilon}}}}} 
\newcommand{\ahead}{a_{h}}

\newcommand{\curlyA}{\mathcal{A}} \newcommand{\curlyB}{\mathcal{B}} 
\newcommand{\curlyC}{\mathcal{C}} \newcommand{\curlyD}{\mathcal{D}}

\newcommand{\curlyP}{\mathcal{P}} 
\newcommand{\curlyQ}{\mathcal{Q}}

\newcommand{\darboux}{\mathbf{D}} 
\newcommand{\cdarboux}[1]{D_{#1}} 

\newcommand{\bom}{\boldsymbol{\omega}}  
\newcommand{\sphase}{\left(\frac{2 \pi s}{\refc{\Lambda}} \right)}
\newcommand{\lphase}{\left(\frac{2 \pi L}{\refc{\Lambda}}\right)}
\newcommand{\dlphase}{\left( 4 \pi n\right)}
\newcommand{\slphasedif}{\frac{2 \pi (s-L)}{\refc{\Lambda}}}
\newcommand{\curvone}{\kappa^{(1)}}
\newcommand{\curvtwo}{\kappa^{(2)}}
\newcommand{\curvature}{\kappa}
\newcommand{\twist}{\tau}
\newcommand{\refc}[1]{{#1}_{0}}       
\newcommand{\uprefc}[1]{{#1}^{(0)}}   
\newcommand{\matd}[1]{\mathbf{d}_{#1}}  

\newcommand{\Finternal}{\mathbf{F}}
\newcommand{\Mbending}{\mathbf{M}}

\newcommand{\vdeltaphi}{\delta\boldsymbol{\phi}}
\newcommand{\forcedens}{\bv{K}}
\newcommand{\torquedens}{\bv{N}}
\newcommand{\resperp}{\zeta_{\perp}}   \newcommand{\respar}{\zeta_{\parallel}}
\newcommand{\resrot}{\zeta_{r}}

\newcommand{\curlyG}{\mathcal{G}}
\newcommand{\Rsinfrac}{\frac{\refc{R}}{\sin\refc{\alpha}}}
\newcommand{\uprmatd}[1]{\mathbf{d}^{(0)}_{#1}}

\newcommand{\vPhi}{\boldsymbol{\Phi}}
\newcommand{\pertzeroend}{\Delta \bv{r}_0}

\newcommand{\curlyAfil}{\mathcal{A}_{fil}} \newcommand{\curlyBfil}{\mathcal{B}_{fil}} 
 \newcommand{\curlyDfil}{\mathcal{D}_{fil}}
\newcommand{\bcurlyAfil}{\bsym{\mathcal{A}}_{fil}} \newcommand{\bcurlyBfil}{\bsym{\mathcal{B}}_{fil}} 
\newcommand{\bcurlyCfil}{\bsym{\mathcal{C}}_{fil}} \newcommand{\bcurlyDfil}{\bsym{\mathcal{D}}_{fil}}
\newcommand{\bMmotor}{\bv{M}_{mot}}
\newcommand{\Mmotor}{M_{mot}}
\newcommand{\bomegafil}{\bsym{\omega}_{fil}}
\newcommand{\bomegahead}{\bsym{\omega}_{h}}
\newcommand{\omegafil}{\omega_{fil}}

\newcommand{\vez}{\mathbf{e}_z}
\newcommand{\ruprefc}{\uprefc{\bv{r}}}
\newcommand{\resist}{\bsym{\mathcal{R}}}
\newcommand{\motil}{\bsym{\mathcal{M}}}
\newcommand{\Nviscous}{\mathbf{N}_{visc}}
\newcommand{\indfun}{\mathbbm{1}}  
\newcommand{\Jclamped}[1]{J_{#1}^{clamp}}
\newcommand{\Jfree}[1]{J_{#1}^{free}}
\newcommand{\Jvrottorq}[1]{J_{#1}^{rot}}
\newcommand{\ClampPertMatr}[1]{\delta \mathcal{#1}^{clamp}} 
\newcommand{\FreePertMatr}[1]{\delta \mathcal{#1}^{free}} 
\newcommand{\VisRotTorqPertMatr}[1]{\delta \mathcal{#1}^{rot}} 
\newcommand{\Utot}{U_{tot}}
\newcommand{\bvrhead}{\bv{r}_{h}}

\begin{document}
	\title{Propulsion by stiff elastic filaments in viscous fluids}
	\author{Panayiota Katsamba}
	\email{Present address: School of Mathematics, 
		University of Birmingham, 
		Edgbaston, Birmingham, 
		B15 2TT, United Kingdom.}	\affiliation{Department of Applied Mathematics and Theoretical Physics, University of Cambridge, Cambridge CB3 0WA, United Kingdom.}
	\author{Eric Lauga}
	\email{e.lauga@damtp.cam.ac.uk}
	\affiliation{Department of Applied Mathematics and Theoretical Physics, University of Cambridge, Cambridge CB3 0WA, United Kingdom.}
	
	\begin{abstract}
	Flexible filaments moving in viscous fluids are ubiquitous in the natural microscopic world. For example, the swimming of bacteria and spermatozoa as well as important physiological functions at organ-level, such as the cilia-induced motion of mucus in the lungs,  or individual cell-level, such as actin filaments or microtubules, all employ flexible filaments moving in viscous fluids. As a result of fluid-structure interactions, a variety of  nonlinear phenomena may arise in the dynamics   of such moving flexible filaments. In this paper we  derive the mathematical tools required to study filament-driven propulsion 	 in the asymptotic limit of stiff filaments.
	Motion in the rigid limit leads to hydrodynamic loads which deform the filament and  impact  the filament propulsion.
	We first derive the general mathematical formulation and then apply it to the  case of a helical filament, a situation relevant for the swimming of flagellated bacteria and for the transport of artificial, magnetically actuated  motors.
	We find that, as a result of flexibility, the helical filament is either stretched or compressed (conforming previous studies) and its axis also bends, a new result which we interpret physically. 
	We then explore and interpret the dependence of the perturbed propulsion speed  due to the deformation on the relevant dimensionless dynamic and geometric parameters.

	\end{abstract}

\date{\today}

\maketitle
\section{Introduction}\label{Intro}
Many biophysical processes and engineering problems exhibit rich nonlinear behaviour due to fluid-structure interactions \cite{DowellHallFluidStructureInteraction2001,HuangAlbenFSIarticle2016}.  
Examples include tall buildings responding to winds \cite{Braun2009}, the flapping of bird wings \cite{Kang2011flapping,WangZhangFlapping2016} and the motion of aircrafts \cite{ThomasDowellHall2002}. 
Physiological flows \cite{ HeilHazelFSIInternalPhysiologicalFlows2011}    provide additional examples, for example  the vibrating 
vocal folds \cite{Horacek2002}, heart valves opening and closing with blood flow \cite{Sacks2009}, lungs expanding and contracting with breathing \cite{Otis1950Breathingmechanics,Dubois1956Lungs}  and pulse propagation in blood vessels  \cite{PedleyBloodFlowArteriesVeins2000}.  Going all the way down to the microscopic world, one comes across elastic structures being deformed due to  hydrodynamic loads from  flows which are, in turn, affected by the deformation. Examples including deforming cilia \cite{BrennenWinet77,Bray2000}, fluctuating actin filaments \cite{Howard2001}, polymerising microtubules \cite{Hawkins2010}, the waving flagella of spermatozoa and the rotating flagellar filaments of bacteria \cite{BergAnderson1973,LaugaPowers2009}.

 \begin{figure}
	\includegraphics[scale=0.5]{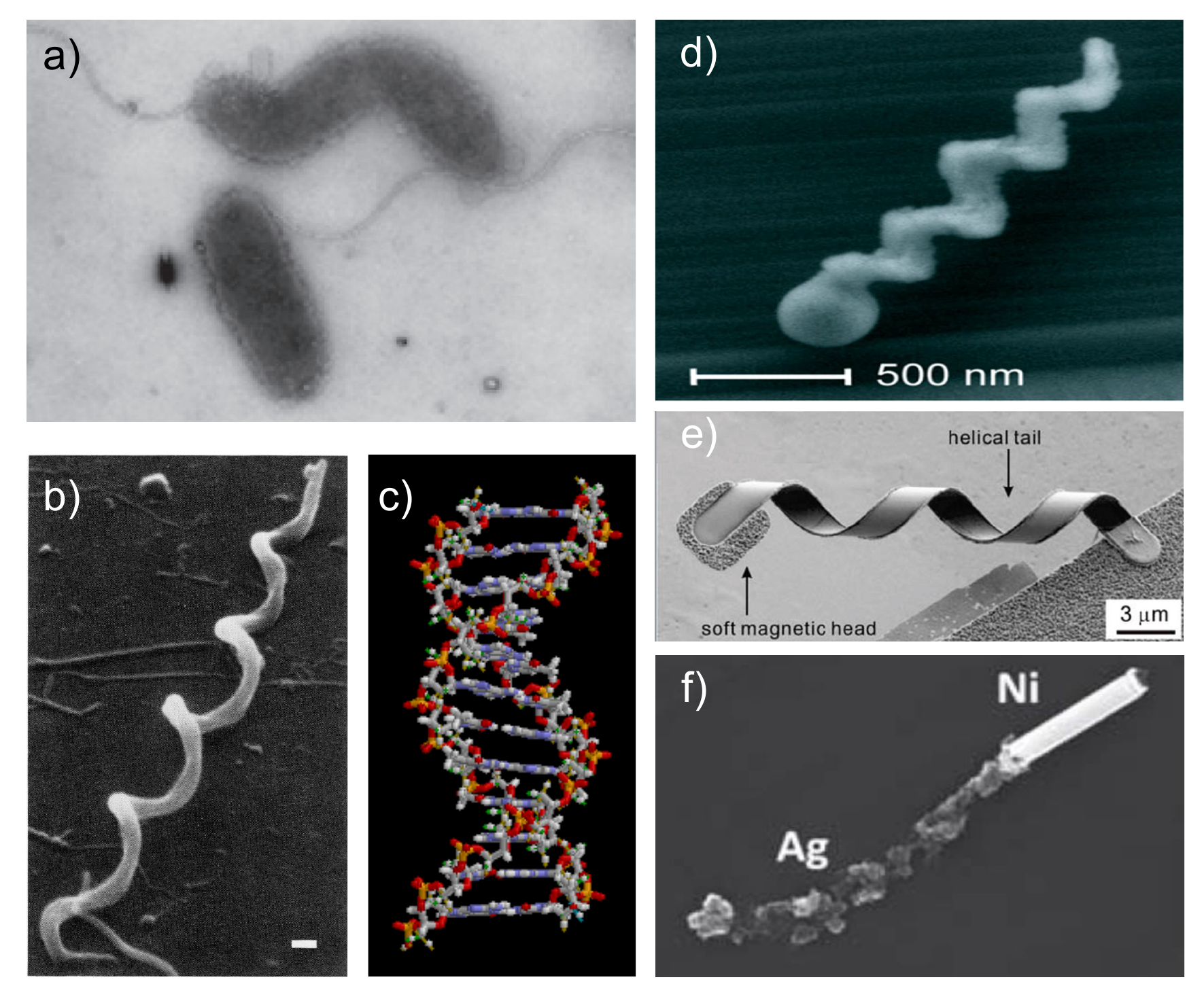}
	\caption{Examples of microscopic helices in the natural and artificial world:  		(A) Polar monotrichous {\it Pseudomonas aeruginosa} bacteria \cite{Fujiietal2008Flagella}; Reprinted from Fujii, Shibata, and  Aizawa, 2008, `Polar, peritrichous, and lateral flagella belong to three distinguishable flagellar families', {\it J. Mol. Biol.}, {\bf 379}, 273-283 with permission from Elsevier; Copyright (2008) Elsevier.  
		(b) Spirochaete bacterium \cite{JohnsonetalSpirochetes1984};	
		Reprinted (adapted) from 
		Johnson, Hyde and Rumpel, 1984, `Taxonomy of the Lyme disease spirochetes', {\it Yale J. Biol. Med.}, {\bf 57}, 529-537 with permission from Yale J. Biol. Med.; Copyright (1984) Yale J. Biol. Med.  
		(c) The structure of part of a DNA double helix, reproduced from Wikimedia Commons; 
		(d) Chiral magnetic propeller \cite{GhoshFischer2009}. 
		Reprinted (adapted) with permission from 
		Ghosh and Fischer, 2009, `Controlled propulsion of artificial magnetic nanostructured propellers', {\it Nano Lett.}, {\bf 9}, 2243-2245  with permission from American Chemical Society; Copyright (2009) American Chemical Society.
		(e) Artificial bacterial flagellum  \cite{Zhang2009}. 
		Reprinted (adapted) with permission from 		Zhang, Abbott, Dong, Peyer, Kratochvil, Zhang, Bergeles and Nelson, 2009,  `Characterizing the swimming properties of artificial bacterial flagella', {\it Nano Lett.}, {\bf 9}, 3663-3667   with permission from  American Chemical Society; Copyright (2009) American Chemical Society.		
		(f) Flexible nanowire motor \cite{Paknanowire2011}. 
		Reprinted (adapted) with permission from Pak, Gao, Wang and Lauga, 2011, `High-speed propulsion of flexible nanowire motors: Theory and experiments', {\it Soft Matter} {\bf 7}, 8169-8181 with permission from The Royal Society of Chemistry; Copyright (2011) The Royal Society of Chemistry.	}\label{Fig1ElasticHelixExamples}
\end{figure}

A prevalent morphology in the microscopic natural world is that of a helix. DNA \cite{BustamanteBryantSmith2003DNAMech,Stricketal1996}, spirochaetes \cite{Lietal2000Spirochete,SFGoldsteinCharon1988spirocheteleptospira}, spiroplasma \cite{YangWolgemuthHuber2009Spiroplasma}, trypanosoma \cite{Trypanosoma2015Stark}  and bacterial flagellar filaments \cite{BergAnderson1973}, all thake the  shape of  helices and spirals, as illustrated in Fig.~\ref{Fig1ElasticHelixExamples}~(A-C). In particular, the chiral shape of a helix is able to couple hydrodynamically rotation to translation. As such, it is used as the propulsive machinery for bacteria swimming in viscous fluids. The helical flagellar filaments of bacteria such as {\it Escherichia coli} ({\it E.~coli}) are several micrometers in length and $20~\rm{nm}$ in radius,  and are rotated at a typical frequency of $100~\rm{Hz}$ by specialised rotary motors \cite{SamuelBerg1996, Lauga2016BactlHydrdyn}.

In the micro-engineering world, different types of externally powered motors have been proposed, studied and constructed. Examples include  rigid helical propellers often termed `artificial bacterial flagella', \cite{Zhang2009,helixpropGhoshFischer2009}, and motors that use flexible filaments \cite{Dreyfus2005,Gaonanowire2010,Paknanowire2011}, both of which are  illustrated in Fig.~\ref{Fig1ElasticHelixExamples}~(D-F).  One of the aspiring application of such artificial microswimmers is non-invasive medicine \cite{Sitti2009voyage,BJNelson2010}, in which swimmers are
to access targeted locations in the body, such as tumours, in order to deliver drugs \cite{Gaocargotowing2012} or perform  delicate surgical tasks \cite{eyesurg2013}.  
In view of the former application, one interesting  possibility is that of the motor being propelled by the drug itself taking the shape of long strips that twist up into a helical shape upon rotation \cite{LiTanZhang2009}.
Magnetically propelled microswimmers with a flexible helix can be also be manufactured \cite{Huangetal2016AddressableShapeControl,Paknanowire2011} and display rich nonlinear dynamics  such as velocity profiles   peaked at certain  operational frequencies \cite{Paknanowire2011}. In addition, nonlinearity can be exploited in achieving selective control of microswimmers in large numbers, a   desirable feature for any practical application \cite{ KatsambaLauga2016,KeiCheang20143bead,Frutiger2010}.

The multitude of biological systems that involve the dynamics of an elastic helix in a viscous fluid environment and the number of micro-engineering applications are compelling evidence for the need to study the elasto-hydrodynamic coupling theoretically.
In the context of bacterial flagellar filaments, there have been several experimental, numerical and analytical investigations to address this problem.  
Theoretical and experimental studies were combined to derive the relationship between the elongation of a flagellar filament (modelled as a chain of segments) and the flow that it is subjected to \cite{HoshikawaKamiya1985}.  
Experimental studies showed different speeds between forward and backwards  swimming of the single polar-flagellated bacterium {\it Vibrio alginolyticus} \cite{Magariyama2001forwardbackwardspeeddiff}. 
These were followed by a numerical analysis for the deformation of a flagellum rotating in a viscous fluid \cite{TakanoGoto2003Numer}. 
Studies of the deformation according to the Kirchhoff rod model combined with the Calladine model of the detailed structure of the filament provided analytical expressions for the bending moment, curvature and torsion of deformed flagellar filaments of swimming bacteria {\it Vibrio alginolyticus} and {\it Salmonella}. The deformation was obtained numerically and a comparison with experimental data provided an estimate of the elastic bending coefficient of the flagellar filament on the order of $EI\approxeq10~\rm{pN}(\mu \rm{m})^2$ \cite{Takanoetal2003HelicFlagel,Takanoetal2005HelicFlagelSalmonella}. 

The extension or compression of an elastic helix by flow and external fields has been studied analytically in the limit of a long, stiff helix \cite{KimPowers2005HelixDeformation}. Force-extension curves were also derived in studies of helical springs subjected to uniaxial  tension \cite{WadaNetz2007}. 
The propulsive force from a rotating, flexible, helical rod in a viscous fluid and the onset of the buckling instability above a critical rotation velocity have been studied by means of experiments and simulations \cite{Vogel2012thesis, VogelStarkBuckling,Jawedetal2015FlexHelixinViscousFluid}. 
In particular, simulations based on a model that uses Kirchhoff's classical elasticity theory for curved rods were used to investigate the transitions between the polymorphic forms of the bacterial 
flagellum \cite{Vogel2012thesis, VogelStarkBuckling}.

In this work, we put forward the mathematical framework necessary to address the steady-state locomotion of stiff elastic, slender filaments in viscous fluids.
The motion of the filament induces a hydrodynamic load that deforms it. This in turn affects the kinematics because the shape has changed. Implementing the overall force and torque balance at leading order involves integrating the hydrodynamic load with the unknown velocity and rotation rate over the new deformed shape as if it were rigid, and inverting the system to solve for the unknown velocity and rotation rate. In order to obtain the perturbation to the rigid kinematics, one needs in particular to perturb the hydrodynamic resistance matrices. In this paper we show how to do this for a long, slender filament of arbitrary shape.  We next apply our analytical framework to study the setup of an elastic helix that is rotating and translating, or equivalently is in the presence of such an external flow of a viscous fluid.
We study the limit where the helix is very stiff, so that any deformation is small, and very long, so that it rotates about its long axis and does not wobble \cite{YiLaugawobbling}. 
This setup is relevant to bacterial flagellar filaments and a popular design for magnetically actuated artificial microswimmers. The latter consists of a flexible helix clamped onto a magnetic head on which an external magnetic torque is exerted. 
We calculate below the full three-dimensional deformation analytically and its feedback on the swimming speed of a bacterium during swimming in a straight line. Our results of the deformation agree with previous analytical results of the extension/compression in Ref.~\cite{KimPowers2005HelixDeformation},
and capture and explain the bending of the helix axis that was observed in the  numerical results of Ref.~\cite{Takanoetal2003HelicFlagel} and whose origin has been unclear so far.

Our paper is organised as follows. 
In \S \ref{General Model} we outline the mathematics framework for the steady motion of a stiff-elastic, slender filament of any shape in a viscous fluid. We show how the  Kirchhoff model for an elastic rod (\S \ref{elasticrod}) combined with resistive-force theory for the hydrodynamic load (\S \ref{hydroload}) lead to the deformation (\S \ref{deform}) and how this in turn perturbs the leading-order kinematics (\S \ref{feedback}). The latter is obtained by implementing the force and torque balance (\S \ref{forcesandtorques}). We formulate this dynamic balance for two specific setups: firstly that of a flexible filament actuated by a magnetic torque  exerted on the head on which it is clamped  (\S \ref{Artifpropsect}) and secondly the case of swimming bacterium rotating a flexible flagellar filament relative to its cell body  (\S \ref{biolocom}). 
Since the prevalent geometry for these two cases is a helical one, in \S \ref{helical} we give the common details of the hydrodynamic load, bending moment and the deformation of the helix (\S \ref{helicaldeform}). We next interpret the bending of the helix axis and proceed to investigate the feedback of the deformation on the kinematics (\S \ref{helicaldeformfeedback}). We then calculate the perturbations to the resistance matrices due to the small deformation. Applying the appropriate forms for the force and torque balance, in both the artificial  (\S \ref{helicartifmotor}) and biological (\S \ref{helicbacteriasec}) setup, we derive the perturbation of the swimming velocity and discuss the physical interpretations of our results.

\section{Propulsion by elastic filaments: General framework} \label{General Model}

We first consider  in this section the propulsion of elastic filaments of arbitrary shapes, and apply it to two  cases:  externally-actuated microswimmer propulsion driven by a rotating magnetic field   and  bacterial propulsion. In both cases, our setup involve a long elastic filament with one end clamped on a head or cell body and which rotates with angular velocity $ \bsym{\omega}$ (in the laboratory frame) and translates with velocity $\bv{U}$. The particular case of helical filaments will be considered in \S\ref{helical}.

In the context of artificial microswimmers, the head is magnetised and actuated by an external magnetic field, $\bv{B}$, rotating about a fixed axis, say the $z$-axis. Assuming the head to have a constant dipole moment, $\bv{m}$, the field exerts a magnetic torque $\bv{M}_{mag}=\mu_0 \bv{m}\wedge\bv{B}$ on the head which thus rotates and, with the proper filament shape, also translates.   
 In the case of bacteria, there is instead a rotary motor embedded in the cell that produces a torque, $\Mmotor$,  actuating the flagellar filament in rotation. The  filament, which in many cases is stiff, is connected to the motor via a very flexible short hook~\cite{BFhookSamateuetalNamba2004}.

 Following the actuation of the magnetised head or the bacterial motor, the elastic filament attached onto it will rotate and (if of the right shape) also  translate. Due to  hydrodynamic loads, it will also start deforming, until it reaches shape equilibrium.  In this work we aim at characterising   the steady state equilibrium configuration of the filament obtained after all  transients, where the shape no longer changes and for which elastic and hydrodynamic stresses balance. The swimming kinematics  are then governed by the force and torque balance over the entire swimming organism (or device), which involve integrating  hydrodynamic loads over the new deformed shape.
The fluid-structure interactions manifests themselves therefore via the deformation  induced by the hydrodynamic load and by the feedback of the deformation on the kinematics via dynamic balance.

In both applications, the magnetised head or cell body and the filament will translate at the same speed in steady state. In the case of the artificial motor, the head and tail will share the same rotation rate since the tail is clamped onto the head.  In our model of bacterium, we do not include details of the hook, but instead assume that its high flexibility allows different rotation rates between the cell body and flagellar filament, the values of which are determined by   torque balance.

\subsection{Elastic Rod}	\label{elasticrod}
	Working in the frame of reference of the filament, let $\bv{r}(s)$ denote the   location of its centreline  and $\{\matd{i}(s)\}_{i=1,2,3}$ the local material frame (which we will take later to coincide with the Frenet-Serret frame)
	so that $\matd{3}$ is tangent to the centreline,
	\begin{equation}
	\partial_{s}\bv{r}=\matd{3}.
	\end{equation}
	The configuration of the material frame along the filament is then described by
	\begin{align}
	\partial_{s}\matd{i}&=\darboux\wedge\matd{i},\label{materialvecsdarboux}\\  \qquad
	\darboux&=\cdarboux{1} \matd{1} + \cdarboux{2} \matd{2} + \cdarboux{3} \matd{3}, \label{darboux}
	\end{align}
	where $\darboux$ is the Darboux vector, measuring the strains in the rod. Its components in the material frame are the material curvatures $(\cdarboux{1}=\curvone, \cdarboux{2}=\curvtwo)$ and the material twist $(\cdarboux{3}=\twist)$ of the rod.
		
	The elastic behaviour of the filament is governed by the classical Kirchhoff equations for a rod \cite{AudolyPomeau}, which give the balance of forces and moments on a cross section
	\begin{align}
	\partial_s \Finternal + \forcedens &=\bv{0},\label{KirchhoffForceBalance}\\ 	\partial_s \Mbending + \matd{3}\wedge\Finternal + \torquedens &=\bv{0}\label{KirchhoffTorqueBalance},
	\end{align}
where 
	$\Finternal$ is the internal force acting on a cross section of the rod, $\Mbending$ is the bending moment and $\forcedens$, $\torquedens$ are the distributed (external) force and torque densities respectively. The constitutive equation for a Hookean material gives   the bending moment as \cite{AudolyPomeau} 
	\begin{equation}\label{bending moment}
	\Mbending =EI^{(1)} (\delta \cdarboux{1})\matd{1} + E I^{(2)}(\delta \cdarboux{2})\matd{2}  + \mu_S J(\delta \cdarboux{3})\matd{3},
	\end{equation}
	where $\delta\cdarboux{i}=\cdarboux{i}-\uprefc{\cdarboux{i}}$ are the deviations of the material curvatures and twist  in the  deformed state from their values in the reference configuration. In the paper, we add a $0$ superscript or subscript to indicate  quantities pertaining to the reference configuration. As appropriate for a linearly elastic  material, the deviations of the   curvatures and twist are therefore  linearly related to the components of the bending moment via Young's modulus, $E$, and the shear modulus, $\mu_S$,  of the material   and depend on three  geometrical coefficients: 
	the principal moments of inertia, $I^{(1)}$ and $I^{(2)}$, and the twist rigidity, $J$, of the rod. 
	For a rod with a circular cross-section of radius $r$ these take the values
	\begin{equation}
		I^{(1)}=I^{(2)}=\pi r^4/4, \quad J=\pi r^4/2. \label{circularprincmomentofinertia}
	\end{equation}

\subsection{Hydrodynamic load}	\label{hydroload}
	\subsubsection{Resistive-force theory}
	The viscous tractions due to the motion of a slender filament in a viscous fluid at low Reynolds number 
	are well captured by resistive-force theory \cite{gray55,Lighthill1976FlagellarHydrodynamics,Rodenborn2012,LaugaPowers2009}. This technique integrates fundamental solutions of the Stokes equations along the centreline of a slender filament  to give an expression for the local hydrodynamic force per unit length, $\forcedens$, exerted on the filament due to its motion in a viscous fluid.  At the position labelled by the contour-length parameter value $s$,  the local hydrodynamic force per unit length exerted on the filament is then given 
	\begin{align}
	\forcedens(s) &=-\resperp[\bv{V_{rel}}(s)-(\matd{3}(s).\bv{V_{rel}}(s))\matd{3}(s)]-\respar (\matd{3}(s).\bv{V_{rel}}(s))\matd{3}(s),  \label{RFT}
	\end{align}
	where $\bv{V_{rel}}(s)$ is the local relative velocity between the filament and the fluid at that position 
	 and $\respar,\resperp$ are the drag coefficients for motion parallel and perpendicular to the local tangent of the filament. 
	There are many forms of the drag coefficients in various geometries \cite{Hancock1953, GrayHancock1955, Lighthill1976FlagellarHydrodynamics, Rodenborn2012} and for the purpose of studying helical filaments in \S\ref{helical}, we will use Lighthill's coefficients given by
	\begin{align}
		\resperp &\approx \frac{4\pi\mu}{\ln(0.18\Lambda/r)+1/2},\quad \respar\approx \frac{2\pi\mu}{\ln(0.18\Lambda/r)}, \label{Lighthill_resistive_coefficients}
	\end{align}
	where $\mu$ is the dynamic viscosity of the fluid and $\Lambda, r$ are the  contour wavelength and cross-sectional radius of a helical filament respectively.
	The ratio of the two coefficients is approximately equal to $1/2$,
	\begin{align}
\respar&= \rho\resperp,
\qquad\rho=\frac{1}{2}\frac{\ln(0.18\Lambda/r)+1/2}{\ln(0.18\Lambda/r)}\approx \frac{1}{2}.\label{Lighthill_resistive_coefficients_ratio}
	\end{align} 
	The fact that the perpendicular drag coefficient is approximately twice the parallel one is the extension to a curved filament of the fact that it is approximately twice as hard to pull a rod through a viscous fluid in a direction perpendicular to its length than lengthwise. This drag anisotropy is at the heart of locomotion of microorganisms and artificial microswimmers. For example, it is the crucial ingredient coupling rotation to translation for bacterial flagellar filaments \cite{Lauga2016BactlHydrdyn}.
	
	There are   two sources of hydrodynamic moments acting on the filament. The first one is the moment due to the distribution of forces in Eq.~\ref{RFT}. The second is the    viscous torque per unit length, $\Nviscous$,   that resists the rotation of an element of the rod about $\matd{3}$ (i.e.~its centreline) given by
	\begin{align}
		\Nviscous=-\resrot (\matd{3}.\bom)\matd{3},
	\end{align} 
where $\resrot$ is the rotational drag coefficient
\begin{align}
	\resrot=4\pi \mu r^2.
\end{align}
This second viscous torque, $\Nviscous$, 
{
can be typically neglected for filaments with sufficiently-small cross-sectional radius $r$.  Consider a helical filament with helical radius $\refc{R}$ in its reference configuration  and compare the magnitude of the   moment (measured with respect to the helical axis) due the hydrodynamic force per unit length, $\forcedens$, with that of $\Nviscous$ in the expression for the bending moment, $\Mbending$, in  Eq.~\ref{BendingMomentIntegralExpr}. Their ratio scales as \cite{KimPowers2005HelixDeformation}
\begin{equation}
\frac{|\Nviscous|}{|\bv{r}\wedge\forcedens|}\sim\frac{\mu \omega r^2 }{\mu \omega R_0^2}\sim\left(\frac{r}{R_0}
\right)^{2}.
\end{equation}
This ratio is  very  small unless one is dealing with nearly straight filament configurations, which is typically not the case for the helical geometry of bacteria and artificial micromotors. The second viscous torque, $\Nviscous$, can therefore be safely neglected.
}

Once steady state has been reached, the filament rotates with uniform angular velocity, $ \bsym{\omega}$, and translates with uniform velocity, $\bsym{U}$.  We may then consider the  frame in which the filament is stationary.
The relative velocity between the filament and the fluid is given by
\begin{align}
\bv{V_{rel}}&=\bv{U}+\bom\wedge\bv{r}(s),
\end{align}
which allows access to the leading-order estimate for the force density, $\forcedens$, in Eq.~\ref{RFT}.

The next step is to calculate the internal force, $\Finternal$,  and bending moment, $\Mbending$, by integrating the Kirchhoff equations, \ref{KirchhoffForceBalance} and \ref{KirchhoffTorqueBalance}. We assume that the  end-point at $s=L$ of the filament is free,
\begin{align}
	\Finternal(L)=\bv{0} ,\quad 
	\Mbending(L)=\bv{0}  \label{BC},
\end{align}
allowing  to obtain explicitly 
\begin{align}
\Finternal(s)&=\int_{s}^{L}\forcedens(s')~\dd s', \label{Fintegral} \\
\Mbending(s)&=\int_{s}^{L} ~\left[\matd{3}(s')\wedge\Finternal(s') + \torquedens_{visc}(s')\right]~\dd s' .
\end{align}
Using that $\matd{3}(s')=\partial_{s'}[\bv{r}(s')]$ we can rewrite $\partial_{s'}[\bv{r}(s')]\wedge\Finternal(s')=\partial_{s'}[\bv{r}(s')\wedge\Finternal(s')]-\bv{r}(s')\wedge\partial_{s'}[\Finternal(s')]$, integrate by parts and use   Eq.~\ref{KirchhoffForceBalance} to obtain
\begin{align}
\Mbending(s) 
=&\left[\bv{r}(s')\wedge\Finternal(s')\right]|_{s}^{L}+\int_{s}^{L}\bv{r}(s')\wedge\forcedens(s') +\Nviscous(s')~\dd s'.\label{BendingMomentIntegralExpr}
\end{align}
Using the boundary conditions of Eq.~\ref{BC}, we then obtain the integral formula
\begin{align}
\Mbending(s)+\bv{r}(s)\wedge\Finternal(s) &=\int_{s}^{L}\bv{r}(s')\wedge\forcedens(s') + \Nviscous(s')~\dd s'. \label{Mintegral}
\end{align}
As we will see in \S \ref{Artifpropsect}, the second term in the left-hand side of Eq.~\ref{Mintegral} arises naturally when one considers the torque balance with respect to the origin, as the  bending moment, $\Mbending(s)$, is  defined with respect to the centre of the cross-section at arclength position $s$ along the filament.

\subsection{Deformation}\label{deform}
As a result of the hydrodynamic forcing, the filament will deform. In the limit of stiff filaments, i.e.~of high Young's modulus, the filament will undergo small deformations and the material frame will be  slightly perturbed.  As the material frame needs to stay orthonormal, this can be represented by a set of three small local rotation vectors, $\delta\boldsymbol{\phi}(s)$, along the rod \cite{Audoly-Pomeau-Elasticity-and-geometry-2000} so that
\begin{align}\label{smallmatframerotation}
\delta\matd{i}(s)=\delta\boldsymbol{\phi}(s) \wedge\uprefc{\matd{i}}(s). 
\end{align}
 
In order to relate the bending moment to the deformation, we   need to relate the perturbations to the components of the Darboux vector, $\delta D_i$, that appear in the constitutive equation, Eq.~\ref{bending moment}, to the small rotations $\delta\boldsymbol{\phi}(s)$ of Eq.~\ref{smallmatframerotation}, following Ref.~\cite{AudolyPomeau}. This is done by considering the first-order perturbations to the two sides of Eq.~\ref{materialvecsdarboux}. 
Perturbing the left-hand side of Eq.~\ref{materialvecsdarboux} and using Eq.~\ref{smallmatframerotation} leads to 
\begin{align}
\delta(\partial_s \matd{i})&=\partial_s (\delta \matd{i})=\partial_s [(\vdeltaphi)\wedge \uprefc{\matd{i}}] =[\partial_s(\vdeltaphi)]\wedge\uprefc{\matd{i}} + (\vdeltaphi)\wedge[\uprefc{\darboux}\wedge\uprefc{\matd{i}}].
\end{align}
Perturbing next the right-hand side of Eq.~\ref{materialvecsdarboux} leads to
$(\delta \darboux)\wedge\uprefc{\matd{i}} + \uprefc{\darboux}\wedge(\delta \matd{i})$ and 
equating the two expressions we obtain
\begin{align}
&[\partial_s(\vdeltaphi)]\wedge\uprefc{\matd{i}} 
+ (\vdeltaphi)\wedge(\uprefc{\darboux}\wedge\uprefc{\matd{i}})-(\delta \darboux)\wedge\uprefc{\matd{i}} -\uprefc{\darboux}\wedge(\vdeltaphi\wedge\uprefc{\matd{i}})
=0. \label{perturbingmaterialvecsdarboux}
\end{align}
Using the Jacobi identity for the second and last terms of Eq.~\ref{perturbingmaterialvecsdarboux} leads to
\begin{align}
[\partial_s(\vdeltaphi) + (\vdeltaphi)\wedge \uprefc{\darboux} - \delta \darboux]\wedge \uprefc{\matd{i}}&=0,
\end{align}
for all $i$. 
We thus have that
\begin{align}
\partial_s(\vdeltaphi) + (\vdeltaphi)\wedge \uprefc{\darboux} - \delta \darboux = 0.
\end{align}	
Rearranging and expressing quantities in the reference configuration material frame gives
\begin{align}
\partial_s(\vdeltaphi) &= \delta \darboux - (\vdeltaphi)\wedge \uprefc{\darboux}\\
&=(\delta D_i)\uprefc{\matd{i}} +  \uprefc{D_i}(\delta\matd{i}) - \uprefc{D_i}(\vdeltaphi \wedge\uprefc{\matd{i}})\\
&=(\delta D_i)\uprefc{\matd{i}},\label{eq:27}
\end{align}
where we used the summation convention for repeated indices and where we used   $\delta \darboux =\delta D_i\uprefc{\matd{i}} + \uprefc{D_i}\delta\uprefc{\matd{i}}$.
Integrating Eq.~\eqref{eq:27} then one obtains the   expression for the infinitesimal rotation $\vdeltaphi(s)$
\begin{align}
(\vdeltaphi)(s) 
&=\int\limits_{0}^{s} (\delta D_i)\uprefc{\matd{i}}\dd s' + \bv{\Phi}. \label{deltaphi}
\end{align}
Note that the lower limit of the integral makes use of the boundary condition 
\begin{align}
	(\vdeltaphi)(0)=\bv{\Phi}.
\end{align}
A clamped boundary condition at   $s=0$  means $\bv{\Phi}=\bv{0}$, however for a free end  at $s=0$  one would have to solve for the value of $\bv{\Phi}$.

Using the classical assumption that $EI^{(1)}=EI^{(2)}=\mu J$  \cite{KimPowers2005HelixDeformation,VogelStark2010, VogelStarkBuckling}, 
we next note that the constitutive equation for the bending moment, Eq.~\ref{bending moment}, simplifies to
\begin{equation}
\Mbending =EI \sum_{i}^{}(\delta D_i)\uprefc{\matd{i}}.
\end{equation}
Thus $\vdeltaphi$ in Eq.~\ref{deltaphi} is obtained by integrating the bending moment along the filament as
\begin{alignat}{4}
(\vdeltaphi)(s) 
&&=&&\frac{1}{EI} \int\limits_{0}^{s} \Mbending (s') \dd s'+ \vPhi. \label{deltaphi_int_M_formula}
\end{alignat}
The perturbation to the tangent vector is given by Eq.~\ref{smallmatframerotation}, hence we obtain
\begin{align}
\delta \matd{3} (s)&=\frac{1}{EI}\left(\int\limits_{0}^{s}\Mbending(s')\dd s'\right) \wedge\uprmatd{3}(s) + \vPhi \wedge\uprmatd{3}(s). \label{perttangent}
\end{align}
Integrating Eq.~\ref{perttangent} next gives access to the deformation 
\begin{align}
\delta \bv{r} (s)=&\int\limits_{0}^{s}\delta\matd{3}(s') \dd s'  + \Delta \bv{r}_0  \label{deform_int_deltad3}\\
=&\frac{1}{EI}\int\limits_{0}^{s} \left(\int\limits_{0}^{s'}\Mbending(s'')\dd s''\right) \wedge\uprmatd{3}(s')\dd s'  
+ \vPhi\wedge\left(\uprefc{\bv{r}}(s) - \uprefc{\bv{r}}(0)\right)
+ \pertzeroend. \label{pertdrwithPhiandDr0}
\end{align}
where the    perturbation to the position of the $s=0$ end-point is denoted $\pertzeroend$; it vanishes if the end-point has fixed position, otherwise for a free end-point, $\pertzeroend$ would have to be solved for.

\subsection{Feedback of the deformation on the kinematics} \label{feedback}
\begin{figure}
\includegraphics[width=0.5\columnwidth]{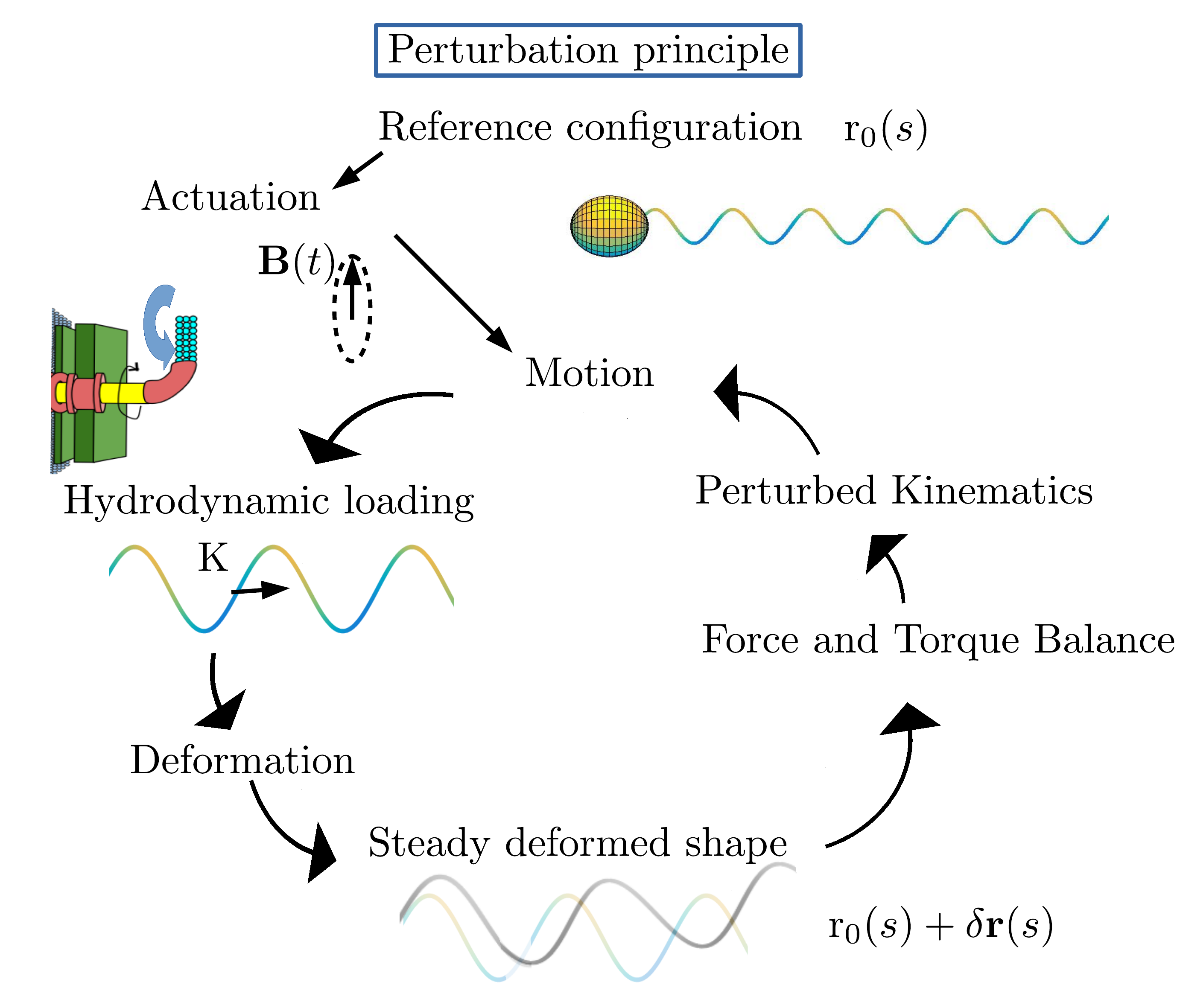}
	\caption{Perturbation principle required to compute the feedback of the deformation on the kinematics: Once the bacterial cell/artificial motor starts moving (due to the flagellar motor actuation or to the rotating magnetic field)  the filament  experiences a hydrodynamic load and  starts deforming until it reaches a steady state shape, which for stiff filaments is slightly perturbed from the reference configuration. The hydrodynamic load then needs to be integrated over this perturbed shape. Thus implementing the force and torque balances leads a system involving the perturbed resistance matrices which are linear functions of the deformation. Once inverted, this gives access to the perturbation of the rigid-body kinematics (diagram of the bacterial motor adapted from  Wikimedia Commons, Mgaetani, 2015).}\label{FigPerturbationPrinciple}
\end{figure}

The swimming kinematics  of the filament-body/head pair are determined by the force and torque balance.
For large values of the filament Young's modulus, the deformations are small and the kinematics are given by those of a rigid filament to leading order. 
In this section we quantity the impact of the deformation $\delta \bv{r}$ on the kinematics, 
i.e.~calculate the next-order effect on both $\bv{U}$ and $\bom$. 
The perturbation principle behind the calculation of the feedback of the deformation on the kinematics is illustrated in  Fig.~\ref{FigPerturbationPrinciple}. 
Once the cell or artificial motor starts moving,  the filament  starts deforming due to the hydrodynamic load until it reaches steady state after which it  undergoes rigid body motion. 
The kinematics will be slightly different from the rigid-body ones, because the hydrodynamic load is integrated along a slightly perturbed shape along which  force and torque balance is enforced. 
We first express the force and torque balance for a general shape of the filament, leading to a linear system involving the hydrodynamic load integrated over the shape of the entire filament. Inputting the reference configuration shape gives the rigid filament kinematics while perturbing the system leads to a relationship linearly relating the perturbed kinematics to the perturbed resistance matrices of the filament,  which themselves are linear functions of the deformation. Once inverted, this procedure gives access to the perturbation in swimming kinematics.
 
\subsection{Relevant forces and torques} \label{forcesandtorques}
Using the term  `head' to mean either the cell body of the bacterium or the magnetised head of the artificial motor, the  hydrodynamic force, $\bv{F}_{h,visc}$,  and torque with respect to the origin, $\bv{M}_{h,visc}$,   on the head  translating with velocity $\bv{U}$ and rotating with rate $\bomegahead$ are given in a Stokes flow by the linear relationship 
 \cite{KimKarrilaMicrohydrodynamicsBook}\begin{align}
&\bv{F}_{h,visc}=\bsym{\alpha}_{h}\cdot\bv{U} + \bsym{\beta}_{h}\cdot\bomegahead, \qquad 	\\
&\bv{M}_{h,visc}=\bsym{\gamma}_{h}\cdot\bv{U}+ \bsym{\delta}_{h}\cdot\bomegahead,
\end{align}
where 
 $\bsym{\alpha}_{h}, \bsym{\beta}_{h}, \bsym{\gamma}_{h}$ and $\bsym{\delta}_{h}$  
 are  the resistance matrices for a spherical head of radius $\ahead$. 
With respect to the centre of the head these are
\begin{align}
\bsym{\alpha}_{h}&=-6\pi \mu \ahead\bsym{1},   &\bsym{\beta}_{h}&=\bsym{0},\\
\bsym{\gamma}_{h}&=\bsym{0},  
&\bsym{\delta}_{h}&=-8\pi \mu \ahead^3\bsym{1},
\end{align}
where $\bv{1}$ denotes the identity matrix.

If the centre of the  head is at  position $\bvrhead$ with respect to the origin,
then the hydrodynamic torque on the head has an extra term given by $\bvrhead\wedge\bv{F}_{h,visc}$. This can be seen by writing the torque with respect to the origin, $\int_S \bv{r} \wedge (\boldsymbol{\sigma} \cdot \bv{n}) \dd S $, where $\boldsymbol{\sigma}$ is the viscous stress tensor and $\bv{n}$ is the normal to the surface $S$ of the head pointing outwards,   as 
$\int_S (\bv{r}-\bvrhead) \wedge (\boldsymbol{\sigma} \cdot \bv{n}) \dd S + \int_S \bvrhead \wedge (\boldsymbol{\sigma} \cdot \bv{n}) \dd S$, with the first integral being the torque with respect to the centre of the head $\bvrhead$, and the second integral simplifying to $\bvrhead\wedge\int_S (\boldsymbol{\sigma} \cdot \bv{n}) \dd S = \bvrhead\wedge\bv{F}_{h,visc}$.
This adds the extra terms 
$\tilde{\bsym{\gamma}}_{h} =\left(\Epsilontensor\cdot\bvrhead\right)\cdot \bsym{\alpha}_{h}$ and $\tilde{\bsym{\delta}}_{h}=\left(\Epsilontensor\cdot\bvrhead\right) \cdot \bsym{\beta}_{h}$
 to $\bsym{\gamma}_{h}$ and $\bsym{\delta}_{h}$ respectively,
 where  the third rank tensor $\Epsilontensor$ is the Levi-Civita tensor with components $\epsilon_{ijk}$ such that $\left(\Epsilontensor\cdot\bv{r}\right)\cdot\bsym{\omega}=\bv{r}\wedge\bsym{\omega}$.  
 In the case of a spherical head, these extra terms become
 $-6\pi \mu \ahead\left(\Epsilontensor\cdot\bvrhead\right)$ and $-8\pi \mu \ahead^3\left(\Epsilontensor\cdot\bvrhead\right)$ respectively.
 We will later focus on the case of a spherical head whose centre lies on the axis of a helical filament that is rotating and translating about its axis, in which case the  cross products of $\bvrhead\wedge \bv{U}$ and $\bvrhead\wedge\bomegahead$ vanish, and hence also the components of the extra terms $\tilde{\bsym{\gamma}}_{h}, \tilde{\bsym{\delta}}_{h}$  along the relevant axis also vanish.

We can write similar expressions for the force and torque (with respect to the origin) acting on the filament using Eqs.~\ref{Fintegral} and \ref{Mintegral},
\begin{align}
\Finternal (0)=\bcurlyAfil.\bv{U} + \bcurlyBfil.\bomegafil,  \\
\Mbending(0) + \bv{r}(0)\wedge \Finternal(0) =\bcurlyCfil.\bv{U} + \bcurlyDfil.\bomegafil, 
\end{align}
where $\bcurlyAfil, \bcurlyBfil, \bcurlyCfil, \bcurlyDfil $ are the resistance matrices of the filament
\begin{align}
\bcurlyAfil&=	
\resperp 
\int_{0}^{L} \left[-\bsym{1} +(1-\rho)\matd{3} \matd{3} \right]\dd s , \label{Aresmatrfil} \\
\bcurlyBfil&=
\resperp \int_{0}^{L} \left[\Epsilontensor\cdot\bv{r}  - (1-\rho)\matd{3} (\matd{3}\wedge \bv{r}) \right] \dd s , \label{Bresmatrfil} \\
\bcurlyCfil&=
\resperp 
\int_{0}^{L} \left[\Epsilontensor\cdot\bv{r}  - (1-\rho)(\matd{3} \wedge \bv{r})\matd{3}\right] \dd s  =  \bcurlyBfil^\intercal ,\label{Cresmatrfil}  \\
\bcurlyDfil&=
\resperp \int_{0}^{L} \left[\bv{r}\bv{r} - |\bv{r}|^2 \bsym{1} + (1-\rho)(\matd{3}\wedge \bv{r})(\matd{3}\wedge \bv{r})\right]\dd s 
 - \resrot \int_{0}^{L} \matd{3} \matd{3} \dd s, \label{Dresmatrfil} 
\end{align}
and the third rank tensor $\Epsilontensor$ in Eqs.~\ref{Bresmatrfil} and \ref{Cresmatrfil} is the Levi-Civita tensor with components $\epsilon_{ijk}$ such that     $\left(\Epsilontensor\cdot\bv{r}\right)\cdot\bsym{\omega}=\bv{r}\wedge\bsym{\omega}$. 
Notice that $\bsym{\curlyA}=\bsym{\curlyA}^\intercal,~\bsym{\curlyD}=\bsym{\curlyD}^\intercal,~\bsym{\curlyB}=\bsym{\curlyC}^\intercal$, as expected for Stokes flows \cite{KimKarrilaMicrohydrodynamicsBook}.

\begin{figure}
	\includegraphics[width=0.5\columnwidth]{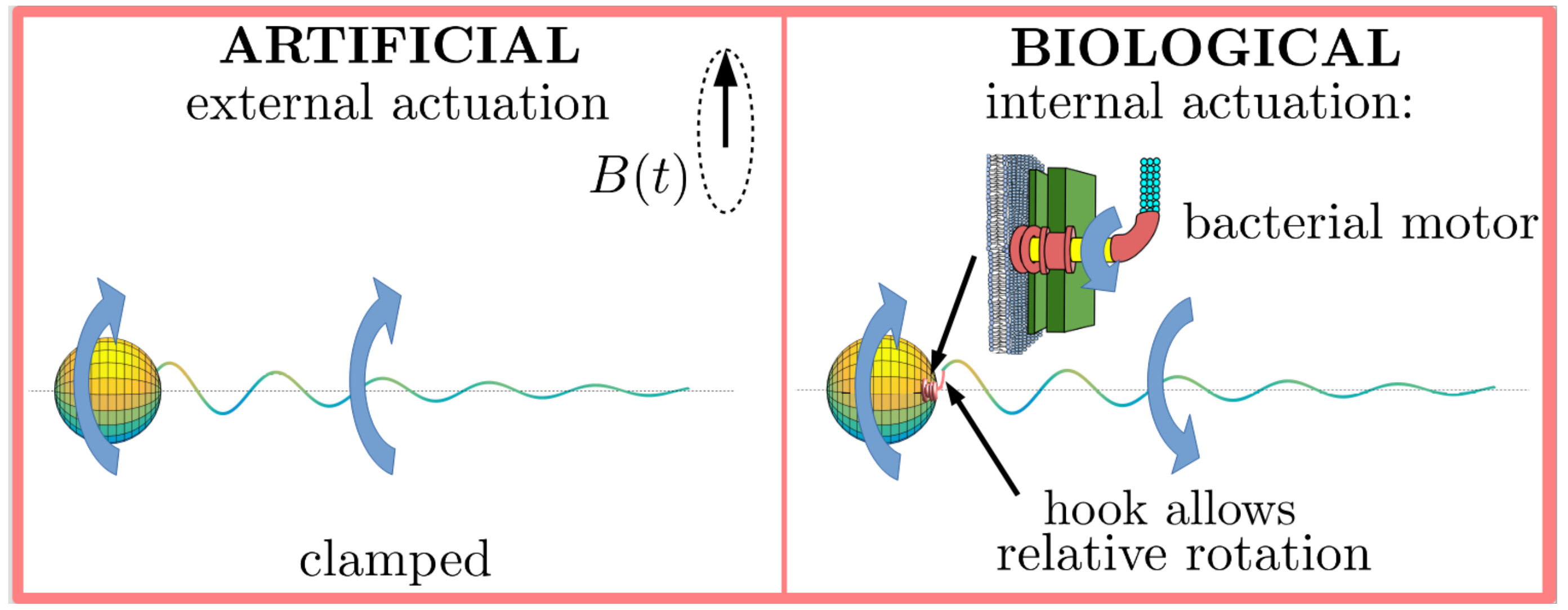}
	\caption{ Two applications of the model developed in this paper. 
	Left: Artificial motors driven by a rotating magnetic field have a magnetised head onto which a flexible elastic filament  is clamped and used for propulsion; the head and filaments translate and rotate together. 
 Right:  Flagellar filaments of bacteria  actuated by a rotary motor  embedded in the cell;  the head and filaments translate  together but they rotate in opposite direction  
  (diagram of the bacterial motor adapted from  Wikimedia Commons, Mgaetani, 2015)
}\label{FigArtificialandBacterialDiagram}
\end{figure}

In the case of artificial propellers, the head and filament rotate at the same rate, so we may  define overall resistance matrices for the head and filament together which are just the sums of the corresponding matrices
\begin{align}
\bsym{\curlyA}&=	
\bcurlyAfil+ \bsym{\alpha}_{h}, \label{Aresmatr} \\
\bsym{\curlyB}&=
\bcurlyBfil+ \bsym{\beta}_{h}, \label{Bresmatr} \\
\bsym{\curlyC}&=
\bcurlyCfil + \bsym{\gamma}_{h} =  \bsym{\curlyB}^\intercal ,\label{Cresmatr}  \\
\bsym{\curlyD}&=
\bcurlyDfil + \bsym{\delta}_{h}. \label{Dresmatr} 
\end{align}
In the case of a bacterium, the rotation rate of the head (i.e.~the cell body) is different from that of the filament; in fact the head rotates in the opposite direction in order to satisfy the overall torque balance.	The difference in   actuation of the filament  in the artificial and biological applications is summarised in Fig.~\ref{FigArtificialandBacterialDiagram} and we now  consider the force and torque balance for each case separately.

\subsection{Artificial propellers with a filament clamped on a magnetised head} \label{Artifpropsect}

We first focus on the artificial propeller  actuated by a rotating magnetic field  exerting a torque on the magnetised head on which the filament is clamped (Fig.~\ref{FigArtificialandBacterialDiagram}, left).
The internal force acting through the cross-section  at $s=0$, transmitted from the filament to the head, needs to balance the viscous hydrodynamic force, $\bv{F}_{h,visc}$, resisting the motion of the head.  Although the same is true for the axial torque balance on the head,  care must be taken    because the bending moment, $\Mbending(0)$,    transmitted through the cross-section at $s=0$ from the filament to the head is  calculated with respect to the centre of the cross-section of the filament while viscous moments on the head and measured by its centre. The torque balance includes therefore the viscous hydrodynamic torque resisting the motion of the head, $\bv{M}_{h,visc}$, the external magnetic moment, both defined with respect to the origin, as well as the bending moment with respect to the origin, $\Mbending(0) + \bv{r}(0)\wedge\Finternal(0)$. 

The dynamic equations are therefore written as
\begin{align}
&\Finternal(0) + \bv{F}_{h,visc}=\bv{0}, \label{ForceBalance} \\
&\Mbending(0) + \bv{r}(0)\wedge\Finternal(0) + \bv{M}_{h,visc} + \bv{M}_{mag} =\bv{0} . \label{TorqueBalance}
\end{align}
Using the results of Sec.~\ref{forcesandtorques}, we see that the system to be solved in order to find $\bv{U},\bsym{\omega}$ in matrix form is
\begin{align}
\begin{pmatrix}
\bsym{\curlyA}  &\bsym{\curlyB}\\
\bsym{\curlyC}  &\bsym{\curlyD}
\end{pmatrix}
\begin{pmatrix}
\bv{U}\\
\bsym{\omega}
\end{pmatrix}
=-
\begin{pmatrix}
\bv{0}\\
\bv{M}_{mag}
\end{pmatrix}. \label{systemforartificial}
\end{align}
The leading-order kinematics are given by inverting the linear system
\begin{align}
\begin{pmatrix}
\uprefc{\bsym{\curlyA}}  &\uprefc{\bsym{\curlyB}}\\
\uprefc{\bsym{\curlyC}}  &\uprefc{\bsym{\curlyD}}
\end{pmatrix}
\begin{pmatrix}
\uprefc{\bv{U}}\\\uprefc{\bsym{\omega}}\end{pmatrix}=-\begin{pmatrix}\bv{0}\\\bv{M}^{(0)}_{mag}\end{pmatrix}. \label{artifmatrixeqnleading}
\end{align}
For the first-order perturbation one needs to evaluate the perturbations to the resistance matrices
\begin{align}\label{perturbofresistance matrices}
\delta \bsym{\curlyA}&=\resperp(1-\rho) 
\int_{0}^{L} \left[\uprmatd{3} (\delta\matd{3}) + (\delta\matd{3}) \uprmatd{3}\right]\dd s,\\
\delta \bsym{\curlyB}&= \resperp \int_{0}^{L} \bigg[\Epsilontensor\cdot\delta\bv{r}   - (1-\rho)(\delta\matd{3}) (\uprmatd{3}\wedge \uprefc{\bv{r}}) - (1-\rho)\uprmatd{3} \delta(\matd{3}\wedge \bv{r}) \bigg] \dd s,\\
\delta \bsym{\curlyC}&=(\delta \bsym{\curlyB})^\intercal,\\
\delta \bsym{\curlyD}
&=\resperp\int_{0}^{L}\bigg\{  \uprefc{\bv{r}}(\delta\bv{r}) +(\delta\bv{r})\uprefc{\bv{r}}  - 2(\uprefc{\bv{r}}.\delta\bv{r}) \bsym{1}\\
&\qquad\qquad\quad + (1-\rho)\left[(\delta(\matd{3}\wedge \bv{r}))(\uprmatd{3}\wedge \uprefc{\bv{r}})
+(\uprmatd{3}\wedge \uprefc{\bv{r}})(\delta(\matd{3}\wedge \bv{r}))\right]\bigg\}\dd s \nonumber\\
&\quad-\resrot\int_{0}^{L} \left[\uprmatd{3} (\delta\matd{3}) + (\delta\matd{3}) \uprmatd{3}\right]\dd s.
\end{align}
The next order perturbation to the force and torque balance in Eq.~\ref{artifmatrixeqnleading} is
\begin{align}
\begin{pmatrix}
\uprefc{\bsym{\curlyA}}  &\uprefc{\bsym{\curlyB}}\\
\uprefc{\bsym{\curlyC}}  &\uprefc{\bsym{\curlyD}}
\end{pmatrix}\begin{pmatrix}\delta\bv{U}\\\delta\bsym{\omega}\end{pmatrix}
&=-\begin{pmatrix}
\delta\bsym{\curlyA}  &\delta\bsym{\curlyB}\\
\delta\bsym{\curlyC}  &\delta\bsym{\curlyD}
\end{pmatrix}\begin{pmatrix}\uprefc{\bv{U}}\\\uprefc{\bsym{\omega}}\end{pmatrix}
-\binom{\bv{0}}{ \delta\bv{M}_{mag}}\\
&=-\begin{pmatrix}
\delta\bsym{\curlyA}  &\delta\bsym{\curlyB}\\
\delta\bsym{\curlyC}  &\delta\bsym{\curlyD}
\end{pmatrix}\begin{pmatrix}
\uprefc{\bsym{\curlyA}}  &\uprefc{\bsym{\curlyB}}\\
\uprefc{\bsym{\curlyC}}  &\uprefc{\bsym{\curlyD}}
\end{pmatrix}^{-1}\begin{pmatrix}\bv{0}\\-\bv{M}^{(0)}_{mag}\end{pmatrix}
-\begin{pmatrix}\bv{0}\\ \delta\bv{M}_{mag}\end{pmatrix}. \label{pertmatreq}
\end{align}
For a more concise notation let us use $\resist$ to denote the large resistance matrix and $\motil$ to denote its inverse, the large motility matrix
\begin{align}
\resist=\begin{pmatrix}
\bsym{\curlyA}  &\bsym{\curlyB}\\
\bsym{\curlyC}  &\bsym{\curlyD}
\end{pmatrix},\quad 
\motil=\resist^{-1}.
\end{align}
Inverting Eqs.~\ref{artifmatrixeqnleading} and \ref{pertmatreq} then leads to
 \begin{align}
\begin{pmatrix}\uprefc{\bv{U}}\\\uprefc{\bsym{\omega}}\end{pmatrix}
&=-\uprefc{\motil}\cdot\begin{pmatrix}\bv{0}\\\bv{M}^{(0)}_{mag}\end{pmatrix},\label{leadingUOmegaartificialgeneral}\\
\begin{pmatrix}\delta\bv{U}\\\delta\bsym{\omega}\end{pmatrix}&=\uprefc{\motil} \cdot(\delta \resist) \cdot	\uprefc{\motil}\cdot\begin{pmatrix}\bv{0}\\\bv{M}^{(0)}_{mag}\end{pmatrix}
-\uprefc{\motil}\cdot \begin{pmatrix}\bv{0}\\ \delta\bv{M}_{mag}\end{pmatrix}. \label{pertUOmegaartificialgeneral}
\end{align}

For analytically tractable calculations, we focus on long chiral filaments that have the $z$-axis as their long axis of rotation. As a result we can assume translation along and rotation about the $z$-axis, i.e.~$\bv{U}=U\vez,~  \bsym{\omega}=\omega\vez$, and only consider the axial components of the force and torque balances. For example, the $z$-component of $\bsym{\curlyA}\cdot\bv{U}$ in Eq.~\ref{systemforartificial} reduces to just $\curlyA_{zz} U$. In this one-dimensional  limit,  we will drop the $zz$ indices for notation convenience and use $\curlyA$ to mean $\curlyA_{zz}$, and similarly for other matrix components.

The leading-order result of  Eq.~\ref{leadingUOmegaartificialgeneral} describes a rigid filament in its reference configuration with both translation and rotation proportional to the external moment, $M_{mag}$, as
\begin{align}
\uprefc{U}&=-\frac{\curlyB}{\curlyA} \uprefc{\omega}, \\
\uprefc{U}&=\frac{\curlyB}{\curlyA\curlyD-\curlyB^2}M_{mag}, \label{rigidU} \\
\uprefc{\omega}&=-\frac{\curlyA}{\curlyA\curlyD-\curlyB^2}M_{mag}, \label{rigidomega} 
\end{align}	
while for the next-order correction we obtain
\begin{align} 
\delta U &=\frac{\big[-\curlyB\curlyD(\delta\curlyA)+(\curlyA\curlyD+\curlyB^2)(\delta \curlyB)- \curlyA\curlyB(\delta\curlyD)\big]}{\left(\curlyA\curlyD-\curlyB^2\right)^2} M_{mag}
,
 \label{perturbvelartificial}\\
\delta \omega &=\frac{\big[\curlyB^2(\delta\curlyA)-2\curlyA\curlyB(\delta \curlyB)+ \curlyA^2(\delta\curlyD)\big]}{\left(\curlyA\curlyD-\curlyB^2\right)^2} 
M_{mag}.
 \label{perturbomegaartificial}
\end{align}

\subsection{Biological locomotion with a filament rotated by a motor} \label{biolocom}
In the case of swimming bacteria (Fig.~\ref{FigArtificialandBacterialDiagram}, right), a motor embedded in the cell wall applies a constant torque, $\Mmotor$, via the short hook in order to rotate a long filament. 
The rotation rate of the filament in the laboratory frame is $\bomegafil$ while 
the head rotates at a different rate denoted by $\bomegahead$. Both the head and the filament translate at the same velocity, $\bv{U}$.

In this case, there are three dynamic balances to consider:  the overall force balance as well as the torque balances on the filament and the head
\begin{align}
&\Finternal(0) + \bv{F}_{h,visc}=0, \label{ForceBalancezBio} \\
&\Mbending(0) + \bv{r}(0)\wedge\Finternal(0) + \bMmotor =0,  \label{TorqueBalancezBioFil}\\
&\bv{M}_{h,visc}  - \bMmotor =0.  \label{TorqueBalancezBioHead}
\end{align}
Substituting in the  terms of the resistance matrices we obtain
\begin{align}
&\left(\bcurlyAfil + \bsym{\alpha}_{h}\right).\bv{U} + \bcurlyBfil.\bomegafil + \bsym{\beta}_{h}.\bomegahead =0,  \label{ForceBalancezBiointermsofResistanceMatrices}\\
&\bcurlyCfil.\bv{U} + \bcurlyDfil.\bomegafil + \bMmotor  =0 , \label{TorqueBalancezBioFilintermsofResistanceMatrices}\\
&\bsym{\gamma}_{h}.\bv{U} + \bsym{\delta}_{h}.\bomegahead - \bMmotor  =0.  \label{omegaheadbio}
\end{align}
Inverting Eq.~\ref{omegaheadbio} for $\bomegahead$ and substituting
in Eq.~\ref{ForceBalancezBiointermsofResistanceMatrices} leads to
\begin{align}
&\left(\bcurlyAfil + \bsym{\alpha}_{h}-\bsym{\beta}_{h}.\bsym{\delta}^{-1}_{h}.\bsym{\gamma}_{h}\right).\bv{U} + \bcurlyBfil.\bomegafil =- \bsym{\beta}_{h}.\bsym{\delta}^{-1}_{h}.\bMmotor.
\end{align}
The system to be solved can thus be written in matrix form as
\begin{align}
&\begin{pmatrix}
\left(\bcurlyAfil + \bsym{\alpha}_{h}-\bsym{\beta}_{h}.\bsym{\delta}^{-1}_{h}.\bsym{\gamma}_{h}\right)& \bcurlyBfil\\\bcurlyCfil&\bcurlyDfil
\end{pmatrix}\cdot 
\begin{pmatrix}\bv{U}\\\bomegafil\end{pmatrix}
=-\begin{pmatrix}\bsym{\beta}_{h}.\bsym{\delta}^{-1}_{h}\bMmotor\\\bMmotor\end{pmatrix},
\end{align}
which in the special case of a spherical head  further reduces to
\begin{align}
\begin{pmatrix}
\bcurlyAfil + \bsym{\alpha}_{h}& \bcurlyBfil\\\bcurlyCfil&\bcurlyDfil
\end{pmatrix}\cdot \begin{pmatrix}\bv{U}\\\bomegafil\end{pmatrix}
=-\begin{pmatrix}\bv{0}\\\bMmotor\end{pmatrix} .\label{systemforbio}
\end{align}

Notice how the system in Eq.~\ref{systemforbio} is mathematically similar to that for the artificial motor in Eq.~\ref{systemforartificial}
if one makes the substitution $\bsym{\curlyA}\to \bsym{\curlyA}$, $\bsym{\curlyB}\to \bcurlyBfil$, $\bsym{\curlyC}\to \bcurlyCfil$, $\bsym{\curlyD}\to \bcurlyDfil$, and $\bv{M}_{mag}\to \bMmotor$. Making these substitutions in the projected one-dimensional versions of the artificial motor equations, 
Eqs.~\ref{rigidU}-\ref{perturbomegaartificial}, allows  to obtain the corresponding ones for a bacterium  translating along, and rotating about, the axis of its long chiral filament. The rigid result is 
\begin{align}
\uprefc{U}&=-\frac{\curlyBfil}{\curlyA} \uprefc{\omegafil}, \\
\begin{pmatrix}\uprefc{U} \\\uprefc{\omegafil}	\end{pmatrix}&=\frac{\Mmotor}{\curlyA\curlyDfil-\curlyBfil^2} 
\begin{pmatrix}\curlyBfil\\-\curlyA\\
\end{pmatrix}, \label{rigidkinembacteria}
\end{align}	
while the first correction is given by
\begin{align} 
\delta U &=\frac{\Mmotor}{\left(\curlyA\curlyDfil-\curlyBfil^2\right)^2}
\left[-\curlyBfil\curlyDfil(\delta\curlyA)
+(\curlyA\curlyDfil+\curlyBfil^2)(\delta \curlyBfil)
- \curlyA\curlyBfil(\delta\curlyDfil)\right], \label{pertUbio}\\
\delta \omegafil &=\frac{\Mmotor\big[\curlyBfil^2(\delta\curlyA)-2\curlyA\curlyBfil(\delta \curlyBfil)+ \curlyA^2(\delta\curlyDfil)\big]}{\left(\curlyA\curlyDfil-\curlyBfil^2\right)^2} \cdot\label{perturbomegabiol}
\end{align}

\section{Propulsion by elastic helical filaments} \label{helical}
Up to this point in the paper, we have formulated everything in terms of an arbitrary shape of the filament in the rigid limit.  
In this section we focus on the case of  filaments with helical geometry, as this is the most common shape in both cases of interest discussed thus far, and we carry out the calculations of the expressions we formulated in \S \ref{General Model}. As done in \S \ref{Artifpropsect} and \ref{biolocom}, in order to keep the analytical calculations tractable we will only study translation along, and rotation about, the $z$-axis of the helical shape, and not the full three-dimensional (3D) motion. This is  an appropriate  limit to consider  if the  filament is long enough   to not wobble \cite{YiLaugawobbling}.

\subsection{Calculating the deformation} \label{helicaldeform}
\subsubsection{Reference configuration: Centreline geometry}

\begin{figure} 
		\includegraphics[width=0.55\columnwidth]{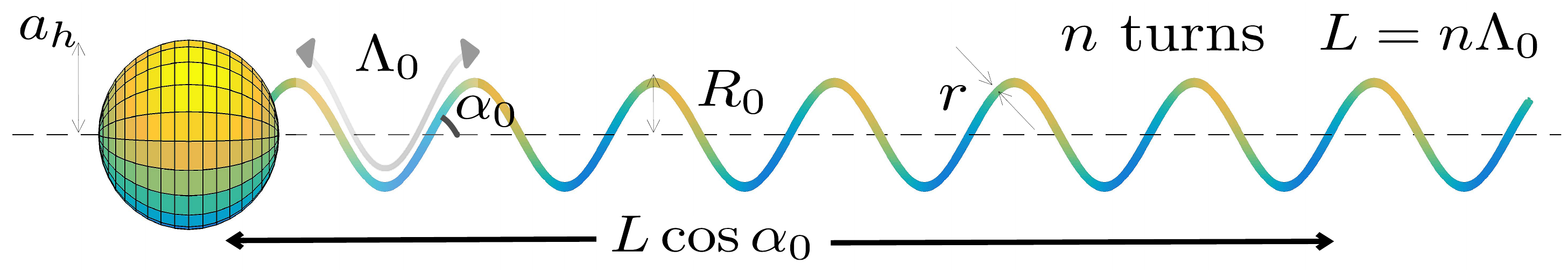}	
		\caption{Reference configuration of a helical filament of helix angle $\refc{\alpha}$, radius $\refc{R}$, $n$ number of turns and filament radius $r$. The total length along the filament is $L$, assumed to be much larger than the size of the head, $\ahead$.}\label{helicalgeometry}
	\end{figure}

	We now compute the steady-state  perturbation results where we assume the actuation from the rotating magnetic field or bacterial motor is weak enough, or the helix stiff enough,  that the helix deforms only slightly. 	The reference configuration is a long, uniform, stress-free rod in the shape of a helix of uniform pitch $\refc{P}$, radius $\refc{R}$, helix angle $\refc{\alpha}$, such that $\tan{\refc{\alpha}}=(2 \pi \refc{R}/\refc{P})$ and its axis is aligned with the $z$ axis. We define the chirality index $h$ which takes the value $\pm 1$  according to whether the helix is right-handed (RH, $h=1$) or left-handed (LH, $h=-1$). 
	In its reference configuration, the centreline of the helix is therefore given by
		\begin{align}
	\refc{\bv{r}}(s)=\bigg[ \refc{R}\cos\sphase, h\refc{R}\sin\sphase, \frac{\refc{P}s}{\refc{\Lambda}} \bigg], \label{centreline}
	\end{align}
	where $\refc{\Lambda}=\sqrt{\refc{P}^2+4\pi^2\refc{R}^2} ~$  is the helix wavelength measured along the arclength $s$.  	This geometry is illustrated in Fig.~\ref{helicalgeometry}.

	Taking the material frame $\{\uprefc{\matd{1}},\uprefc{\matd{2}},\uprefc{\matd{3}}\}$ to coincide with the Serret-Frenet frame $(\bv{n}, \bv{b}, \bv{t})$ similarly to Ref.~\cite{KimPowers2005HelixDeformation}, we 
	use $\partial_s\refc{\bv{r}}(s)=\bv{t}=\uprefc{\matd{3}}$, $\partial _s \bv{t}=\refc{\curvature} \bv{n}$ and $\bv{b}=\bv{t}\wedge\bv{n}=\uprefc{\matd{2}}$
	to obtain 	
	\begin{align}
	\uprefc{\matd{3}}&=\bigg( -\sin\refc{\alpha}\sin\sphase ,  h\sin\refc{\alpha}\cos\sphase, \cos\refc{\alpha} \bigg),\\
	\uprefc{\matd{1}}&=\bigg(-\cos\sphase,-h\sin\sphase, 0\bigg),\\
	\uprefc{\matd{2}}&=\bigg( h\cos\refc{\alpha}\sin\sphase,-\cos\refc{\alpha}\cos\sphase, h\sin\refc{\alpha}\bigg),
	\end{align}
\normalsize
	with
\begin{align}
\refc{\curvature}=\sin^2 \refc{\alpha} /\refc{R},
\end{align}
and, using $\partial_s\bv{b}=-\refc{\twist}\bv{n}$,
\begin{align}
\refc{\twist}= h\sin\refc{\alpha}\cos\refc{\alpha}/\refc{R}.  
\end{align}
	\normalsize
	Identifying the Serret-Frenet equations 
	\begin{align}
	\partial_s \begin{pmatrix}\uprefc{\matd{1}}\\\uprefc{\matd{2}}\\\uprefc{\matd{3}}\end{pmatrix}
	=\begin{pmatrix}0& \refc{\twist}& -\refc{\curvature} \\ -\refc{\twist}&  0&  0\\\refc{\curvature}&  0&  0&\end{pmatrix} \begin{pmatrix}\uprefc{\matd{1}}\\\uprefc{\matd{2}}\\\uprefc{\matd{3}}\end{pmatrix},
	\end{align}
	with $\partial_s \uprefc{\matd{i}}=\uprefc{\darboux}\wedge\uprefc{\matd{i}}$, 
	gives the components of the Darboux vector
	\begin{align}
	\uprefc{\darboux}&=\sum\limits_{i}^{}\uprefc{\cdarboux{i}}\uprefc{\matd{i}},\\
	\uprefc{\cdarboux{1}}&=0,  \qquad \uprefc{\cdarboux{2}}=\refc{\curvature}, \qquad \uprefc{\cdarboux{3}}=\refc{\twist}.
	\end{align}

\begin{figure}[t] 
\includegraphics[width=0.3\columnwidth]{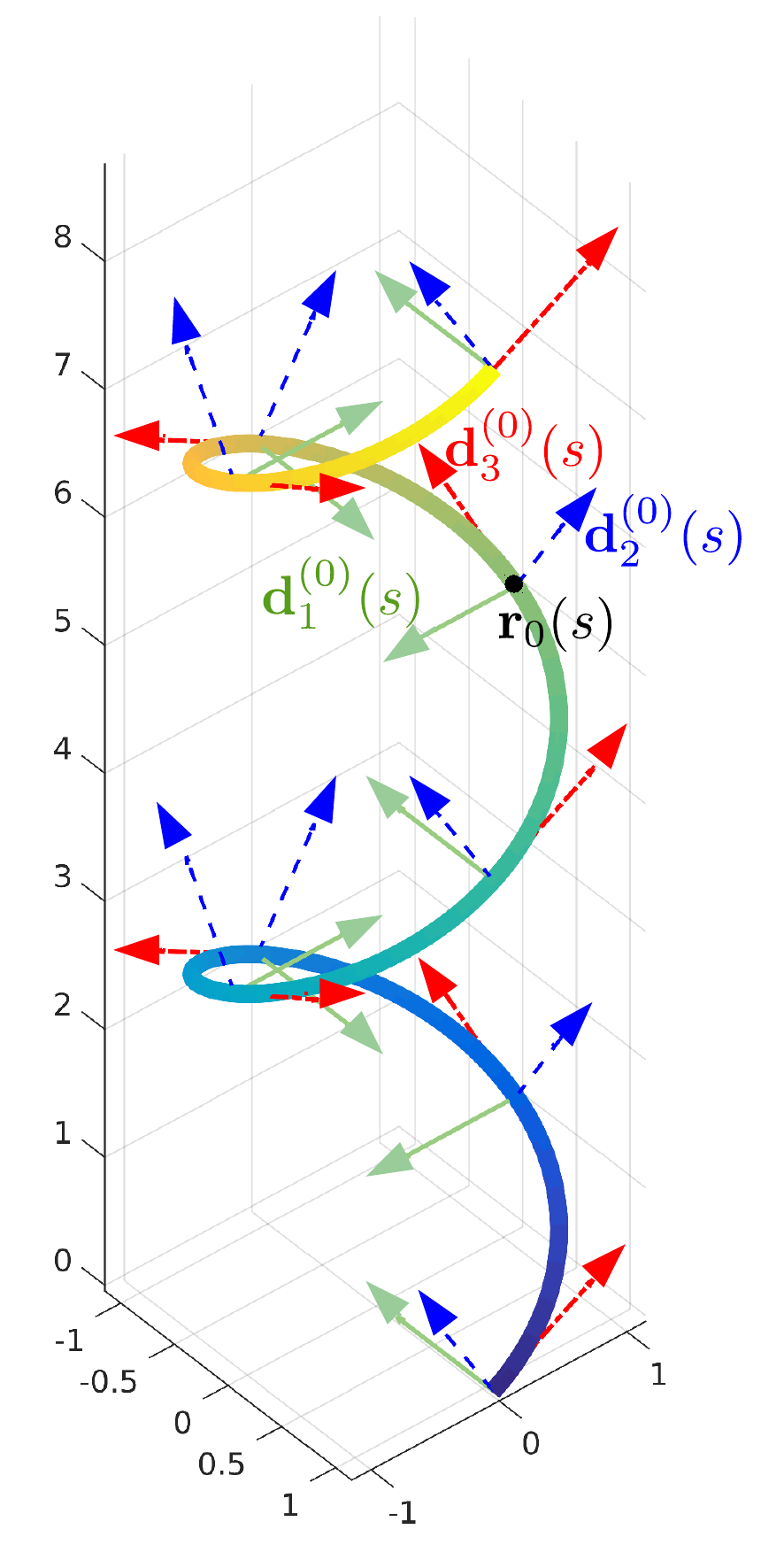}	
		\caption{The material frame of the rigid helix shown at various positions along the filament. The  arrows show the tangent to the centreline ($\matd{3}$, dash-dotted red), the  normal ($\matd{1}$, solid green) and the binormal ($\matd{2}$, dashed blue).}\label{tnbhelixsurfplot}
	\end{figure}

	The material frame at various positions along the filament is illustrated in Fig.~\ref{tnbhelixsurfplot}, with the  tangent, normal and binormal vectors shown in dash-dotted red, solid green, and dashed blue respectively.

	\subsubsection{Hydrodynamic load}
	Actuated by the magnetic field, once steady state has been reached the helix rotates with uniform angular velocity $\omegafil \bv{e_z}$ and translates at uniform velocity $U\bv{e_z}$. 
While it is deformed from its reference configuration shape due to the forces and torques acting on it,  it does undergo rigid body motion as its shape is no longer changing. We consider the frame in which the helix is stationary.
	The relative velocity between the helix and the fluid is given by
	\begin{align}
	\bv{V_{rel}}
	&=U\bv{e_z}+\bom\wedge\bv{r}(s)=\begin{pmatrix}-\omega y(s) \\ \omega x(s) \\U\end{pmatrix}
	\approx \begin{bmatrix}-\omega h\refc{R}\sin\sphase \\ \omega \refc{R}\cos\sphase \\U\end{bmatrix}.
	\end{align}
	where we approximated the  shape to the reference helical one and used the rigid helix kinematics. This allows us to calculate the leading-order   force density using  Eq.~\ref{RFT} and noting that $\uprefc{\matd{3}}\cdot\uprefc{\bv{V}_{rel}}=\uprefc{\omegafil} h \refc{R} \sin\refc{\alpha} + \uprefc{U} \cos\refc{\alpha}$,
	\begin{align}
\forcedens(s)=& \resperp\begin{bmatrix}\uprefc{\omegafil} h\refc{R}\sin\sphase \\ -\uprefc{\omegafil} \refc{R}\cos\sphase \\-\uprefc{U}\end{bmatrix}  
+ \resperp (1-\rho)\left[\uprefc{\omegafil} h \refc{R} \sin\refc{\alpha} + \uprefc{U} \cos\refc{\alpha}\right] \begin{bmatrix}
-\sin\refc{\alpha}\sin\sphase \\ \quad h\sin\refc{\alpha}\cos\sphase \\ \cos\refc{\alpha}
\end{bmatrix}.	\end{align}
	Rearranging gives
	\begin{align}
	\forcedens(s)\approx\begin{bmatrix} \quad A_x \sin\sphase \\ -h A_x	\cos\sphase \\ A_z	\end{bmatrix},
	\end{align}
	where
\begin{align}
A_x&=\resperp \left[-(1-\rho)\sin\refc{\alpha}\cos\refc{\alpha}\uprefc{U} + (\cos^2\refc{\alpha} + \rho \sin^2\refc{\alpha})\uprefc{\omegafil} h \refc{R} \right], \label{Ax}\\
A_z&=\resperp \left[-(\sin^2\refc{\alpha}+\rho \cos^2\refc{\alpha})\uprefc{U} + (1-\rho)\sin\refc{\alpha}\cos\refc{\alpha}\uprefc{\omegafil} h \refc{R} \right]. \label{Az}
\end{align}
 
	Integrating the first Kirchhoff equation, $\partial_s \Finternal + \forcedens=0$, with the boundary condition that $\Finternal(L)={\bf 0}$ at the free end gives access to  the distribution of  internal force as
	\begin{align}
	\Finternal(s)&=-\int_{}^{s}\dd s'\uprefc{\forcedens}
	=\begin{pmatrix} &\frac{\refc{R}}{\sin\refc{\alpha}}& A_x &&\left[\cos\sphase - \cos\lphase\right] \\
h&\frac{\refc{R}}{\sin\refc{\alpha}}& A_x &&\left[\sin\sphase - \sin\lphase\right] \\
 &-& A_z &&(s-L)\hfill\hfill                          \end{pmatrix}.
\end{align}
Similarly, integrating the second Kirchhoff equation, $\partial_s\Mbending+\matd{3}\wedge\Finternal+\Nviscous=\bv{0}$,  
with the free-end boundary condition, $\Mbending(L)=\bf 0$, allows to compute  the bending moment as
\begin{alignat}{4}
\Mbending(s)
&&=&& A_x\refc{R}&
\begin{bmatrix}
h\cot\refc{\alpha}\left(-\Rsinfrac[\cos\sphase-\cos\lphase]-(s-l)\sin\lphase    \right) \\
-\cot\refc{\alpha}\left(\quad\Rsinfrac[\sin\sphase-\sin\lphase]-(s-L)\cos\lphase
\right) \\
h\left[(s-L)\quad-\quad\Rsinfrac\sin\slphasedif \right]
\end{bmatrix}\nonumber\\
&&~&&+ A_z \refc{R}&
\begin{bmatrix}
h&\left[\Rsinfrac\left(\cos\sphase-\cos\lphase\right)+(s-L)\sin\sphase\right]\\
&\left[\Rsinfrac\left(\sin\sphase-\sin\lphase\right)-(s-L)\cos\sphase \right] \\
& 0
\end{bmatrix}\nonumber\\
&&~&&+ \resrot \omega \cos\refc{\alpha}&
\begin{bmatrix}
&\refc{R}\left(\cos\sphase-\cos\lphase\right)\\
h& \refc{R}\left(\sin\sphase-\sin\lphase\right) \\
& (s-L)\cos\refc{\alpha}
\end{bmatrix}.
\label{Mbendinglongexpression}
\end{alignat}

\subsubsection{Leading-order kinematics}\label{leadingorderkinematics} 
For a stiff elastic filament, the leading-order kinematics are given by the rigid limit.
The $z$-components of the expressions for the total force and torque give the $zz$-components of the resistance matrices for a rigid helical filament,
\begin{align}
\curlyAfil&=-\resperp L (\sin^2\refc{\alpha}+\rho \cos^2\refc{\alpha}),   \label{Afil} \\
\curlyBfil&=\resperp h\refc{R}L (1-\rho)\sin\refc{\alpha}\cos\refc{\alpha}, \label{Bfil}\\
\curlyDfil&=- \resperp \refc{R}^2 L (\cos^2\refc{\alpha}+\rho \sin^2\refc{\alpha}) - \resrot L \cos^2\refc{\alpha}. \label{Dfil}
\end{align}
The $zz$-components of the overall resistance matrices for both the helical filament and the spherical head are
\begin{align}
\curlyA&=-\left[ \resperp L (\sin^2\refc{\alpha}+\rho \cos^2\refc{\alpha}) + 6\pi\mu \ahead \right], \label{A}\\
\curlyB&=\resperp h\refc{R}L (1-\rho)\sin\refc{\alpha}\cos\refc{\alpha}, \label{B}\\
\curlyD&=-\left[ \resperp \refc{R}^2 L (\cos^2\refc{\alpha}+\rho \sin^2\refc{\alpha})+ \resrot L \cos^2\refc{\alpha} + 8\pi\mu \ahead^3 \right].  \label{D}
\end{align}

\paragraph{Unified approach.} 
We now make use of the similarity in the expressions for the kinematics and their perturbation for the artificial helical motors and the swimming bacterium,
noting that the formulae for the artificial motors can be changed into those for bacteria by replacing
$\curlyD$ by $\curlyDfil$, and ${M}_{mag}$ by $\Mmotor$.
We thus write 
\begin{align}
\uprefc{U}&=\frac{M}{\curlyA\curlyD_*-\curlyB^2}\curlyB, \label{rigidkinemHelixGenForm_U} \\
\uprefc{\omega}&=-\frac{M}{\curlyA\curlyD_*-\curlyB^2}\curlyA,    \label{rigidkinemHelixGenForm_omega}
\end{align}	
where $M$ is to be substituted by either $M_{mag}$ or $M_{mot}$ for artificial motors and bacteria respectively,
and write 
\begin{align}
\curlyD_*=&-\left[ \resperp \refc{R}^2 L (\cos^2\refc{\alpha}+\rho \sin^2\refc{\alpha})+ \resrot L \cos^2\refc{\alpha} + 8\pi\mu \ahead^3\indfun \right], \label{Dstar}
\\
\curlyA\curlyD_*-\curlyB^2=&\rho\resperp^2\refc{R}^2 L^2 + 6\pi\mu \resperp \refc{R}^2 L \ahead (c^2+\rho s^2)+ \resrot L c^2 \left[ \resperp L (s^2+\rho c^2) + 6\pi\mu \ahead \right]
\nonumber\\
&
+ 8\pi \mu\ahead^3\left[ \resperp L  (s^2+\rho c^2) 
+ 6\pi\mu\ahead  \right]\indfun, \label{ADstar_B2}
\end{align}
where we use the indicator function $\indfun$ to take the value $1$ for the case of artificial motors and $0$ for bacteria. The coefficients $\curlyA$ and $ \curlyB$ are defined as in Eqs.~\ref{A} and \ref{B}.

\paragraph{Artificial motors.} 
The leading-order kinematics, given by Eqs.~\ref{rigidkinemHelixGenForm_U} and \ref{rigidkinemHelixGenForm_omega}, are obtained as
\begin{align}
\uprefc{U}&=\frac{\resperp h\refc{R}L (1-\rho)\sin\refc{\alpha}\cos\refc{\alpha} M_{mag}}{\curlyA\curlyD-\curlyB^2} , \label{rigidUhelixartifful} \\
\uprefc{\omega}&=\frac{\left[ \resperp L (\sin^2\refc{\alpha}+\rho \cos^2\refc{\alpha}) + 6\pi\mu \ahead \right]M_{mag}}{\curlyA\curlyD-\curlyB^2}, \label{rigidomegahelixartifful} 
\end{align}	
where the denominator is given explicitly  by
\begin{align}
\curlyA\curlyD-\curlyB^2 
&=\rho\resperp^2\refc{R}^2 L^2 + 6\pi\mu \resperp \refc{R}^2 L \ahead (\cos^2\refc{\alpha}+\rho \sin^2\refc{\alpha}) \nonumber\\
&\qquad+ 8\pi \mu \resperp L \ahead^3 (\sin^2\refc{\alpha}+\rho \cos^2\refc{\alpha})
+ 48\pi^2\mu^2\ahead^4 \nonumber\\
&\qquad + \resrot L \cos^2\refc{\alpha} \left[ \resperp L (\sin^2\refc{\alpha}+\rho \cos^2\refc{\alpha}) + 6\pi\mu \ahead \right]. \label{AD-B2}
\end{align}

We note that the factor $\sin\refc{\alpha}\cos\refc{\alpha}$ present in the numerator of Eq.~\ref{rigidUhelixartifful} gives vanishing speeds for helix angles $0$ or $\pi/2$; this is expected by symmetry since chirality is lost in these two limits. 
In particular, we note  that taking the limit of the helix angle to zero while keeping the total contour length $L$ and the number of turns $n$ fixed means that the helical radius $\refc{R}=L\sin\refc{\alpha}/(2\pi n)$ is also shrinking to zero. This limit gives vanishing speeds since the expression for the denominator given in Eq.~\ref{AD-B2} is non-zero due to the presence of the terms involving the head and the viscous rotational torque coefficient ($\resrot$).

The leading-order results in the limit of a filament which is slender ($r\ll L$) and long compared to the size of the magnetised head $(\ahead\ll L)$, and for a  non-vanishing helix angle are
\begin{align}
\uprefc{U}&=\frac{h (1-\rho) \sin\refc{\alpha} \cos\refc{\alpha} M_{mag}}{\rho \resperp \refc{R} L} , \label{rigidUhelixartif} \\
\uprefc{\omega}&=\frac{ (\sin^2\refc{\alpha} + \rho \cos^2 \refc{\alpha}) M_{mag}}{\rho \resperp \refc{R}^2 L},\label{rigidomegahelixartif} 
\end{align}	
with relative errors of order ${\mathcal O}\left(r^2/L^2, \ahead/L\right)$.
\paragraph{Bacteria.}
In the case of a swimming bacterium, the leading-order kinematics, given by Eqs.~\ref{rigidkinemHelixGenForm_U} and \ref{rigidkinemHelixGenForm_omega}, are
\begin{align}
\begin{pmatrix}\uprefc{U} \\\uprefc{\omegafil}	\end{pmatrix}&=\frac{\Mmotor}{\curlyA\curlyDfil-\curlyBfil^2} 
\begin{pmatrix}\resperp h\refc{R}L (1-\rho)\sin\refc{\alpha}\cos\refc{\alpha}\\\resperp L (\sin^2\refc{\alpha}+\rho \cos^2\refc{\alpha}) +6\pi\mu\ahead 
\end{pmatrix},
\label{bacteriakinemleading}
\end{align}
where the denominator is given by
\begin{align}
\curlyA\curlyDfil-\curlyBfil^2=&\rho\resperp^2\refc{R}^2 L^2 + 6\pi\mu \resperp \refc{R}^2 L \ahead (\cos^2\refc{\alpha}+\rho \sin^2\refc{\alpha})
\\
&+ \resrot L \cos^2\refc{\alpha} \left[ \resperp L (\sin^2\refc{\alpha}+\rho \cos^2\refc{\alpha}) + 6\pi\mu \ahead \right].\label{ADfil_Bfil2}
\end{align}
Here also  the speed vanishes in the limiting cases of helix angles $0$ or $\pi/2$ as expected. 
When the cell body is small  compared to the length of the slender flagellar filament, $\ahead \ll  L$, in the case of a non-vanishing helix angle we can further simplify Eq.~\ref{bacteriakinemleading} as 
\begin{align}
\uprefc{U}
&=\frac{h\Mmotor}{\resperp\refc{R}L}  \frac{(1-\rho)\sin\refc{\alpha}\cos\refc{\alpha}}{\rho}, \label{rigidUhelixbacteria} \\
\uprefc{\omegafil}
&=\frac{\Mmotor}{\resperp\refc{R}^2L}  \frac{\left(\sin^2\refc{\alpha}+\rho \cos^2\refc{\alpha}\right)}{\rho}
. \label{rigidomegahelixbacteria}
\end{align}
with relative errors of order ${\mathcal O}\left(r^2/L^2, a_h/L\right)$.

\subsubsection{Deformation for clamped Helix}
The full details of the deformation of the helix are shown in Appendix \ref{AppendixCalcDeform}. 
Under the assumption that $EI^{(1)}=EI^{(2)}=\mu J$,
we obtain at location $s/\refc{R}\gg 1$
\begin{alignat}{4} 
EI\delta\bv{r}(s)
&&\approx&& \frac{\cos^2 \refc{\alpha}}{\sin \refc{\alpha}}(A_x\refc{R})&
\begin{pmatrix}
(\frac{s^3}{6} - L\frac{s^2}{2})\cos \lphase +\Rsinfrac \frac{s^2}{2}\sin \lphase + {\mathcal O}(\refc{R}^2 s)\\
h(\frac{s^3}{6} - L\frac{s^2}{2})\sin \lphase  -h\Rsinfrac \frac{s^2}{2} \cos \lphase + {\mathcal O}(\refc{R}^2 s)\\
0
\end{pmatrix}\nonumber\\
&&~&&+ \cos\refc{\alpha}(A_z\refc{R})&
\begin{pmatrix}
-\Rsinfrac \frac{s^2}{2} \sin \lphase  + {\mathcal O}(\refc{R}^2 s)\\
h\Rsinfrac \left[\frac{s^2}{2} \cos\lphase +Ls\right]+ {\mathcal O}(\refc{R}^2 s) \\
0
\end{pmatrix}\nonumber\\
&&~&&+ \sin\refc{\alpha}(A_x\refc{R})&
\begin{pmatrix}
-\Rsinfrac (\frac{s^2}{2}-Ls) \sin \sphase  + {\mathcal O}(\refc{R}^2 s)\\
-h\Rsinfrac (\frac{s^2}{2}-Ls) \cos\sphase+ {\mathcal O}(\refc{R}^2 s) \\
0
\end{pmatrix}\nonumber\\
&&~&&- \refc{R}\begin{pmatrix}0\\0\\1\end{pmatrix} &\left\{\left(
\frac{s^2}{2}-Ls \right) \left[(A_x\refc{R})\cot \refc{\alpha} \cos \slphasedif + (A_z\refc{R})\right] + {\mathcal O}(\refc{R}^2 s)\right\}  
\nonumber\\
&&~&&+ \resrot\uprefc{\omegafil}\refc{R}\cos\refc{\alpha} &
\begin{pmatrix}
	-h\cos\refc{\alpha}\left[\frac{s^2}{2}\left(\sin\lphase+\sin\sphase\right)-L s \sin\sphase\right]  + {\mathcal O}(\refc{R} s)\\~~	~~~
	\cos\refc{\alpha}\left[\frac{s^2}{2}\left(\cos\lphase+\cos\sphase\right)-L s \cos\sphase\right]+ {\mathcal O}(\refc{R} s) \\
	-h\refc{R}~s \sin\slphasedif +{\mathcal O}(\refc{R}^2)
\end{pmatrix}.
 \label{deformation}
\end{alignat}
\normalsize

 The $z$-component of the deformation in Eq.~\ref{deformation} giving the extension/compression is in agreement with the results obtained in Ref.~\cite{KimPowers2005HelixDeformation}.
In addition, the non-zero $x,y$ components of the deformation lead to bending of the helical axis. Such bending of the axis was observed in the numerical results of Ref.~\cite{Takanoetal2003HelicFlagel}, but its origin remained unclear. We now discuss and interpret these two aspects of the deformation.

\subsubsection{Compression or extension?}\label{CompressionOrExtension}

The total amount of extension / compression is given by the $z$-component of the deformation evaluated at $s=L$,
\begin{align}
\delta z(L)
&= 
\frac{\resperp\refc{R}^2 }{EI}\left[-\uprefc{U} 
+ \left(\uprefc{\omegafil} h \refc{R}\cot\refc{\alpha}\right) \right]
\bigg[\frac{L^2}{2}-\left(\Rsinfrac\right)  L\sin\lphase\\
&\qquad\qquad\qquad\qquad\qquad\qquad\qquad\qquad
+\left(\Rsinfrac\right)^2\left(1 - \cos\lphase\right)\bigg], \\
&=
\frac{\resperp\refc{R}^2 L^2}{2 EI} \left[-\uprefc{U} + (\uprefc{\omegafil} h\refc{R}) \cot \refc{\alpha}\right]\left[1 + {\mathcal O}\left(\frac{\refc{R}}{L}\right)\right]. \label{extensioncompression}
\end{align}
\normalsize
This  recovers  exactly the result of Ref.~\cite{KimPowers2005HelixDeformation} which investigated the compression/extension of a clamped helical filament subject to a uniform translating flow or a rotating flow. 
The contribution from the axial flow is the motion of the microswimmer propelling with a positive axial velocity, $\uprefc{U}>0$, which is equivalent to the swimmer being fixed and subject to a flow $-\uprefc{U}\vez$. This  leads to compression, regardless of the helix handedness, as expected from intuition. 
In contrast, the part due to the rotational flow predicts that a helix rotating in the positive sense  about the $z$-axis $(\refc{\omega}>0)$ will be extended if it is right-handed (RH, $h>0$) and compressed if left-handed (LH, $h<0$). 

In the cases we are investigating, the kinematics are set by   the force and torque balances and $\uprefc{U}$ and $\uprefc{\omegafil}$ are not independent. Instead both are proportional to the actuating torque, as per Eqs.~\ref{rigidkinemHelixGenForm_U} and \ref{rigidkinemHelixGenForm_omega}.
A helix actuated by a positive 
 torque, $M>0$, will rotate in the positive sense ($\uprefc{\omegafil}>0$) regardless of its handedness, but will translate with a positive velocity along the $z$-axis if RH and negative if LH. 
The effect of the translation is then that a RH helix ($h>0$) actuated with a positive magnetic torque will have a positive velocity, $\uprefc{U}>0$, and hence be compressed, whereas a LH helix will move in the opposite direction, $\uprefc{U}<0$, and will thus be extended.  In contrast, the effect of rotation is the opposite to that of translation. Helices of both handedness will rotate in the positive sense, and a RH helix is extended as it is experiencing a rotating flow in the direction that  `uncoils' it, whereas a LH helix is compressed.  As a result,    translation and rotation have opposite effects in terms of   extension/compression. 

Substituting Eqs.~\ref{rigidkinemHelixGenForm_U} and \ref{rigidkinemHelixGenForm_U} in Eq.~\ref{extensioncompression} gives
	\begin{align}
	\delta z(L)&= 
	\frac{h\resperp\refc{R}^3L^2 M  \left(\rho\resperp L + 6\pi\mu \ahead  \right)\cot\refc{\alpha}}{2EI\left(\curlyA\curlyD_*-\curlyB^2\right)}~
	\curlyG, 
	\end{align}
where
\begin{align}
	\curlyG=&1
	-2\left(\frac{\refc{R}}{L\sin\refc{\alpha}}\right)  \sin\lphase
		+2\left(\frac{\refc{R}}{L\sin\refc{\alpha}}\right)^2\left(1 - \cos\lphase\right),
\end{align}
and $(\curlyA\curlyD_*-\curlyB^2)$ in the denominator is given in  Eq.~\ref{ADstar_B2}. Under the long and slender approximations and for small head, this simplifies to
\begin{align}
\delta z(L)&= 
\frac{hM\refc{R}L\cot\refc{\alpha}}{2EI} \cdot
\end{align}
As a result,  the effect of rotation dominates and a RH helix ($h>0$) is always extended whereas a LH helix is compressed.

\subsubsection{Bending of the helix axis}
Far from the clamped end, i.e.~for $s\sim L\gg \refc{R}$,  the deformation is given by
\begin{alignat}{4}
EI\delta\bv{r}(s)&&\approx&& \frac{\cos^2 \refc{\alpha}}{\sin \refc{\alpha}}(A_x\refc{R})&
\begin{pmatrix}
(\frac{s^3}{6} - L\frac{s^2}{2})\cos \lphase \\
h(\frac{s^3}{6} - L\frac{s^2}{2})\sin \lphase  \\
0
\end{pmatrix}, \label{deformationbendingofaxis}
\end{alignat}
with relative errors of order 
${\mathcal O}\left({\refc{R}}/{L},{r}/{L} \right)$.
We can see from Eq.~\ref{deformationbendingofaxis} that the helix axis bends parallel to the direction 
$\left(\cos \lphase, h\sin \lphase,0\right)$, i.e.~the $xy$ projection of the position vector of the free end-point $\refc{\bv{r}}(s=L)$ in the reference configuration. In order to physically interpret this result we evaluate the forces and torques  acting at the clamped end of the filament
\begin{align}
	\Finternal(0)=\int\limits_{0}^{L}\bv{K}(s)\dd s = \begin{pmatrix}
	\Rsinfrac A_x (1-\cos \lphase) \\ -h \Rsinfrac A_x \sin \lphase \\ A_z L
	\end{pmatrix},
\end{align}
and
\begin{align} 
\Mbending(0) + \bv{r}(0)\wedge \Finternal(0)=
& -h(A_x\refc{R})\cot\refc{\alpha}
\begin{bmatrix}
\left(\Rsinfrac[1-\cos\lphase]-L\sin\lphase    \right) \\
h\left(-\Rsinfrac \sin\lphase +L\cos\lphase \right) \\
 L\tan\refc{\alpha}
\end{bmatrix}\nonumber\\
~&~~~+ (A_z \refc{R})
\begin{bmatrix}
h&\left[\Rsinfrac\left(1-\cos\lphase\right)\right]\\
&\left[-\Rsinfrac \sin\lphase  \right] \\
& 0
\end{bmatrix}.
\end{align}
Hence for $\refc{R}/L \ll 1$,
\begin{align} 
\Mbending(0) + \bv{r}(0)\wedge \Finternal(0)
&= (A_x\refc{R})L\cot\refc{\alpha}
\left[\begin{pmatrix}
h\sin\lphase \\
-\cos\lphase \\
-h\tan\refc{\alpha}
\end{pmatrix}  + {\mathcal O}\left(\frac{\refc{R}}{L}\right) \right].
\end{align}

\begin{figure} 
		\includegraphics[width=0.25\columnwidth]{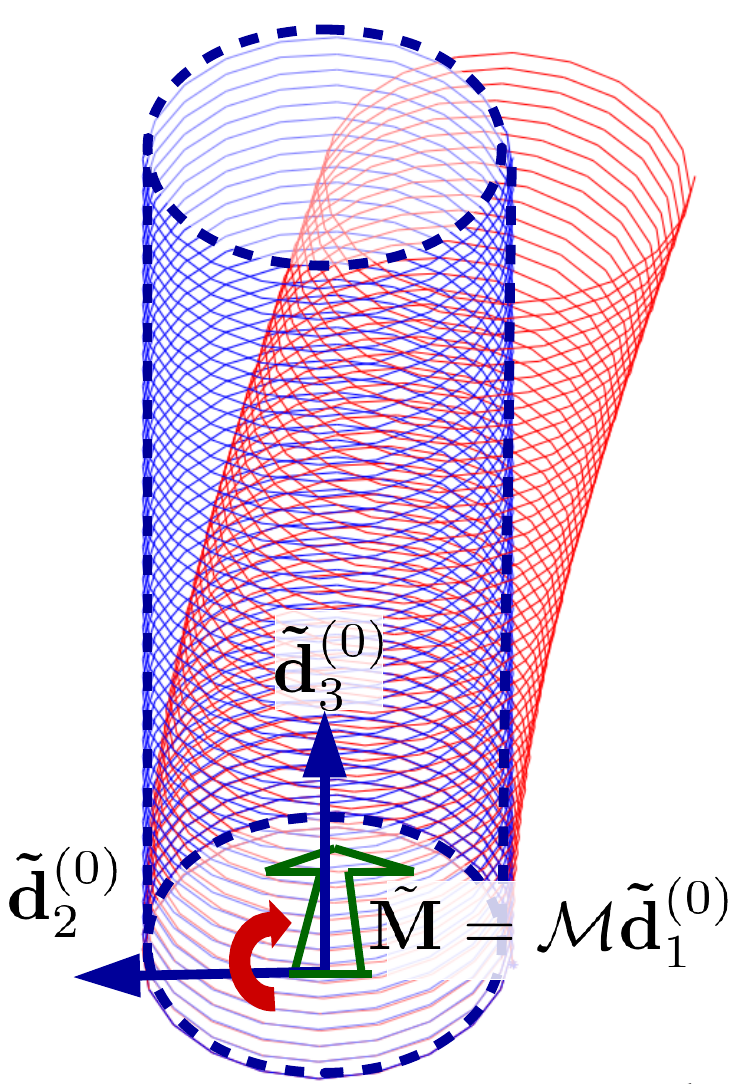}	
		\caption{Bending of the helix due to its translation / rotation. The effective rod in the reference configuration is shown in the blue dashed line while the bent helix is shown in the red solid line. }\label{FigEffectiveTorque}
	\end{figure}

We can now explain the bending of the axis of the helix using an  `effective' rod analogy, as has  been used in a variety of settings \cite{BiezenoGrammel1939,Haringx1942,Haringx1948, Vogel2012thesis}. 
Consider an effective rod around which the helix is wind, following the shape of the helix axis. This is 
 a straight rod clamped at the origin and parallel with the $z$-axis in its reference configuration, so that the material frame of the effective rod is
\begin{align}
\left(\uprefc{\tilde{\matd{1}}},\uprefc{\tilde{\matd{2}}},\uprefc{\tilde{\matd{3}}}\right)=(\bv{e}_x,\bv{e}_y,\bv{e}_z ).	
\end{align} 
A bending torque, $\tilde{\Mbending}=\mathcal{M}\uprefc{\tilde{\matd{1}}}$  with $\mathcal{M}=(A_x\refc{R})\cot\refc{\alpha}L$, exerted at the clamped end of the rod will lead to a perturbation to the Darboux vector of the rod with only non-zero component
\begin{align}
	\delta \darboux_1=\frac{\mathcal{M}}{EI}.
\end{align}
The resulting infinitesimal rotation is given by
 \begin{align}
 	\delta\bsym{\phi}=\int\limits_{0}^{s}\frac{\mathcal{M}}{EI}\uprefc{\tilde{\matd{1}}}\dd s=\frac{\mathcal{M}s}{EI}\uprefc{\tilde{\matd{1}}}.
 \end{align}
 Calculating the perturbations to the material frame of the  rod as
  \begin{align}
  \delta \tilde{\matd{i}} =\delta\bsym{ \phi} \wedge \uprefc{\tilde{\matd{i}}}=\frac{\mathcal{M}s}{EI}\uprefc{\tilde{\matd{1}}}\wedge \uprefc{\tilde{\matd{i}}},
  \end{align}
  we find
\begin{align}
\delta\tilde{\matd{1}}=\bv{0},~ \delta\tilde{\matd{2}}=\frac{\mathcal{M}s}{EI}\uprefc{\tilde{\matd{3}}},~ \delta\tilde{\matd{3}}=-\frac{\mathcal{M}s}{EI}\uprefc{\tilde{\matd{2}}}.
\end{align}
The perturbed tangent vector is given by
\begin{align}
\tilde{\matd{3}}= \uprefc{\tilde{\matd{3}}} + \delta \tilde{\matd{3}}=-\frac{\mathcal{M}}{EI}s\uprefc{\tilde{\matd{2}}}+\uprefc{\tilde{\matd{3}}},  
\end{align}
which gives the deformed shape as
\begin{align}
\tilde{\bv{r}}(s)= \int\limits_{0}^{s} \tilde{\matd{3}}=
-\frac{\mathcal{M}}{EI}\frac{s^2}{2}\uprefc{\tilde{\matd{2}}}+s \uprefc{\tilde{\matd{3}}}.
\end{align}
This has the same quadratic bending away from the $z$-axis as the helix axis in Eq.~\ref{deformationbendingofaxis} and  captures the bending of the axis in the direction perpendicular to the bending torque exerted at its clamped end and also the quadratic terms in Eq.~\ref{deformationbendingofaxis} for  s  such that $(\refc{R}\ll s\ll L)$. 
This bending of the helix axis is   illustrated in Fig.~\ref{FigEffectiveTorque}.

\subsection{Feedback of the deformation on the kinematics}\label{helicaldeformfeedback}

\subsubsection{Scalings} \label{Scalings}
Having computed the deformation of the helical filament, we now proceed to investigate its feedback on the locomotion kinematics.
Since we assumed that the helix was long enough so as not to wobble,  we may take the velocity and rotation rate to be axial. We   also assume that the perturbed velocity and rotation rate are along the long axis. Although there are  non-axial deformations due to the bending of the axis, they average out to zero upon a complete rotation of the swimmer. 

In order to find the expected order of magnitudes for the perturbed quantities, we consider the following scaling arguments. From Eqs.~\ref{Ax} and  \ref{Az} we obtain the scalings $A_x, A_z\sim \resperp \uprefc{U}\sim \resperp \uprefc{\omegafil} \refc{R}$. The leading-order kinematics in Eq.~\ref{rigidkinemHelixGenForm_U} give $\uprefc{U}\sim h M/ \resperp \refc{R} L$ where $M$ stands $M_{mag}$ or $M_{mot}$ 
for artificial motors and bacteria respectively. 
The small rotations of the material frame due to its deformation defined by Eq.~\ref{smallmatframerotation}, scale as $\delta \phi\sim \curlyP$ where $\curlyP$ is a small dimensionless number
\begin{align}
	\curlyP
	\sim \frac{\text{viscous}}{\text{elastic}}
	\sim \frac{A_x \refc{R} L^2}{EI}
	\sim \frac{ M }{EI/L}\cdot
\end{align}
Such scaling  can be seen in Eq.~\ref{deltaphilongexpression} of Appendix \ref{AppendixCalcDeform} for example. Thus we have $\delta \matd{3}\sim \curlyP$ and the deformation scales as $\delta \bv{r} \sim \curlyP L$. We note that the bending of the axis is quantified by an angle $\beta$ between the  end of the bent axis  and the $z$ axis, where $\tan \beta \sim |\delta \bv{r}_{xy~proj}|/L\cos \refc{\alpha}$, hence $\beta\sim \curlyP$.
The effect of the bending of the axis to the kinematics will therefore be an ${\mathcal O}(\curlyP)$ velocity component perpendicular to the $z$ axis that, due to rotation of the swimmer,  will average to 0 at ${\mathcal O}(\curlyP)$, and similarly for $\uprefc{\omegafil}$. 
It is thus appropriate to project the force and torque balances along the $z$-axis and assume that $\delta\bv{U}$ and $\delta \bsym{\omega}$ are aligned with the $z$-axis. As $\delta \matd{3} \sim \curlyP$, from Eqs.~\ref{perturbvelartificial}, \ref{perturbomegaartificial}, \ref{pertUbio} and \ref{perturbomegabiol} we therefore expect $\delta U$ and $\delta \omega$ to be of order ${\mathcal O}(\curlyP)$.

\subsubsection{Perturbing the resistance matrices }
The leading-order kinematics in $\curlyP$ are given in \S~\ref{leadingorderkinematics}. 
In this section we perform an asymptotic analysis for the next-order kinematics, i.e.~the feedback of the deformation on the kinematics.
We start by calculating the projections along the $z$-axis of the perturbations to the resistance matrices
\begin{align}\label{zzprojectionperturbofresistance matrices}
\delta \curlyA_{zz}&=2\resperp(1-\rho) \cos\refc{\alpha}
\int_{0}^{L} (\delta\matd{3})_{z} \dd s,\\
\delta \curlyB_{zz}&= \resperp(1-\rho)\bigg\{ h\refc{R}\sin\refc{\alpha}\int_{0}^{L}   (\delta\matd{3})_z \dd s
- \cos\refc{\alpha}\int_{0}^{L} [\delta(\matd{3}\wedge \bv{r})]_z  \dd s\bigg\},\\
\delta \curlyC_{zz}&=\delta \curlyB_{zz},\\
\delta \curlyD_{zz}&=2\resperp\bigg\{  
\cos\refc{\alpha}\int_{0}^{L} s(\delta\bv{r})_z \dd s
-\int_{0}^{L}(\bv{r}.\delta\bv{r}) \dd s
-(1-\rho)h \refc{R} \sin \refc{\alpha} \int_{0}^{L}[\delta(\matd{3}\wedge \bv{r})]_z \dd s\bigg\} \nonumber\\
&\quad-2\resrot\cos\refc{\alpha}\int_{0}^{L}  (\delta\matd{3})_z \dd s.
\end{align}

We may write these results as linear combinations of four integrals, $J_i$ ($1\leq i\leq 4$),
\begin{align}\label{zzprojectionperturbofresistance matricesJs}
\delta \curlyA_{zz}&=2\resperp(1-\rho) \cos\refc{\alpha}
 J_1,\\
\delta \curlyB_{zz}&= \resperp(1-\rho)\left\{ h\refc{R}\sin\refc{\alpha} J_1- \cos\refc{\alpha}J_2\right\},\\
\delta \curlyC_{zz}&=\delta \curlyB_{zz},\\
\delta \curlyD_{zz}&=2\resperp\left\{  
\cos\refc{\alpha}J_3-J_4-(1-\rho)h \refc{R} \sin \refc{\alpha} J_2\right\}
 - 2\resrot\cos\refc{\alpha} J_1,
\end{align}
where the four integrals are given by
\begin{align}
	J_1&=\int_{0}^{L} (\delta\matd{3})_{z} \dd s, \\
	J_2&=\int_{0}^{L} [\delta(\matd{3}\wedge \bv{r})]_z  \dd s
	=\vez \cdot\int_{0}^{L} \left[\delta\matd{3}\wedge \ruprefc +\uprmatd{3}\wedge \delta\bv{r} \right]  \dd s,\\
	J_3&=\int_{0}^{L} s(\delta\bv{r})_z \dd s, \\
	J_4&=\int_{0}^{L}(\ruprefc\cdot\delta\bv{r}) \dd s,
\end{align}
which depend on the  deformation integrated along the entire length of the helix.
The details of this long calculation are given in Appendices
\ref{Clamped_Appendix} and \ref{Appfreeendpoint} where 
the cases of either clamped or free $s=0$ ends are  addressed, as well as the effects of the viscous rotational torque.
 
In the case of a clamped end at $s=0$  ($\vPhi,\pertzeroend=\bv{0}$), the perturbations to the resistance matrices projected along the $z$ axis  are
\begin{align}
\delta \curlyA_{zz}^{clamped}&=\resperp(1-\rho) \cos\refc{\alpha}
\frac{1}{EI}\left[\refc{R}^2 L^2 + {\mathcal O}(\refc{R}^3 L)\right] \left[A_x\cot\refc{\alpha} + A_z\right],  \label{deltaAclamped}\\
\delta \curlyB_{zz}^{clamped}&= \resperp(1-\rho)\left\{ h\refc{R}\sin\refc{\alpha} J_1- \cos\refc{\alpha}J_2\right\} \nonumber\\
&=\resperp(1-\rho)\frac{h\refc{R}^3 L^2}{2EI}\bigg\{  ~~A_x\cos\refc{\alpha}\left[1-\left(\cot^2\refc{\alpha} -\cos\left(\frac{2L}{\refc{R}/\sin\refc{\alpha}}\right)\right)\right] \nonumber\\
&\qquad\qquad\qquad\qquad~~+ A_z\sin\refc{\alpha}\enspace\left[1-\cot^2\refc{\alpha}\left(\frac{3}{2} + \cos\left(\frac{2L}{\refc{R}/\sin\refc{\alpha}}\right)\right)\right] ~~ \bigg\} \nonumber\\
&~~\quad+{\mathcal O}\left( \frac{\refc{R}^4LA_{x,z}}{EI} \right) \label{deltaBclamped},\\
\delta \curlyC_{zz}^{clamped}&=\delta \curlyB_{zz}^{clamped},\\
\delta \curlyD_{zz}^{clamped}&= 2\resperp\left[ \cos\refc{\alpha}J_3-J_4-(1-\rho)h \refc{R} \sin \refc{\alpha} J_2\right] \nonumber\\
&=2\resperp\frac{\refc{R}^4 L^2}{EI}\bigg\{  \qquad\frac{\cos\refc{\alpha}}{\sin^2\refc{\alpha}}A_z\left[1+\cos\left(\frac{2L}{\refc{R}/\sin\refc{\alpha}}\right)\right] \nonumber\\
&\qquad\qquad~~~~~
-(1-\rho)\sin\refc{\alpha} \bigg[0.5A_x \left(\cot^2\refc{\alpha} -\cos\left(\frac{2L}{\refc{R}/\sin\refc{\alpha}} \right)\right) \nonumber\\ 
&\qquad\qquad\qquad\qquad\qquad\qquad~+A_z\cot\refc{\alpha}\left( \frac{3}{2} + \cos\left(\frac{2L}{\refc{R}/\sin\refc{\alpha}}\right) \right)~~\bigg]~~
\bigg\}  \nonumber\\
&~~+{\mathcal O}\left( \frac{\refc{R}^5LA_{x,z}}{EI} \right) . \label{deltaDclamped}
\end{align}
Notably,  the viscous rotational torque, $\Nviscous$, does not contribute to any of the perturbations of the resistance matrices.

\subsubsection{Unified Approach}
We once  again invoke the unified approach of section \S\ref{leadingorderkinematics}.
The leading-order kinematics are given by Eqs.~\ref{rigidkinemHelixGenForm_U} and \ref{rigidkinemHelixGenForm_omega} and the perturbation to the swimming velocity is given by
\begin{align}
\delta U &=\frac{M}{\left(\curlyA\curlyD_*-\curlyB^2\right)^2}\left[-\curlyB\curlyD_*(\delta\curlyA)+(\curlyA\curlyD_*+\curlyB^2)(\delta \curlyB)- \curlyA\curlyB(\delta\curlyD)\right], \label{deltaUHelixGenform}
\end{align}	
where $M$ stands for either $M_{mag}$ or $M_{mot}$ for artificial motors and bacteria respectively, and where 
$\curlyA$, $\curlyB$ and 
 $\curlyD_*$ are given by Eqs.~\ref{A}, \ref{B} and \ref{Dstar} respectively. The denominator $\curlyA\curlyD_*-\curlyB^2$ is given by Eq.~\ref{ADstar_B2}.
After substituting in the values of the resistance matrices, the quantities $A_x,A_z$ become
\begin{align}
A_x&
=\frac{h \refc{R}\resperp M}{\curlyA\curlyD_*-\curlyB^2}\left[\rho\resperp L +(\cos^2\refc{\alpha} + \rho \sin^2\refc{\alpha})6\pi\mu \ahead 
\right],\\ 
A_z&=\frac{h \refc{R}\resperp M  }{\curlyA\curlyD_*-\curlyB^2}6\pi\mu \ahead (1-\rho)\sin\refc{\alpha}\cos\refc{\alpha} ,\\
A_x\cot\refc{\alpha}+A_z&=\frac{h\refc{R}\resperp M}{\curlyA\curlyD_*-\curlyB^2}
\left[\rho\resperp L  +  6\pi\mu \ahead \right]\cot\refc{\alpha}.
\end{align}
Considering only the clamped contribution, the perturbations to the resistance matrices in Eqs.~\ref{deltaAclamped}-\ref{deltaDclamped} which are linear in $A_x$ and $A_z$ can be calculated (with details shown in Appendix \ref{FeedbacktoKinem}).  
The perturbation $\delta U$ is finally obtained as
\begin{align}
\delta U=\frac{\refc{R}^2 L^2\resperp^2 M^2}{EI\left(\curlyA\curlyD_*-\curlyB^2\right)^3}(1-\rho)\refc{R}^2\curlyQ  \label{deltaUgenformCalculated},
\end{align}
where $\curlyA\curlyD_*-\curlyB^2$ is given by Eq.~\ref{ADstar_B2} and 
\footnotesize
\begin{align}
\curlyQ
=&\resperp  L(1-\rho) c^3\left[ \resperp \refc{R}^2 L (c^2+\rho s^2)+ \resrot L c^2 + 8\pi\mu \ahead^3 \indfun \right]
\left[\rho\resperp L  +  6\pi\mu \ahead \right]
\nonumber\\
&+\frac{1}{2}\begin{pmatrix}
[\rho + 2(1-\rho)^2s^2c^2]\resperp^2 \refc{R}^2L^2\\
+ 6\pi\mu\resperp\refc{R}^2 L\ahead (c^2+\rho s^2) \\
+ \resrot L c^2 \left[ \resperp L (s^2+\rho c^2) + 6\pi\mu \ahead \right]\\
+8\pi\mu \ahead^3\indfun\left[ \resperp L (s^2+\rho c^2) + 6\pi\mu \ahead \right]
\end{pmatrix}
\begin{pmatrix}
\left[\rho\resperp L +(c^2 + \rho s^2)6\pi\mu \ahead 
\right] c\left[1-\left(\cot^2\refc{\alpha} -\cos\dlphase\right)\right] \\
+   6\pi\mu \ahead (1-\rho)s^2c\enspace\left[1-\cot^2\refc{\alpha}\left(\frac{3}{2} + \cos\dlphase\right)\right]
\end{pmatrix}
\nonumber\\
&-2\resperp L\refc{R}^2(1-\rho) sc\left[ \resperp L (s^2+\rho c^2) + 6\pi\mu \ahead \right] 
\begin{pmatrix}
-\frac{c^2}{s}   6\pi\mu \ahead \left[1+\cos\dlphase\right] \\
+0.5 \left[\rho\resperp L +(c^2 + \rho s^2)6\pi\mu \ahead 
\right]  s\left(\cot^2\refc{\alpha} -\cos\dlphase\right) \\ 
\qquad\qquad
+   6\pi\mu \ahead (1-\rho)s c^2\left( \frac{3}{2} + \cos\dlphase \right)
\end{pmatrix},			 \label{QgenformCalculated}	 
\end{align}
\normalsize
with relative errors of order ${\mathcal O}\left(\refc{R}/L\right)$ and where we have   used the shorthand notation $s\equiv\sin\refc{\alpha}$ and $c\equiv\cos\refc{\alpha}$. 
The total velocity, $\Utot$, is simply $\Utot=\uprefc{U}+\delta U$.
The details of the calculation are given in Appendix \ref{FeedbacktoKinem}. 

We now proceed by considering the applications of this unified approach to artificial motors and bacteria separately.

 \subsubsection{Artificial motors}\label{helicartifmotor}

 \paragraph*{A  stiff helical microswimmer.}
\begin{figure}[t]
	\includegraphics[width=0.8\columnwidth]{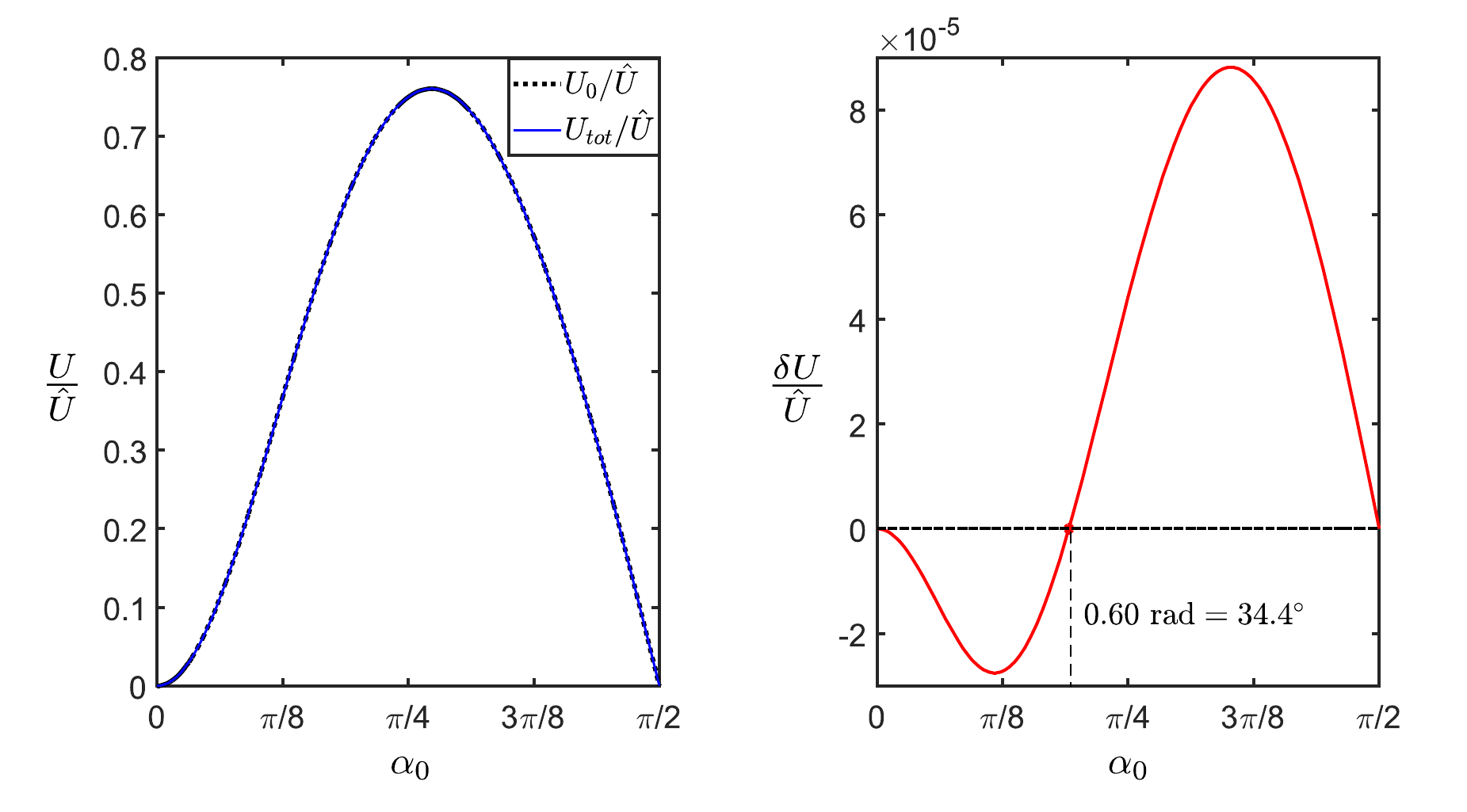}
	\caption{Left: Dimensionless speeds in the reference configuration, $\uprefc{U}/\hat{U}$,  and after the deformation, $\Utot/\hat{U}$, as a function of the reference helix angle, $\refc{\alpha}$, obtained from Eqs.~\ref{rigidkinemHelixGenForm_U}.  Right: Perturbation in velocity,  $\delta U/\hat{U}$,  as a function of $\refc{\alpha}$ from Eq.~\ref{deltaUHelixGenform}. In all cases, we use the full expressions for the resistance matrices (within the limit $\refc{R}/L\ll1$), for a RH artificial bacterial flagellum with $\curlyP=0.0350$. The microswimmer has the parameters $\ahead=2~\mu \rm{m}$, $r=50~\rm{nm}$,
	$\Lambda=5~\mu\rm{m}$, $n=8$, $L=40~\mu\rm{m}$, $M_{mag}=4.3~\times 10^{-17}~\rm{Nm}$, and $E=10^{10}~\rm{Pa}$ and is moving in water with dynamic viscosity $\mu=10^{-3}~\rm{Pa s}$. 
	The sign of the perturbation $\delta \hat{U}$ transitions at an angle $\refc{\alpha}\approx0.60~\rm{rad}  ~(34.4^{\circ})$. 
	The non-dimensionalisation uses the velocity scale	$\hat{U}=M_{mag} / \mu L^2$, that takes the value of $27 ~\mu\rm{m~ s}^{-1}$.
	}\label{Figure_Lighthill_Artif_U0UtotdeltaU}
\end{figure}

Typical profiles of the dimensionless  velocities  $\uprefc{U}/\hat{U}$ in the reference configuration and $\Utot/\hat{U}$ after the deformation (left) and the perturbation $\delta U/\hat{U}$ due to the deformation (right) are shown in Fig.~\ref{Figure_Lighthill_Artif_U0UtotdeltaU} as a function of the reference helix angle $\refc{\alpha}$ (the choice of speed scale $\hat{U}$ is given below).  These   results are obtained using Eqs.~\ref{rigidkinemHelixGenForm_U} and \ref{deltaUHelixGenform} respectively, using the full expressions for the resistance matrices (within the limit $\refc{R}/L\ll1$), for a RH artificial bacterial flagellum with the choice $\curlyP\equiv  M_{mag}L/EI=0.035$.
For the non-dimensionalisation we use the length scale $L$, time scale $T=\mu L^3/M_{mag}$  and   force scale $\mu L^2 / T = M_{mag} / L$  (or, equivalently,   mass scale $\mu LT=\mu^2 L^4/M_{mag}$), so that the scale for the speed is 
$\hat{U}=M_{mag} / \mu L^2$.
We use values for the geometrical parameters   similar in order of magnitude to those of the rigid `artificial bacterial flagellum' of Ref.~\cite{Zhang2009} with the choice $n=8$ for the number of turns so that we can satisfy the requirement that $\refc{R}/L\ll 1$. 
As such, our model microswimmer has $\ahead=2~\mu \rm{m}$, $r=50~\rm{nm}$,
$\Lambda=5~\mu\rm{m}$, $n=8$, $L=40~\mu\rm{m}$, $M_{mag}=4.3~\times 10^{-17}~\rm{Nm}$  
and is moving in water with dynamic viscosity $\mu=10^{-3}~\rm{Pa s}$.
As a result, the scale for speed, $\hat{U}$, takes the value of $27 ~\mu\rm{m~ s}^{-1}$.
We set the Young's modulus to be $E=10^{10}~\rm{Pa}$, as an indicative example of a stiff helical microswimmer, according to the   `equivalent' Young's modulus of nanowire filaments of existing microswimmer designs \cite{Paknanowire2011}. This value is also relevant for \rm{ZnO} nanohelices \cite{GaoMaiWang2006}.

Importantly,  the perturbation in velocity, $\delta U$, changes sign at an angle $\refc{\alpha}^*\approx0.60~\rm{rad}  ~(34.4^{\circ})$, being negative for smaller values of $\refc{\alpha}$ and positive for larger ones.
Note that inside the resistance matrices used in  Eq.~\ref{deltaUHelixGenform} we have kept all  terms involving $\ahead$ and $\resrot$, and have not   imposed   the ratio $\ahead/L$  to be   small.

\paragraph*{Approximation.}
We   proceed by further approximating the expression for $\curlyQ$  in Eq.~\ref{QgenformCalculated}, assuming that both the helical radius and the head are negligible in size compared to the total contour length of the helix, such that $\refc{R}$ and $\ahead$ satisfy  the limits
\begin{equation}
\refc{R}/L,~\ahead/L,~\ahead^3/\refc{R}^2L,~\ahead^4/\refc{R}^2L^2\ll 1.
\end{equation}
The resulting expression for $\delta U$ in Eq.~\ref{deltaUgenformCalculated} simplifies to
\begin{align}
	&\delta U=
	\frac{\resperp^5M^2_{mag}\refc{R}^6L^5\rho(1-\rho)}{EI(\curlyA\curlyD-\curlyB^2)^3}\Gamma(\refc{\alpha}), \label{approxdeltaUfulldenominator}
\end{align}
where $\curlyA\curlyD-\curlyB^2$ is given by Eq.~\ref{AD-B2} and 
\begin{align}
\Gamma(\refc{\alpha})=&\cos\refc{\alpha}\bigg[
-\frac{\rho\cos(2\refc{\alpha})}{2\sin^2\refc{\alpha}} +\left[0.5\rho + (1-\rho)\sin^2\refc{\alpha}\right]\cos\dlphase\bigg]. \label{Gamma}
\end{align}

When reducing the ratios $\refc{R}/L$ and $\ahead/L$, for non-vanishing helix angles for which the ratio $\refc{R}/L$ is kept small, the profile of $\delta U$ vs.~$\refc{\alpha}$ that we obtain from the full expression of Eq.~\ref{deltaUHelixGenform} converges to the theoretical approximation of Eq.~\ref{approxdeltaUfulldenominator}.
Indeed, if we keep all parameter values as above and  only decrease the size of the head, the angle $\alpha_0^*$ at which $\delta U$ vanishes converges to $33.1^{\circ}$, which is well approximated by the root of $\Gamma(\refc{\alpha})$ in Eq.~\ref{Gamma} equal to $32.8^{\circ}$.

For non-vanishing $\refc{\alpha},\refc{R}$,
\begin{align}
\delta U 
&=\frac{M^2_{mag}}{\resperp EIL} \frac{(1-\rho)}{\rho^2}
\Gamma(\refc{\alpha}),
\end{align}
with a relative error of order of magnitude
${\mathcal O}\left(\refc{R}/L,~\ahead/L,~\ahead^3/\refc{R}^3L,~\ahead^4/\refc{R}^2L^2\right)$, 
so that the corrected velocity is 
\begin{align}
\Utot= \uprefc{U} + \delta U 
=\frac{h  M_{mag}}{ \resperp \refc{R} L}\frac{(1-\rho)}{\rho}\bigg[ \sin\refc{\alpha}\cos\refc{\alpha} + \frac{\epsilon h}{\rho}\Gamma(\refc{\alpha})\bigg],
\end{align} 
with the   dimensionless parameter $\epsilon$ defined as 
\begin{equation}
\epsilon=\frac{M_{mag}}{EI/\refc{R}} =\curlyP\frac{\refc{R}}{L},\label{smallparameter}
\end{equation}
where $\curlyP$ is the   parameter defined in \S \ref{Scalings}. We therefore see that a stiff elastic helix has a quadratic  perturbation  in $M_{mag}$ to the classical linear relation between $U$ and $M_{mag}$ for a rigid helix. This is of course valid for a weak enough actuation, or a stiff enough helix,  that $\epsilon$ is kept small.

In our numerical results, $\resperp$ and $\rho$ are given by Eqs.~\ref{Lighthill_resistive_coefficients} and \ref{Lighthill_resistive_coefficients_ratio}. 
For the purpose of simplicity and to allow  physical interpretation, we can further assume an integer number of turns for the helix and take the approximate value $\rho=0.5$ for the ratio of drag coefficients, so that the expressions for $\delta U$ and the corrected velocity $\Utot$  further simplify to
\begin{align}
\delta U 
&=\frac{M^2_{mag}}{\resperp EIL} \frac{\cos\refc{\alpha}}{2 \sin^2\refc{\alpha}}
\left(2 \sin^4\refc{\alpha} + 3 \sin^2\refc{\alpha} -1\right), \label{deltaUartifmotor} \\
\Utot&=\frac{h  M_{mag}}{ \resperp \refc{R} L}\bigg[ \sin\refc{\alpha}\cos\refc{\alpha} +\epsilon h \frac{\cos\refc{\alpha}}{2 \sin^2\refc{\alpha}}
\left(2 \sin^4\refc{\alpha} + 3 \sin^2\refc{\alpha} -1\right)\bigg]. \label{Utotal}
\end{align} 
 
\paragraph*{Feedback to the speed of propulsion.}
We may now investigate whether the small amount of elasticity speeds up or slows down a stiff elastic helical filament compared to its rigid kinematics, which is governed by the  sign of $\delta U$. 
We first note that the sign of $\uprefc{U}$ is opposite for right-handed (RH) and left-handed (LH) helices, as shown by the presence of the chirality index $h$ in its expression. In contrast, the expression for  $\delta U$ in  Eqs.~\ref{deltaUgenformCalculated} and \ref{QgenformCalculated}  does not involve the value of  $h$. 
Thus, if all other geometrical parameters are kept constant, the magnitude of the speed for a stiff helix, $|\uprefc{U} + \delta U|$, will be increased or decreased according to the handedness of the helix. In addition, as discussed above, the sign of $\delta U$ changes at a critical angle $\refc{\alpha}^*$ (with a value which  depends on the geometry). 
The case of our typical stiff helical swimmer above had $\refc{\alpha}^*=34.4^{\circ}$.
For a microswimmer with a smaller head but otherwise same geometrical parameters, we obtained instead $\refc{\alpha}^*=33.1^{\circ}$. 
Thus the perturbation to the speed changes sign according to a subtle interplay between the handedness and the reference helix angle.

\begin{figure}
\includegraphics[width=0.6\columnwidth]{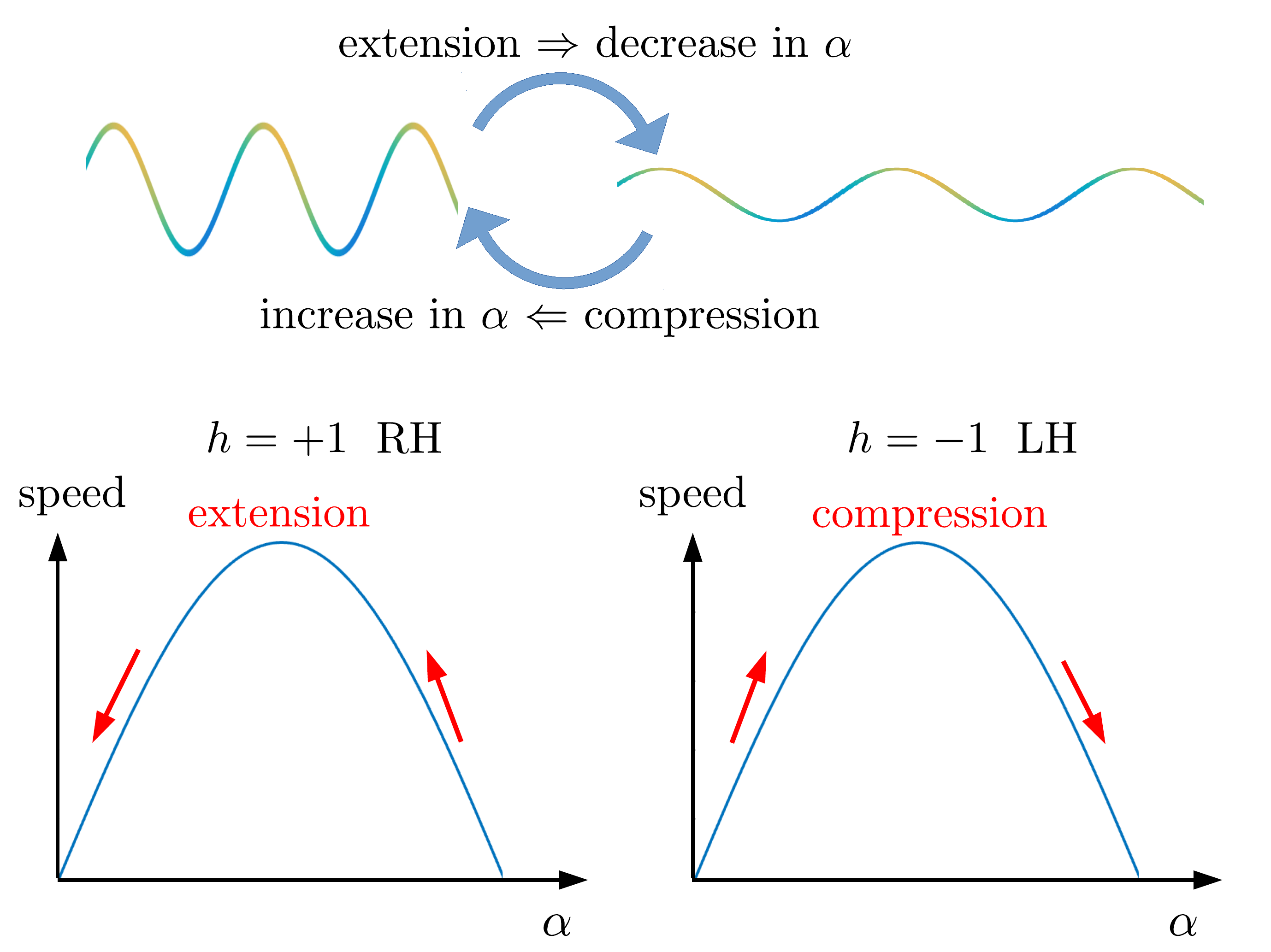}
\caption{Understanding the feedback of the deformation on the speed of a stiff, elastic helix. 
	Top panel: The extension (resp.~compression) of a right- (resp.~left-) handed helix leads to decrease (resp.~increase) in the helix angle. Bottom panel: If the value of the original helix angle  is greater than the optimal angle maximising the speed profile of a rigid helix as a function of the helix angle, then the shift in the effective helix angle due to extension (or compression) will give rise to an increase (resp.~a decrease) in the speed for a RH (resp.~LH) helix. Similarly, if the value of the original helix angle is less than the optimal angle maximising the speed profile, then the shift in the effective helix angle due to extension 
	(resp.~compression) will give rise to a decrease (resp.~increase) in the speed for a RH (resp.~LH) helix.
}\label{feedbackprinciple}
\end{figure}

We can interpret the change in the sign of $\delta U$ by approximating the new, steady-state, perturbed shape by an effective   helix over which the extension or compression has been uniformly distributed. As illustrated in the top panel of Fig.~\ref{feedbackprinciple}, the extension (or compression) of a RH (or LH)  helix can be interpreted  as a decrease (or increase) in the helix angle.
This provides us with an intuitive reasoning for the sign of the perturbation to the speed.

Consider the rigid speed profile as a function of the reference helix angle.
As illustrated in the bottom panel of Fig.~\ref{feedbackprinciple}, if we start at an angle greater than the optimum, a RH helix that extends will reduce its effective helix angle, thereby moving towards the maximum of the speed profile and as a result will speed up. In contrast, a LH helix with a reference helix angle greater than the optimum will move away from the optimum, as its effective helix angle increases due to the compression.
If we start at an angle less than the optimum, a RH helix that extends will reduce its effective helix angle, thereby moving away from the maximum of the speed profile, and will thus slow down. In contrast, a LH helix with a reference helix angle less than the optimum will move towards the maximum and speed up, as its effective helix angle increases due to the compression.
	We think that the quantitive discrepancy between the angle for which $\delta U$ vanishes and the optimum value of the uniform rigid helix velocity profile is likely due to the nonuniformity of the extension/compression along the filament and the bending of the helix axis.

\subsubsection{Bacteria} \label{helicbacteriasec}
Swimming bacteria will have, in principle, a contribution in their speed from the values of  $\vPhi, \pertzeroend$ (in Eq.~\ref{pertdrwithPhiandDr0}) as the end of the flagellar filament   connected to the hook. This is due to the flexibility of the  hook at the base of the flagellum. In fact, for some  single-flagellated bacteria such as {\it V.~alginolyticus}, the  buckling of the hook after the transition from a backward to a forward swimming period  is a mechanism to change the angle between the cell body and the flagellar filament axis \cite{SonGuastoStockeronHook2013}. Our study  focuses however on  the swimming   along straight lines and hence for our purposes we will neglect buckling of  the hook and will assume that both $\vPhi$ and $ \pertzeroend$ are zero.

\begin{figure}[t]
	\includegraphics[width=0.8\columnwidth]{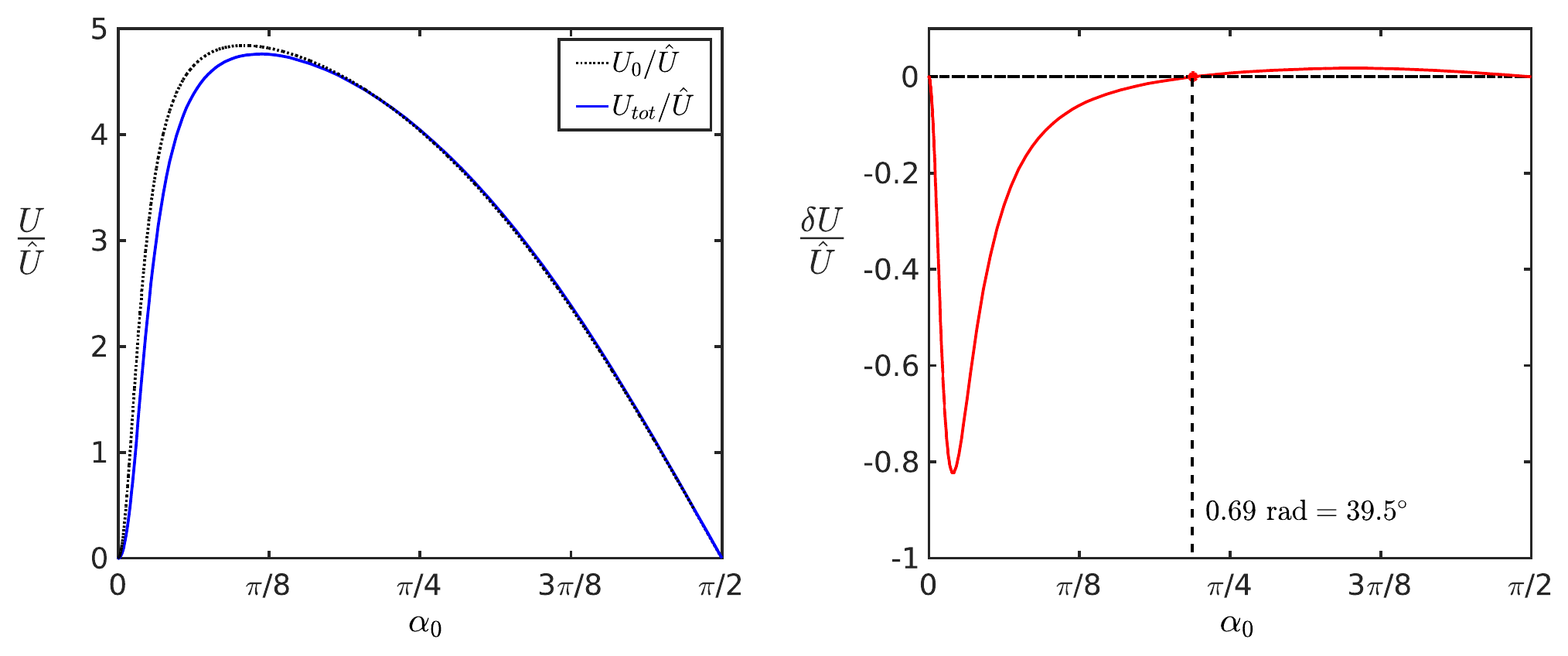}
	\caption{
	Left: Dimensionless speeds in the reference configuration, $\uprefc{U}/\hat{U}$,  and after the deformation, $\Utot/\hat{U}$, as a function of the reference helix angle, $\refc{\alpha}$, obtained from Eqs.~\ref{bacteriakinemleading}.  Right: Perturbation in velocity,  $\delta U/\hat{U}$,  as a function of $\refc{\alpha}$ from Eq.~\ref{pertUbio}. In all cases, we use the full expressions for the resistance matrices (within the limit $\refc{R}/L\ll1$), for a RH bacterial flagellar filament with $\curlyP\equiv M_{mot}L /EI=0.19$. 
	The bacterium has parameters $\ahead=1~\mu \rm{m}=10^{-6}~\rm{m}$, $r=11.5~\rm{nm}=1.15\times10^{-8}~\rm{m}$, $\Lambda=2.3~\mu\rm{m}$, $n=6$, $L=13.8~\mu\rm{m}$, $M_{mot}=2~\times 10^{-18}~\rm{Nm}$, and Young's modulus $E=1.04\times10^{10}~\rm{Pa}$ so that its bending stiffness is $EI=1.43\times10^{-22}~\rm{Nm^2}$ and is moving in water with dynamic viscosity $\mu=10^{-3}~\rm{Pa s}$.  
	The sign of the perturbation $\delta \hat{U}$ transitions at an angle $\refc{\alpha}\approx0.69~\rm{rad}  ~(39.5^{\circ})$.  
	The non-dimensionalisation uses the velocity scale $\hat{U}=M_{mot} / \mu L^2$, that takes the value of $1.05\times10^{-5} ~\rm{m s}^{-1}=10.5~\mu\rm{m~ s}^{-1}$. 	
	}\label{Figure_Lighthill_Bact_U0UtotdeltaU}
\end{figure}

In the case of bacteria, we can make use of Eqs.~\ref{deltaUgenformCalculated} and \ref{QgenformCalculated}, where the indicator function now takes the value $0$ and thus the terms multiplied by it are removed. Typical profiles of the dimensionless velocities  in the reference configuration, $\uprefc{U}/\hat{U}$,  and after the deformation,  $\Utot/\hat{U}$,  are shown in  Fig.~\ref{Figure_Lighthill_Bact_U0UtotdeltaU} (left) as a function of the reference helix angle, $\refc{\alpha}$, for a RH bacterial flagellar filament with $\curlyP\equiv M_{mot} L/EI=0.19$ and using the full expressions for the resistance matrices (within the limit $\refc{R}/L\ll1$) (Eqs.~\ref{bacteriakinemleading}). We also plot in   Fig.~\ref{Figure_Lighthill_Bact_U0UtotdeltaU} (right)  the perturbation in swimming speed due to the deformation, $\delta U/\hat{U}$, as a function of $\refc{\alpha}$. Here the non-dimensionalisation used the velocity scale
 $\hat{U}=M_{mot} / \mu L^2$. The parameter values are chosen as the typical values for {\it Salmonella} bacteria given in Refs.~\cite{NambaVonderviszt1997, KamiyaAsakura1976} (and  references therein), so that we   have $\ahead=1~\mu \rm{m}$, $r=11.5~\rm{nm}$, $\Lambda=2.3~\mu\rm{m}$, $n=6$, $L=13.8~\mu\rm{m}$, $M_{mot}=2~\times 10^{-18}~\rm{Nm}$ and take the viscosity to be that of water,  $\mu=10^{-3}~\rm{Pa s}$. 
The scale for speed, $\hat{U}$, takes as a result the value of $10.5~\mu\rm{m~ s}^{-1}$. Further
we  take the Young's modulus to be $E=1.04\times10^{10}~\rm{Pa}$ as in 
Ref.~\cite{TrachtenbergHammel1992} so that the bending stiffness is $EI=1.43\times10^{-22}~\rm{Nm^2}$. 
We obtain as a result that the   perturbation in swimming speed,  $\delta \hat{U}$,   transitions in sign at the angle $\refc{\alpha}\approx0.69~\rm{rad}  ~(39.5^{\circ})$, being negative for smaller values of $\refc{\alpha}$ and positive for larger ones.

Importantly, the reported values for the elasticity of flagellar filaments in the literature show large variations, which naturally leads to a wide  range of values for the bending stiffness, $EI$ \cite{KimPowers2005HelixDeformation}. Trachtenberg and Hammel reported values of the Young's modulus ranging from $E=1.04\times 10^{10} ~\rm{Pa}$ to $E=1.77\times10^{11}\rm{Pa}$ for a variety of filament types \cite{TrachtenbergHammel1992}, which correspond to bending stiffness of $EI=10^{-22}~\rm{Nm^2}$ and $2\times10^{-21}~\rm{Nm^2}$, and with  
the parameter values above, the dimensionless number $\curlyP$ would range from $\curlyP=0.19$ to $\curlyP=0.011$.    
Other reported values of the filament elasticity include $(2-4)\times 10^{-24}~\rm{Nm^{2}}$ for the bending stiffness $EI$ of {\it Salmonella typhimurium} filaments \cite{FujimeMaruyamaAsakura1972} 
and $10^{10}~\rm{N/m^{2}}$ for the shear modulus of {\it S.~typhimurium} flagellar filament which gives the value $\mu_S J=10^{-22}~\rm{Nm^{2}}$ for the twist modulus of a filament of radius of $10~\rm{nm}$ \cite{HoshikawaKamiya1985}.
Takano et al.~estimated $EI=10^{-24}~\rm{Nm^2}$ for {\it Vibrio alginolyticus} \cite{Takanoetal2003HelicFlagel} whereas Kim and Powers estimated $EI=10^{-24}~\rm{Nm^2}$ for {\it S.~typhimurium} \cite{KimPowers2005HelixDeformation} by reinterpreting the data of  Hoshikawa and Kamiya \cite{HoshikawaKamiya1985}.

Our analysis mainly focuses on the asymptotics and feedback of the deformation to the kinematics, and to first order, the perturbation $\delta U$ scales linearly with $1/EI$. In particular, we note that within the `stiff' filament regime and first-order analysis presented here, the value of the reference angle $\refc{\alpha}$ for which the sign of $\delta U$ changes is independent of the Young's modulus of the filament.

The resulting expression for $\delta U$ in Eq.~\ref{deltaUgenformCalculated} simplifies to
\begin{align}
&\delta U=
\frac{\resperp^5M^2_{mot}\refc{R}^6L^5\rho(1-\rho)}{EI(\curlyA\curlyDfil-\curlyBfil^2)^3}\Gamma(\refc{\alpha}), 
\end{align}
where $\curlyA\curlyDfil-\curlyBfil^2$ and  $\Gamma(\refc{\alpha})$ are given by Eqs.~\ref{ADfil_Bfil2} and \ref{Gamma} respectively. For non-vanishing $\refc{\alpha},\refc{R}$, in the limit of small cell body and helical radius compared to the length of the flagellar filament, the expression for $\delta U$ reduces to
\begin{align}
\delta U 
&=\frac{M^2_{mot}}{\resperp EIL} \frac{(1-\rho)}{\rho^2}
\Gamma(\refc{\alpha}),
\end{align}
with a relative error of order of magnitude ${\mathcal O}\left(\refc{R}/L,~\ahead/L,~\ahead^3/\refc{R}^3L,~\ahead^4/\refc{R}^2L^2\right)$, 
similarly to the result for artificial motors in Eq.~\ref{deltaUartifmotor}. Thus, similar conclusions can be drawn as for the case of artificial motors.

\section{Discussion}\label{concldiscuss}

In this paper, motivated by the locomotion of bacteria and artificial  microswimmers, we have developed the mathematical framework to study the steady-state motion of  individual, stiff elastic filaments   attached on a rigid body and propelling in a viscous fluid. The richness of this fluid-structure interaction problem comes from the hydrodynamic loads that induce deformation and whose integrated effect along the whole deformed shape leads to perturbed swimming kinematics.

Our analytical approach consisted in first integrating the hydrodynamic load along the initial reference configuration geometry and obtaining the induced deformation via the linear constitutive equation for the bending moment given by the classical Kirchhoff equations.   
For a helical geometry, the axis bends and the helix is extended or compressed if it is right- or left-handed, respectively. 
Our mathematical expressions describe the full three-dimensional deformation and an effective rod analogy provides an explanation for the bending of the helix axis whose origin was previously unclear.
Imposing the force and torque balance along the new, deformed shape linearly relates the velocity perturbation to the deformation integrated along the entire length of the filament. As a result, the    propulsion speed acquires  a quadratic correction in the actuation. 

Our analytical expression for the correction to the speed for stiff elastic helical filaments changes sign according to a subtle interplay between the handedness and reference helix angle.  Approximating the new, steady-state, perturbed shape as an `effective' uniform helix, the extension (or compression) of a right- (or left-) handed helix can be rationalised as due to  a decrease (or increase) in the helix angle. This interpretation  provides us with an intuitive reasoning for the sign of the  speed perturbation. If the value of the original helix angle in the reference configuration is greater than the optimal angle maximising the speed profile of a rigid helix as a function of the helix angle, then the shift in the effective helix angle due to extension (or compression) will give rise to increased (or decreased) speeds depending on  the handedness of the helix.

We note that our mathematical formulation allows to tackle any slender filament shape, and need not be limited to helices. Furthermore, one could include in it a model for the activity of the filament. A natural extension  of this type would address   magnetised tails such as  the recently-manufactured  hydrogels with embedded magnetic nanoparticles which not only guide self-folding during fabrication  \cite{Huangetal2016Nature} but could also display  interesting non-linear   relationship between the external actuation and the  propulsion speed.

\section*{Acknowledgements}
This project has received funding from the EPSRC (PK) and the European Research Council (ERC) under the European Union's Horizon 2020 research and innovation programme  (grant agreement 682754 to EL).

\bibliographystyle{unsrt}
\bibliography{ElasticHelixRefBib}
\newpage 
\onecolumngrid
\appendix

\normalsize
\section{Calculation of the Deformation for helical filaments} \label{AppendixCalcDeform}
The small rotations of the material frame due to its deformation (defined by Eq.~\ref{smallmatframerotation}) are calculated as the integral of the bending moment (given in Eq.~\ref{Mbendinglongexpression}), in accordance to Eq.~\ref{deltaphi_int_M_formula},
\begin{alignat}{4}
&&&&EI (\vdeltaphi)(s)\nonumber\\
&&=&& A_x\refc{R}&
\begin{bmatrix}
-h\cot\refc{\alpha}\left[\left(\Rsinfrac\right)^2\sin\sphase - \Rsinfrac s \cos\lphase +(s^2/2-Ls)\sin\lphase    \right] \\
\cot\refc{\alpha}\left[\left(\Rsinfrac\right)^2\left( \cos \sphase -1\right) + \Rsinfrac s \sin\lphase +(s^2/2-Ls)\cos\lphase    \right] \\
h\left[(s^2/2-Ls)\quad+\quad\left(\Rsinfrac\right)^2\left(\cos\slphasedif - \cos\lphase\right) \right]
\end{bmatrix}\nonumber\\
&&~&&+ A_z \refc{R}&
\begin{bmatrix}
h&\left[2\left(\Rsinfrac\right)^2\sin\sphase - \Rsinfrac s \cos\lphase -\Rsinfrac\left[(s-L)\cos\sphase +L\right] \right] \\
& -2\left(\Rsinfrac\right)^2(\cos\sphase-1)- \Rsinfrac s \sin\lphase -\Rsinfrac(s-L)\sin\sphase\\
& 0
\end{bmatrix}\nonumber\\
&&~&&+ \resrot \omega \cos\refc{\alpha}&
\begin{bmatrix}
&\refc{R}\left[ \Rsinfrac\sin\sphase-s\cos\lphase\right]\\
& h\refc{R}\left[-\Rsinfrac\left(\cos\sphase-1\right)-s\sin\lphase\right] \\
& (s^2/2-Ls)\cos\refc{\alpha}
\end{bmatrix}.\label{deltaphilongexpression}
\end{alignat}
The perturbation to the tangent vector is given by Eq.~\ref{smallmatframerotation} as $\delta\matd{3}=\delta\boldsymbol{\phi} \wedge\uprefc{\matd{3}}$, which gives
\small
\begin{alignat}{4} 
&&&&EI\delta\matd{3}\nonumber\\
&&=&& \cos\refc{\alpha}(A_x\refc{R})&
\begin{bmatrix}
\cot\refc{\alpha}\left[\left(\Rsinfrac\right)^2\left( \cos \sphase -1\right) + \Rsinfrac s \sin\lphase+(s^2/2-Ls)\cos\lphase \right] \\
h\cot\refc{\alpha}\left[\left(\Rsinfrac\right)^2\sin\sphase - \Rsinfrac s \cos\lphase +(s^2/2-Ls)\sin\lphase    \right] \\
0
\end{bmatrix}\nonumber\\
&&~&&+ \cos\refc{\alpha}(A_z\refc{R})&
\begin{bmatrix}
& -2\left(\Rsinfrac\right)^2(\cos\sphase-1)- \Rsinfrac s \sin\lphase -\Rsinfrac(s-L)\sin\sphase\\
-h&\left[2\left(\Rsinfrac\right)^2\sin\sphase - \Rsinfrac s \cos\lphase -\Rsinfrac\left[(s-L)\cos\sphase +L\right] \right] \\
& 0
\end{bmatrix}\nonumber\\
&&~&&+ \sin\refc{\alpha}(A_x\refc{R})&
\begin{bmatrix}
&-\left\{(s^2/2-Ls)\cos\sphase + \left(\Rsinfrac\right)^2\left[\cos\sphase \cos\slphasedif - \cos\sphase \cos\lphase   \right] \right\}\\
& h\left\{(s^2/2-Ls)\sin\sphase + \left(\Rsinfrac\right)^2\left[\sin\sphase \cos\slphasedif - \sin\sphase \cos\lphase   \right] \right\}\\
& 0
\end{bmatrix}\nonumber\\
&&~&&+ \sin\refc{\alpha}&\begin{pmatrix}0\\0\\1\end{pmatrix}
\begin{bmatrix} -(A_x\refc{R})\cot\refc{\alpha} &\bigg[ & \left(\Rsinfrac\right)^2 \sin\sphase -\Rsinfrac s \cos\slphasedif&& \\
&&\hfill- (s^2/2-Ls)\sin\slphasedif \bigg]&& \\ +(A_z\refc{R})&\bigg[& 2\left(\Rsinfrac\right)^2 \sin\sphase - \Rsinfrac s \cos\slphasedif &&\\ 
&&\hfill-\Rsinfrac [s-L + L\cos \sphase]\bigg]&&\end{bmatrix} \nonumber\\
&&~&&+ \resrot \omega \sin\refc{\alpha}\cos\refc{\alpha}&
\begin{bmatrix}
h\left[\refc{R}\cot\refc{\alpha}\left(-\Rsinfrac\left( \cos \sphase -1\right) - s \sin\lphase \right) - \cos\refc{\alpha}(s^2/2-Ls)\cos\sphase\right] \\
\left[-\refc{R}\cot\refc{\alpha}\left(\Rsinfrac\sin \sphase - s \cos\lphase \right) - \cos\refc{\alpha}(s^2/2-Ls)\sin\sphase \right]\\
h\refc{R}\left[\Rsinfrac\sin\sphase -  s \cos\slphasedif \right]
\end{bmatrix}.\label{deltad3longexpression}
\end{alignat}
\normalsize
The deformation of the helix is given be Eq.~\ref{deform_int_deltad3}, as the integral of the perturbation to the tangent vector.

Integrating the expression for $\delta \matd{3}(s)$ in Eq.~\ref{deltad3longexpression} gives
\scriptsize
\begin{alignat}{4} 
&&&&EI\delta \bv{r} (s)=EI\int\limits_{0}^{s}\delta\matd{3}(s') \dd s'&\\
&&=&& \frac{\cos^2 \refc{\alpha}}{\sin \refc{\alpha}}(A_x\refc{R})&
\begin{pmatrix}
\left[  (\frac{s^3}{6} - L\frac{s^2}{2})\cos \lphase + 0.5\left(\Rsinfrac\right) s^2\sin\lphase - \left(\Rsinfrac\right)^2 s + \left(\Rsinfrac\right)^3\sin\sphase   \right]\\
..................................................................\\
h\left[  (\frac{s^3}{6} - L\frac{s^2}{2})\sin \lphase  -0.5\left(\Rsinfrac\right) s^2\cos\lphase - \left(\Rsinfrac\right)^3 \left(\cos\sphase -1 \right) \right]\\
..................................................................\\
0
\end{pmatrix}\nonumber\\
&&~&&+ \cos\refc{\alpha}(A_z\refc{R})&
\begin{pmatrix}
\left[  - 0.5\left(\Rsinfrac\right) s^2\sin\lphase + \left(\Rsinfrac\right)^2 \left[2s + (s-L)\cos\sphase + L\right] -3 \left(\Rsinfrac\right)^3\sin\sphase   \right]\\
..................................................................\\
h\left[ \left(\Rsinfrac\right) \left(0.5 s^2\cos\lphase + Ls\right) 
+\left(\Rsinfrac\right)^2 (s-L)\sin\sphase
+ 3\left(\Rsinfrac\right)^3 \left(\cos\sphase -1 \right) \right]\\
..................................................................\\
0
\end{pmatrix}\nonumber\\
&&~&&+ \sin\refc{\alpha}(A_x\refc{R})&
\begin{pmatrix}
-\bigg\{ 
&&\hspace{-2cm}
\Rsinfrac\left(\frac{s^2}{2}-Ls\right)\sin \sphase
 +\left(\Rsinfrac\right)^2\left[(s-L)\cos\sphase +L  + \frac{1}{2} s \cos\lphase\right] \\
&& +\left(\Rsinfrac\right)^3  \left[\frac{1}{4}\left(\sin\left(\frac{2s-L}{\Lambda/2\pi}\right) + \sin\lphase\right)  -  \sin\sphase \left(1+\cos\lphase\right)\right]    \bigg\} \\
&&.............................................................\\
h\bigg\{&&  \hspace{-1cm}
 \left[-\Rsinfrac(\frac{s^2}{2}-Ls) \cos \sphase + \left(\Rsinfrac\right)^3 \left(\cos\sphase - 1 \right) + \left(\Rsinfrac\right)^2 (s-L) \sin\sphase \right]\\
&& +\frac{1}{2} \left(\Rsinfrac\right)^2s \sin\lphase -\frac{1}{4}\left(\Rsinfrac\right)^3 \left(\cos\left(\frac{2s-L}{\Lambda/2\pi}\right) - \cos\lphase\right)   \\&&\hspace{6cm} +\left(\Rsinfrac\right)^3 \left(\cos\sphase-1\right) \cos\lphase   \bigg\}\\
&&.............................................................\\
&& 0
\end{pmatrix}\nonumber\\
&&~&&+ \sin\refc{\alpha}\begin{pmatrix}0\\0\\1\end{pmatrix}
\bigg\{
(A_x\refc{R})\cot\refc{\alpha}&
\begin{bmatrix}
-\Rsinfrac\left(\frac{s^2}{2}-Ls\right)\cos\slphasedif\\
~~~~~+\left(\Rsinfrac\right)^2\left[s \sin\slphasedif + (s-L) \sin\slphasedif - L\sin\lphase\right]\\
+\left(\Rsinfrac\right)^3 \left[\cos\sphase -1 + 2\left(\cos\slphasedif - \cos\lphase\right)\right]
\end{bmatrix}\nonumber\\
&&~&&- (A_z\refc{R})&
\begin{bmatrix}
\Rsinfrac(\frac{s^2}{2}-Ls) \\
+\left(\Rsinfrac\right)^2 \left[L\sin\sphase + s \sin\slphasedif \right]\\
+\left(\Rsinfrac\right)^3\left[\cos\slphasedif - \cos\lphase +2\left(\cos\sphase -1\right) \right]
\end{bmatrix}\bigg\}\nonumber\\
\qquad\nonumber\\
%
%
%
&&~&&+ \resrot \omega \refc{R}\cos\refc{\alpha} &
\begin{bmatrix}
-h\cos\refc{\alpha} \bigg\{
  (s^2/2) \sin\lphase 
     + \left(\frac{1}{2}s^2-L s\right)  \sin\sphase
     +  \left(\Rsinfrac\right) (s-L) \left(\cos\sphase-1\right) 
    \bigg\} \\
..................................................................\\
+\cos\refc{\alpha} \bigg\{  \frac{s^2}{2} \cos\lphase  + \left( \frac{1}{2}s^2 -Ls\right)  \cos\sphase 
- \left(\Rsinfrac\right) (s-L)\sin\sphase \bigg\} \\
..................................................................\\
-h\refc{R} \left\{ s\sin\slphasedif+\left(\Rsinfrac\right)\left[\cos\sphase-1 +   \cos\slphasedif-\cos\lphase\right]   \right\}
\end{bmatrix}.
\end{alignat}

\pagebreak
\normalsize
\section{Clamped $s=0$ end-point contribution} \label{Clamped_Appendix}
In this appendix, we calculate the leading-order estimates of the quantities $\Jclamped{1}-\Jclamped{4}$ for our helical geometry, clamped at one end.

\subsection{Calculation of $\Jclamped{1}-\Jclamped{4}$} \label{AppJ1J4}
We first calculate the contributions to  $\Jclamped{1}-\Jclamped{4}$, without the terms due to the viscous rotational torque, as those will be calculated separately in the next subsection.
Straightforward integrations give
\begin{align}
\Jclamped{1}&=\int_{0}^{L} (\delta\matd{3})_{z} \dd s =\frac{1}{2EI}\left[\refc{R}^2 L^2 + {\mathcal O}(\refc{R}^3 L)\right] \left[A_x\cot\refc{\alpha} + A_z\right],\\
\Jclamped{3}&=\int_{0}^{L} s(\delta\bv{r})_z \dd s =\frac{5}{24EI}A_z\left[\refc{R}^2 L^4 + {\mathcal O}(\refc{R}^3 L^3)\right].
\end{align}
We can rewrite $\Jclamped{4}$ and $\Jclamped{2}$ as 
\begin{align}
\Jclamped{4}&=\int_{0}^{L}(\bv{r}.\delta\bv{r}) \dd s =\refc{R}\int_{0}^{L}\left[ \cos\sphase \delta x + h \sin\sphase \delta y\right]\dd s + \cos \refc{\alpha}J_3 ,\\
\Jclamped{2}&=\int_{0}^{L} [\delta(\matd{3}\wedge \bv{r})]_z  \dd s =-h\sin\refc{\alpha}\int_{0}^{L}\left[ \cos\sphase \delta x + h \sin\sphase \delta y\right]\dd s \nonumber\\
&\qquad\qquad\qquad\qquad\qquad+ h\refc{R}\int_{0}^{L}\left[ \sin\sphase (\delta\matd{3} )_x - h \cos\sphase (\delta\matd{3} )_y\right]\dd s.
\end{align}
We thus need to calculate the two integrals
$I_1=\int_{0}^{L}\left[ \cos\left(2\pi s/\Lambda_0\right) \delta x + h \sin\left(2\pi s/\Lambda_0\right) \delta y\right]\dd s$ and $I_2=\int_{0}^{L}\left[ \sin\left(2\pi s/\Lambda_0\right) (\delta\matd{3} )_x - h \cos\left(2\pi s/\Lambda_0\right) (\delta\matd{3} )_y\right]\dd s$.
Straightforward integrations lead to
\begin{align}
I_2&=\int_{0}^{L}\left[ \sin\sphase (\delta\matd{3} )_x - h \cos\sphase (\delta\matd{3} )_y\right]\dd s \nonumber\\
&= \frac{1}{2EI} \frac{\refc{R}^2 L^2}{\sin\refc{\alpha}} \left[A_x \left(\frac{\cos^2\refc{\alpha}}{\sin\refc{\alpha}} -\sin\refc{\alpha}\cos\left(\frac{2L}{\refc{R}/\sin\refc{\alpha}}\right)\right) + A_z\cos \refc{\alpha}\right] + {\mathcal O}(\refc{R}^3 L A_{x,z}/EI).
\end{align}
Using integration by parts, we obtain
\begin{align}
I_1&=\int_{0}^{L}\left[ \cos\sphase \delta x + h \sin\sphase \delta y\right]\dd s \nonumber\\
&=\frac{\refc{R}}{\sin\refc{\alpha}}\bigg\{\sin\lphase \delta x|_{L}-h\cos\lphase \delta y|_{L} \nonumber\\
&\qquad\qquad\quad-\int_{0}^{L}\left[ \sin\sphase (\delta\matd{3} )_x - h \cos\sphase (\delta\matd{3} )_y\right]\dd s  \bigg\}. 
\end{align}

The boundary terms are 
\begin{align}
\sin\lphase \delta x|_{L}-h\cos\lphase \delta y|_{L}\qquad\qquad\qquad\qquad&
 \nonumber\\ 
=\frac{1}{EI} \frac{\refc{R}^2 L^2}{\sin\refc{\alpha}} \bigg[\frac{1}{2}A_x \left(\frac{\cos^2\refc{\alpha}}{\sin\refc{\alpha}} -\sin\refc{\alpha}\cos\left(\frac{2L}{\refc{R}/\sin\refc{\alpha}}\right)\right)&   \nonumber\\ 
- A_z\cos \refc{\alpha}\left(\frac{1}{2}+ \cos\lphase\right)\bigg]&
\left[1+{\mathcal O}\left(\refc{R}/L\right)\right].
\end{align}
Thus
\begin{align}
I_1&=\int_{0}^{L}\left[ \cos\sphase \delta x + h \sin\sphase \delta y\right]\dd s \nonumber\\&=-\frac{\refc{R}^3 L^2}{EI} \frac{\cos\refc{\alpha}}{\sin^2\refc{\alpha}} A_z\left[1+ \cos\lphase\right] + {\mathcal O}(\refc{R}^4 L A_{x,z}/EI), 
\end{align} and
\begin{align}
\Jclamped{4}&=\frac{5}{24EI}A_z\cos\refc{\alpha}\left[\refc{R}^2 L^4 + {\mathcal O}(\refc{R}^3 L^3)\right]\\
\Jclamped{2}&=h\frac{\refc{R}^3 L^2}{EI}\left[ \frac{1}{2}A_x \left(\cot^2\refc{\alpha}- \cos\left(\frac{2L}{\refc{R}/\sin\refc{\alpha}}\right)  \right) + A_z \cot\refc{\alpha}\left(\frac{3}{2} + \cos\left(\frac{2L}{\refc{R}/\sin\refc{\alpha}}\right) \right)  \right] \nonumber \\
&+ {\mathcal O}(\refc{R}^4 L A_{x,z}/EI)
\end{align}
Putting it all together we obtain
\begin{align}
\Jclamped{1}&=\frac{1}{2EI}\left[\refc{R}^2 L^2 + {\mathcal O}(\refc{R}^3 L)\right] \left[A_x\cot\refc{\alpha} + A_z\right]\\
\Jclamped{2}&=h\frac{\refc{R}^3 L^2}{EI}\left[ \frac{1}{2}A_x \left(\cot^2\refc{\alpha}- \cos\left(\frac{2L}{\refc{R}/\sin\refc{\alpha}}\right)  \right) + A_z \cot\refc{\alpha}\left(\frac{3}{2} + \cos\left(\frac{2L}{\refc{R}/\sin\refc{\alpha}}\right) \right)  \right]  \nonumber \\
&+ {\mathcal O}(\refc{R}^4 L A_{x,z}/EI)\\
\Jclamped{3}&=\frac{5}{24EI}A_z\left[\refc{R}^2 L^4 + {\mathcal O}(\refc{R}^3 L^3)\right]\\
\Jclamped{4}&=\frac{5}{24EI}A_z\cos\refc{\alpha}\left[\refc{R}^2 L^4 + {\mathcal O}(\refc{R}^3 L^3)\right],
\end{align}
with the terms arising from the viscous rotational torque to be calculated in the next subsection.

 \normalsize
\subsection{Contribution from the viscous rotational torque} \label{AppJ1J4viscrottorq}
In this subsection we calculate the contributions $\Jvrottorq{1}-\Jvrottorq{4}$ arising from the viscous rotational torque to $\Jclamped{1}-\Jclamped{4}$.
\begin{align}
&\Jvrottorq{1}=\int_{0}^{L} (\delta\matd{3})_{z} \dd s \equiv 0\\
&\Jvrottorq{4}=\int_{0}^{L}(\bv{r}.\delta\bv{r}) \dd s \\
&\quad=\refc{R}\int_{0}^{L}\left[ \cos\sphase \delta x + h \sin\sphase \delta y\right]\dd s + \cos \refc{\alpha}\Jvrottorq{3},
\end{align}
where
\begin{align}
&\Jvrottorq{3}=\int_{0}^{L} s(\delta\bv{r})_z \dd s \nonumber\\
&=-\frac{h \resrot \omega \refc{R}^2\cos\refc{\alpha}}{EI}\int_{0}^{L}\Bigg\{ s^2\sin\slphasedif
\nonumber\\&
\qquad\qquad\qquad\qquad\qquad\qquad+\left(\Rsinfrac\right)s\left[\cos\sphase-1 +   \cos\slphasedif-\cos\lphase\right]   \Bigg\}\dd s
\\
&=\frac{h \resrot \omega \refc{R}^3\cot\refc{\alpha}}{EI}
\Bigg\{
\frac{L^2}{2} \left(3 + \cos\lphase\right) 
\nonumber\\&
\qquad\qquad\qquad\qquad
- L\left(\Rsinfrac\right)\sin\lphase   + 2 \left(\Rsinfrac\right)^2 \left(\cos\lphase-1\right) 
\Bigg\},
\end{align}
and
\begin{align}
&\int_{0}^{L}\left[ \cos\sphase \delta x + h \sin\sphase \delta y\right]\dd s =0. \label{Appresult}
\end{align}
\normalsize
The calculation to obtain the result of Eq.~\ref{Appresult} is the following:
\normalsize
\begin{align}
&\int_{0}^{L}\left[ \cos\sphase \delta x + h \sin\sphase \delta y\right]\dd s  \nonumber\\
&= h\resrot \omega \refc{R}\cos^2\refc{\alpha} \times \nonumber\\
&\int_{0}^{L}
\bigg[ 
-\cos\sphase  \bigg\{
\frac{s^2}{2} \sin\lphase 
+ \left(\frac{s^2}{2}-L s\right)  \sin\sphase
\nonumber\\&
\qquad\qquad\qquad\qquad\qquad
+  \left(\Rsinfrac\right) (s-L) \left(\cos\sphase-1\right) 
\bigg\} \nonumber
\\&\quad\quad
+  \sin\sphase \bigg\{  \frac{s^2}{2} \cos\lphase  + \left( \frac{s^2}{2} -Ls\right)  \cos\sphase 
- \left(\Rsinfrac\right) (s-L)\sin\sphase \bigg\}\bigg]\dd s
\nonumber\\
&= h\resrot \omega \refc{R}\cos^2\refc{\alpha}\int_{0}^{L}
\bigg[   \frac{s^2}{2} \sin\slphasedif   
- \left(\Rsinfrac\right) (s-L)\left(1-\cos\sphase\right) \bigg]\dd s
\nonumber\\
&= h\resrot \omega \refc{R}\cos^2\refc{\alpha}
\Bigg\{   -\frac{1}{2}  \left[L^2 + 2 \left(\Rsinfrac\right)^2 \left[\cos\lphase-1\right]\right] \nonumber\\
&\qquad\qquad\qquad\qquad\quad
-  \left[-\frac{L^2}{2} + \left(\Rsinfrac\right)^2\left(1 - \cos\lphase\right)\right]\Bigg\}
\nonumber\\
&=0.
\end{align}
\normalsize
Hence 
\begin{align}
\Jvrottorq{4}&= \cos \refc{\alpha}\Jvrottorq{3} .
\end{align}
Integrating by parts gives 
\begin{align}
&\int_{0}^{L}\left[ \cos\sphase \delta x + h \sin\sphase \delta y\right]\dd s  \nonumber\\
& =\frac{\refc{R}}{\sin\refc{\alpha}}\Bigg\{\sin\lphase \delta x|_{L}-h\cos\lphase \delta y|_{L} 
\nonumber\\
&\qquad\qquad\quad
-\int_{0}^{L}\left[ \sin\sphase (\delta\matd{3} )_x - h \cos\sphase (\delta\matd{3} )_y\right]\dd s  \Bigg\} 
\end{align}
Hence
\begin{align}
&\int_{0}^{L}\left[ \sin\sphase (\delta\matd{3} )_x - h \cos\sphase (\delta\matd{3} )_y\right]\dd s  \nonumber\\
&\qquad\qquad=\sin\lphase \delta x|_{L}-h\cos\lphase \delta y|_{L} \nonumber\\
&\qquad\qquad=0,
\end{align}
with the last equality arising by substitution of the formulae for $\delta x|_{L}, \delta y|_{L}$ terms arising from the viscous rotational torque.
\begin{align}
\Jvrottorq{2}=&\int_{0}^{L} [\delta(\matd{3}\wedge \bv{r})]_z  \dd s\nonumber\\
 =&-h\sin\refc{\alpha}\int_{0}^{L}\left[ \cos\sphase \delta x + h \sin\sphase \delta y\right]\dd s \nonumber\\
&\quad+ h\refc{R}\int_{0}^{L}\left[ \sin\sphase (\delta\matd{3} )_x - h \cos\sphase (\delta\matd{3} )_y\right]\dd s  \nonumber\\
&=0
\end{align}
In summary, $\Jvrottorq{1}=0$,
$\Jvrottorq{2}=0$ and $\Jvrottorq{4}=\cos\refc{\alpha}\Jvrottorq{3}$, with

\begin{align}
\Jvrottorq{3}
&=\frac{h \resrot \omega \refc{R}^3\cot\refc{\alpha}}{EI}
\bigg\{
\frac{L^2}{2} \left(3 + \cos\lphase\right) - L\left(\Rsinfrac\right)\sin\lphase \nonumber \\
&\qquad\qquad\qquad\qquad\qquad\qquad\qquad\qquad\qquad\quad  + 2 \left(\Rsinfrac\right)^2 \left(\cos\lphase-1\right) 
\bigg\}.
\end{align}
These values of the contributions to the $J_1-J_4$ integrals,
$\Jvrottorq{1}=0,~
\Jvrottorq{2}=0,~
\Jvrottorq{4}=\cos\refc{\alpha}\Jvrottorq{3}$, mean that the viscous rotational torque does not contribute to the perturbations to the resistance matrices,
\begin{align}
\VisRotTorqPertMatr{A}&=2\resperp(1-\rho) \cos\refc{\alpha} \Jvrottorq{1}
=0,\\
\VisRotTorqPertMatr{B}
&= \resperp(1-\rho)\left\{ h\refc{R}\sin\refc{\alpha} \Jvrottorq{1}- \cos\refc{\alpha}\Jvrottorq{2}\right\}=0,\\
\VisRotTorqPertMatr{C}
&=\VisRotTorqPertMatr{B}=0,\\
\VisRotTorqPertMatr{D}
&=2\resperp\left\{  
\cos\refc{\alpha}\Jvrottorq{3}-\Jvrottorq{4}-(1-\rho)h \refc{R} \sin \refc{\alpha} \Jvrottorq{2}\right\} - 2\resrot\cos\refc{\alpha} \Jvrottorq{1}=0,
\end{align}
We summarize the resulting $J_1-J_4$ for the clamped case, ($\vPhi,\pertzeroend=\bv{0}$), including the terms arising from the viscous rotational torque
\begin{align}
\Jclamped{1}&=\frac{1}{2EI}\left[\refc{R}^2 L^2 + {\mathcal O}(\refc{R}^3 L)\right] \left[A_x\cot\refc{\alpha} + A_z\right],\\
\Jclamped{2}
&=h\frac{\refc{R}^3 L^2}{EI}\left[ \frac{1}{2}A_x \left(\cot^2\refc{\alpha}- \cos\left(\frac{2L}{\refc{R}/\sin\refc{\alpha}}\right)  \right) + A_z \cot\refc{\alpha}\left(\frac{3}{2} + \cos\left(\frac{2L}{\refc{R}/\sin\refc{\alpha}}\right) \right)  \right] \nonumber \\
&\quad+ {\mathcal O}(\refc{R}^4 L A_{x,z}/EI), \\
\Jclamped{3}
&=\frac{5}{24EI}A_z\left[\refc{R}^2 L^4 + {\mathcal O}(\refc{R}^3 L^3)\right] \nonumber\\
&\quad+\frac{h \resrot \omega \refc{R}^3\cot\refc{\alpha}}{EI}
\bigg[
\frac{L^2}{2} \left(3 + \cos\lphase\right) \nonumber\\
&\qquad\qquad\qquad\qquad\qquad  - L\left(\Rsinfrac\right)\sin\lphase   + 2 \left(\Rsinfrac\right)^2 \left(\cos\lphase-1\right) 
\bigg],\\
\Jclamped{4}&=\frac{5}{24EI}A_z\cos\refc{\alpha}\left[\refc{R}^2 L^4 + {\mathcal O}(\refc{R}^3 L^3)\right]\nonumber\\
&\quad+\frac{h \resrot \omega \refc{R}^3\cos^2\refc{\alpha}}{EI\sin\refc{\alpha}}
\bigg[
\frac{L^2}{2} \left(3 + \cos\lphase\right) - L\left(\Rsinfrac\right)\sin\lphase \nonumber\\
&\qquad\qquad\qquad\qquad\qquad\qquad\qquad\qquad  \qquad \quad
+ 2 \left(\Rsinfrac\right)^2 \left(\cos\lphase-1\right) 
\bigg].
\end{align}

\normalsize
The resulting perturbations to the resistance matrices (projected along the z axis) are
\begin{align}
\ClampPertMatr{A}
&=\resperp(1-\rho) \cos\refc{\alpha}
\frac{1}{EI}\left[\refc{R}^2 L^2 + {\mathcal O}(\refc{R}^3 L)\right] \left[A_x\cot\refc{\alpha} + A_z\right],  \label{App_deltaAclamped}\\
\ClampPertMatr{B}&= \resperp(1-\rho)\left\{ h\refc{R}\sin\refc{\alpha} \Jclamped{1}- \cos\refc{\alpha}\Jclamped{2}\right\} \nonumber\\
&=\resperp(1-\rho)\frac{h\refc{R}^3 L^2}{2EI}\bigg\{  ~~A_x\cos\refc{\alpha}\left[1-\left(\cot^2\refc{\alpha} -\cos\left(\frac{2L}{\refc{R}/\sin\refc{\alpha}}\right)\right)\right] \nonumber\\
&\qquad\qquad\qquad\qquad~~+ A_z\sin\refc{\alpha}\enspace\left[1-\cot^2\refc{\alpha}\left(\frac{3}{2} + \cos\left(\frac{2L}{\refc{R}/\sin\refc{\alpha}}\right)\right)\right] ~~ \bigg\} \nonumber\\
&~~\quad+{\mathcal O}\left( \frac{\refc{R}^4LA_{x,z}}{EI} \right) \label{App_deltaBclamped},\\
%
\ClampPertMatr{C}&=\ClampPertMatr{B},\\
%
\ClampPertMatr{D}
&= 2\resperp\left\{ \cos\refc{\alpha}\Jclamped{3}-\Jclamped{4}-(1-\rho)h \refc{R} \sin \refc{\alpha}\Jclamped{2} \right\} \nonumber\\
&=2\resperp\frac{\refc{R}^4 L^2}{EI}\bigg\{  \qquad\frac{\cos\refc{\alpha}}{\sin^2\refc{\alpha}}A_z\left[1+\cos\left(\frac{2L}{\refc{R}/\sin\refc{\alpha}}\right)\right] \nonumber\\
&\qquad\qquad~~~~~
-(1-\rho)\sin\refc{\alpha} \bigg[0.5A_x \left(\cot^2\refc{\alpha} -\cos\left(\frac{2L}{\refc{R}/\sin\refc{\alpha}} \right)\right) \nonumber\\ 
&\qquad\qquad\qquad\qquad\qquad\qquad~+A_z\cot\refc{\alpha}\left( \frac{3}{2} + \cos\left(\frac{2L}{\refc{R}/\sin\refc{\alpha}}\right) \right)~~\bigg]~~
\bigg\}  \nonumber\\
&~~+{\mathcal O}\left( \frac{\refc{R}^5LA_{x,z}}{EI} \right), \label{App_deltaDclamped}
\end{align}
where we note again that the viscous rotational torque, $\Nviscous$, does not contribute to any of the perturbations of the resistance matrices.

\pagebreak

\normalsize
\section{Free $s=0$ end-point contribution} \label{Appfreeendpoint}
In this Appendix we calculate the contributions of $\vPhi,\pertzeroend$ to the perturbations to the resistance matrices. We first find the contributions to $\Jfree{1}-\Jfree{4}$ from $\vPhi,\pertzeroend$ for the case of free $s=0$ end-point,
i.e.~with $\vPhi,\pertzeroend\neq\bv{0}$.
\normalsize
The calculation is as follows:
\begin{align}
\Jfree{1}&=\int\limits_{0}^{L} \vez\cdot\left(\vPhi
\wedge\uprmatd{3}\right) \dd s= -\vPhi\cdot\left[\vez\wedge\left(\ruprefc(s)-\ruprefc(0)\right)\right]= -\vPhi\cdot
\begin{pmatrix}
-h\refc{R}\sin\lphase\\ \refc{R}\left(\cos\lphase -1\right)\\ 0
\end{pmatrix}\\
\Jfree{2}&=\vez \cdot\int_{0}^{L} \left[\left(\vPhi\wedge\uprmatd{3}(s)\right)\wedge \ruprefc(s) +\uprmatd{3}(s)\wedge \left(\vPhi\wedge\left(\ruprefc(s) - \ruprefc(0)\right)\right)+ \pertzeroend \right]  \dd s \nonumber\\
&=\vez \cdot\int_{0}^{L} \bigg\{\left[\uprmatd{3} \left(\vPhi\cdot\ruprefc(s) \right) -\ruprefc(s) \left(\uprmatd{3}\cdot\vPhi\right)\right]
\nonumber\\
&\qquad\qquad\qquad  - \uprmatd{3}\cdot\left[\ruprefc(0) ~\vPhi - \vPhi~\ruprefc(0)\right] + \uprmatd{3} \wedge\pertzeroend\bigg\} \dd s\nonumber\\
&=\vez \cdot\bigg\{\vPhi\cdot\int_{0}^{L} \left[\ruprefc(s)~\uprmatd{3}(s)-\uprmatd{3}(s)\ruprefc(s)\right]\dd s\nonumber\\ &\qquad\qquad-\vPhi\left[\ruprefc(0)\cdot\left(\ruprefc(L)-\ruprefc(0)\right)\right] 
\nonumber\\
&\qquad\qquad + \ruprefc(0)\left[\vPhi\cdot\left(\ruprefc(L) -\ruprefc(0)\right)\right] + \left[\ruprefc(L)-\ruprefc(0)\right]\wedge\pertzeroend\bigg\}\nonumber\\
&=\vPhi\cdot\int_{0}^{L} \left[\ruprefc(s)~\left(\vez \cdot\uprmatd{3}(s)\right)-\uprmatd{3}(s)\left(\vez \cdot\ruprefc(s)\right)\right]\dd s\\ &\qquad\qquad-\left(\vez \cdot\vPhi\right)\left[\ruprefc(0)\cdot\left(\ruprefc(L)-\ruprefc(0)\right)\right]\nonumber\\&\qquad\qquad + \left(\vez \cdot\ruprefc(0)\right)\left[\vPhi\cdot\left(\ruprefc(L) -\ruprefc(0)\right)\right] +\vez \cdot \left(\left[\ruprefc(L)-\ruprefc(0)\right]\wedge\pertzeroend\right)\nonumber\\
&=\cos\refc{\alpha}\vPhi\cdot\int_{0}^{L} \left[\ruprefc(s)-s~\uprmatd{3}(s)\right]\dd s 
-\left(\vez\cdot\vPhi\right)\refc{R}^2\left[\cos\lphase - 1 \right]  \nonumber\\
&\qquad\qquad\qquad 
+\pertzeroend  \left(\vez\wedge\left[\ruprefc(L)-\ruprefc(0)\right]\right)\nonumber\\
&=\vPhi\cdot\begin{pmatrix}
\cos\refc{\alpha}\left[2\refc{R}\Rsinfrac\sin\lphase -L\refc{R}\cos\lphase\right]\\\cos\refc{\alpha}\left\{-2\refc{R}\Rsinfrac\left[\cos\lphase - 1\right] - hL\refc{R}\sin\lphase\right\}\\-\refc{R}^2\left[\cos\lphase - 1 \right]
\end{pmatrix}  \nonumber\\
&\quad+\pertzeroend \cdot \begin{pmatrix}
-h\refc{R}\sin\lphase\\\refc{R}\left[\cos\lphase - 1\right]\\0
\end{pmatrix}
\end{align}

\begin{align}
\Jfree{3}&=\vez\cdot\int\limits_{0}^{L} s\left[\vPhi\wedge\left(\ruprefc(s)-\ruprefc(0)\right) + \pertzeroend\right]\dd s
\nonumber\\
&=\vez\cdot\left\{\vPhi\wedge\int\limits_{0}^{L} s\left(\ruprefc(s)-\ruprefc(0)\right)\dd s 
\right\}
+ \vez\cdot\pertzeroend \frac{L^2}{2} \nonumber\\
&=-\vPhi\cdot\left\{\vez\wedge\int\limits_{0}^{L} s\left(\ruprefc(s)-\ruprefc(0)\right)\dd s 
\right\}+ \vez\cdot\pertzeroend \frac{L^2}{2} \nonumber\\
&=-\vPhi\cdot
\begin{pmatrix}
-h\refc{R}\left[\left(\Rsinfrac\right)^2 \sin\lphase  - \left(\Rsinfrac\right)L\cos\lphase\right]\\
\refc{R}\left[\Rsinfrac L\sin\lphase + \left(\Rsinfrac\right)^2\left(\cos\lphase - 1\right) - \frac{L^2}{2}\right]\\
0	\end{pmatrix} + \vez\cdot\pertzeroend \frac{L^2}{2}
\end{align}

\begin{align}
\Jfree{4}&= \int\limits_{0}^{L} \ruprefc(s)\cdot\left[\vPhi\wedge\left(\ruprefc(s) - \ruprefc(0)\right) + \pertzeroend\right] \dd s =\left(\int\limits_{0}^{L} \ruprefc(s) \dd s\right)\cdot\left[\pertzeroend - \vPhi\wedge\ruprefc(0)\right] \nonumber\\
&=\left(\int\limits_{0}^{L} \ruprefc(s)\right)
\cdot\pertzeroend - \vPhi\cdot\left(\ruprefc(0)\wedge\int\limits_{0}^{L} \ruprefc(s)\dd s\right)	 \nonumber\\
&=\begin{pmatrix}
\refc{R}\Rsinfrac\sin\lphase\\-\refc{R}\Rsinfrac\left(\cos\lphase -1 \right)\\\cos\refc{\alpha} \frac{L^2}{2}
\end{pmatrix}
\cdot\pertzeroend -\refc{R} \vPhi\cdot
\begin{pmatrix}
0\\-\cos\refc{\alpha} \frac{L^2}{2}\\-\refc{R}\Rsinfrac\left(\cos\lphase -1 \right)
\end{pmatrix}  \nonumber\\
&=\begin{pmatrix}
\refc{R}\Rsinfrac\sin\lphase\\-\refc{R}\Rsinfrac\left(\cos\lphase -1 \right)\\\cos\refc{\alpha} \frac{L^2}{2}
\end{pmatrix}
\cdot\pertzeroend + \refc{R}\vPhi\cdot
\begin{pmatrix}
0\\\cos\refc{\alpha} \frac{L^2}{2}\\\refc{R}\Rsinfrac\left(\cos\lphase -1 \right)
\end{pmatrix}
\end{align}
\normalsize
With the contributions $\Jfree{1}-\Jfree{4}$ of $\Phi, \Delta\refc{\bv{r}}$ to $J_1-J_4$ calculated, we now proceed to find the contribution of $\Phi, \Delta\refc{\bv{r}}$ to the perturbations to the resistance martices.
\begin{align}
&\FreePertMatr{A}\nonumber\\
&=2\resperp(1-\rho) \cos\refc{\alpha}\refc{R}
\vPhi\cdot
\begin{pmatrix}
h\sin\lphase\\ -\left(\cos\lphase -1\right)\\ 0
\end{pmatrix}\\
&\frac{\FreePertMatr{B}}{\resperp(1-\rho)}\nonumber\\
&= \left\{ h\refc{R}\sin\refc{\alpha} \Jfree{1}- \cos\refc{\alpha}\Jfree{2}\right\}\nonumber\\
&= \vPhi\cdot\begin{pmatrix}
\refc{R}^2\sin\refc{\alpha}\sin\lphase-\cos^2\refc{\alpha}\left[2\refc{R}\Rsinfrac\sin\lphase -L\refc{R}\cos\lphase\right]\\
-h\sin\refc{\alpha}\refc{R}^2\left(\cos\lphase -1\right)-\cos^2\refc{\alpha}\left\{-2\refc{R}\Rsinfrac\left[\cos\lphase - 1\right] - hL\refc{R}\sin\lphase\right\}\\
\cos\refc{\alpha}\refc{R}^2\left[\cos\lphase - 1 \right]
\end{pmatrix}\nonumber\\
&\quad- \cos\refc{\alpha}\pertzeroend \cdot \begin{pmatrix}
-h\refc{R}\sin\lphase\\\refc{R}\left[\cos\lphase - 1\right]\\0
\end{pmatrix}\nonumber\\
&= \vPhi\cdot\begin{pmatrix}
L\refc{R}\cos^2\refc{\alpha}\cos\lphase\\
hL\refc{R}\cos^2\refc{\alpha}\sin\lphase\\
\cos\refc{\alpha}\refc{R}^2\left[\cos\lphase - 1 \right]
\end{pmatrix}\quad
- \refc{R} \cos\refc{\alpha}\pertzeroend \cdot \begin{pmatrix}
-h\sin\lphase\\\left[\cos\lphase - 1\right]\\0
\end{pmatrix}\\
&\FreePertMatr{C}
=\FreePertMatr{B}
\end{align}

\normalsize
\begin{align}
&\frac{\FreePertMatr{D}}{2\resperp}\nonumber\\
&=    
\cos\refc{\alpha}\Jfree{3}-\Jfree{4}-(1-\rho)h \refc{R} \sin \refc{\alpha} \Jfree{2}  \nonumber\\
&=  
\cos\refc{\alpha}\left[\vPhi\cdot
\begin{pmatrix}
h\refc{R}\left[\left(\Rsinfrac\right)^2 \sin\lphase  - \left(\Rsinfrac\right)L\cos\lphase\right]\\
-\refc{R}\left[\Rsinfrac L\sin\lphase + \left(\Rsinfrac\right)^2\left(\cos\lphase - 1\right) - \frac{L^2}{2}\right]\\
0	\end{pmatrix} + \frac{L^2}{2}\pertzeroend\cdot\vez \right] \nonumber\\
& \qquad-\left[\vPhi\cdot
\begin{pmatrix}
0\\\cos\refc{\alpha} \refc{R}\frac{L^2}{2}\\\refc{R}^2\Rsinfrac\left(\cos\lphase -1 \right)
\end{pmatrix}
+\pertzeroend\cdot\begin{pmatrix}
\refc{R}\Rsinfrac\sin\lphase\\-\refc{R}\Rsinfrac\left(\cos\lphase -1 \right)\\\cos\refc{\alpha} \frac{L^2}{2}
\end{pmatrix}\right]\nonumber\\
&\qquad -(1-\rho)h \refc{R} \sin \refc{\alpha}
\begin{bmatrix}
\bigg[\vPhi\cdot\begin{pmatrix}
\cos\refc{\alpha}\left[2\refc{R}\Rsinfrac\sin\lphase -L\refc{R}\cos\lphase\right]\\\cos\refc{\alpha}\left\{-2\refc{R}\Rsinfrac\left[\cos\lphase - 1\right] - hL\refc{R}\sin\lphase\right\}\\-\refc{R}^2\left[\cos\lphase - 1 \right]
\end{pmatrix}\nonumber\\
+\pertzeroend \cdot \begin{pmatrix}
-h\refc{R}\sin\lphase\\\refc{R}\left[\cos\lphase - 1\right]\\0
\end{pmatrix}
\end{bmatrix} \nonumber\\
&= 
\vPhi\cdot
\begin{pmatrix}
\cos\lphase\\
h\sin\lphase\\
\cos\lphase 
\end{pmatrix}h\frac{\refc{R}^2L\cos\refc{\alpha} + {\mathcal O}(\refc{R}^3)}{\sin\refc{\alpha}} \left[-1 +(1-\rho) \sin^2\refc{\alpha} \right]
+{\mathcal O}\left(\refc{R}^2|\pertzeroend|\right)
 \nonumber\\
&= \frac{ h \refc{R}^2L\cos\refc{\alpha} }{\sin\refc{\alpha}}
\left[\vPhi\cdot
\begin{pmatrix}
\cos\lphase\\
h\sin\lphase\\
\cos\lphase 
\end{pmatrix} \left[-1 +(1-\rho) \sin^2\refc{\alpha} \right]
+{\mathcal O}\left(\frac{\refc{R}}{L},\frac{|\pertzeroend|}{L}\right)
\right]
\end{align}

\normalsize
In summary, we have
\begin{align}
\Jfree{1}&=\vPhi\cdot
\begin{pmatrix}
h\refc{R}\sin\sphase\\ -\refc{R}\left(\cos\sphase -1\right)\\ 0
\end{pmatrix},\\
\Jfree{2}
&=\vPhi\cdot\begin{pmatrix}
\cos\refc{\alpha}\left[2\refc{R}\Rsinfrac\sin\lphase -L\refc{R}\cos\lphase\right]\\\cos\refc{\alpha}\left\{-2\refc{R}\Rsinfrac\left[\cos\lphase - 1\right] - hL\refc{R}\sin\lphase\right\}\\-\refc{R}^2\left[\cos\lphase - 1 \right]
\end{pmatrix} \nonumber 
\\
&\quad+\pertzeroend \cdot \begin{pmatrix}
-h\refc{R}\sin\lphase\\\refc{R}\left[\cos\lphase - 1\right]\\0
\end{pmatrix},\\
\Jfree{3}
&=\vPhi\cdot
\begin{pmatrix}
h\refc{R}\left[\left(\Rsinfrac\right)^2 \sin\lphase  - \left(\Rsinfrac\right)L\cos\lphase\right]\\
-\refc{R}\left[\Rsinfrac L\sin\lphase + \left(\Rsinfrac\right)^2\left(\cos\lphase - 1\right) - \frac{L^2}{2}\right]\\
0	\end{pmatrix} + \frac{L^2}{2}\pertzeroend\cdot\vez ,\\
\Jfree{4}
&=
\vPhi\cdot
\begin{pmatrix}
0\\ \cos\refc{\alpha}\refc{R} \frac{L^2}{2}\\\refc{R}^2\Rsinfrac\left(\cos\lphase -1 \right)
\end{pmatrix}
+\pertzeroend\cdot\begin{pmatrix}
\refc{R}\Rsinfrac\sin\lphase\\-\refc{R}\Rsinfrac\left(\cos\lphase -1 \right)\\\cos\refc{\alpha} \frac{L^2}{2}
\end{pmatrix},
\end{align}
\normalsize
and the resulting contributions to the perturbations of the resistance matrices from $\vPhi,\pertzeroend$ are
\begin{align}
\FreePertMatr{A}
&=2\resperp(1-\rho) \cos\refc{\alpha}\refc{R}
\vPhi\cdot
\begin{pmatrix}
h\sin\lphase\\ -\left(\cos\lphase -1\right)\\ 0
\end{pmatrix},\\
\FreePertMatr{B}
&=\resperp(1-\rho)\vPhi\cdot\begin{pmatrix}
L\refc{R}\cos^2\refc{\alpha}\cos\lphase\\
hL\refc{R}\cos^2\refc{\alpha}\sin\lphase\\
\cos\refc{\alpha}\refc{R}^2\left[\cos\lphase - 1 \right]
\end{pmatrix}
\nonumber\\
&
\quad
-\resperp\refc{R}(1-\rho)\cos\refc{\alpha}\pertzeroend \cdot \begin{pmatrix}
-h\sin\lphase\\\left[\cos\lphase - 1\right]\\0
\end{pmatrix},\\
\FreePertMatr{C}
&= \FreePertMatr{B},\\
\FreePertMatr{D}
& = \frac{2h\resperp\refc{R}^2L\cos\refc{\alpha} }{\sin\refc{\alpha}}
\left[\vPhi\cdot
\begin{pmatrix}
\cos\lphase\\
h\sin\lphase\\
\cos\lphase 
\end{pmatrix} \left[-1 +(1-\rho) \sin^2\refc{\alpha} \right]
+{\mathcal O}\left(\frac{\refc{R}}{L},\frac{|\pertzeroend|}{L}\right)
\right].
\end{align}

\pagebreak
\normalsize
\section{Feedback to the kinematics} \label{FeedbacktoKinem}
We note that throughout this Appendix, we will use the shorthand notation $s=\sin\refc{\alpha}$ and $c=\cos\refc{\alpha}$.
Using the rigid, leading order kinematics of Eqs.~\ref{rigidkinemHelixGenForm_U}-\ref{rigidkinemHelixGenForm_omega},  and Eqs.~\ref{Ax}-\ref{Az}, we can express $A_x$ and $A_z$ as 
\begin{align}
&A_x=\frac{-\resperp M}{\curlyA\curlyD-\curlyB^2} \left[(1-\rho)sc\curlyB + (c^2 + \rho s^2) h \refc{R}\curlyA \right]=\frac{h \refc{R}\resperp M}{\curlyA\curlyD-\curlyB^2}\left\{\rho\resperp L +(c^2 + \rho s^2)6\pi\mu \ahead 
\right\},\\ %
&A_z=\frac{-\resperp M}{\curlyA\curlyD-\curlyB^2} \left[(s^2+\rho c^2)\curlyB + (1-\rho)sc h \refc{R}\curlyA \right]=\frac{h \refc{R}\resperp M  }{\curlyA\curlyD-\curlyB^2}6\pi\mu \ahead (1-\rho)sc ,\\
&A_x\cot\refc{\alpha}+A_z=\frac{-\resperp M}{\curlyA\curlyD-\curlyB^2}
\left[\curlyB + \cot\refc{\alpha} h \refc{R}\curlyA \right]=\frac{h\refc{R}\resperp M}{\curlyA\curlyD-\curlyB^2}
\left[\resperp L \rho +  6\pi\mu \ahead \right]\cot\refc{\alpha}.
\end{align}
Substituting into Eqs.~\ref{deltaAclamped}-\ref{deltaDclamped} (and using the simplified notation $\delta\curlyA$ instead of $\delta \curlyA_{zz}^{clamped}$ etc) we obtain 
\begin{align}
\delta \curlyA
&=
\frac{h\refc{R}^3 L^2\resperp^2 M}{EI\left(\curlyA\curlyD-\curlyB^2\right)} 
\left[\rho\resperp L  +  6\pi\mu \ahead \right](1-\rho)\frac{c^2}{s} , \\
\delta \curlyB
&=\frac{\refc{R}^4 L^2\resperp^2 M}{EI\left(\curlyA\curlyD-\curlyB^2\right)}(1-\rho)\frac{1}{2}\bigg\{  ~~\left\{\rho\resperp L +(c^2 + \rho s^2)6\pi\mu \ahead 
\right\} c\left[1-\left(\cot^2\refc{\alpha} -\cos\dlphase\right)\right] \nonumber\\
&\qquad\qquad\qquad\qquad\qquad\qquad~~+  6\pi\mu \ahead (1-\rho)s^2 c\enspace\left[1-\cot^2\refc{\alpha}\left(\frac{3}{2} + \cos\dlphase\right)\right] ~~ \bigg\}, \\
\delta \curlyD
&=\frac{h\refc{R}^5 L^2\resperp^2 M}{EI\left(\curlyA\curlyD-\curlyB^2\right)} 2(1-\rho)\bigg\{    6\pi\mu \ahead\frac{c^2}{s} \left[1+\cos\dlphase\right] \nonumber\\
&\qquad\qquad\qquad\qquad\qquad\qquad~~
-s \bigg[0.5\left\{\rho\resperp L +(c^2 + \rho s^2)6\pi\mu \ahead 
\right\}  \left(\cot^2\refc{\alpha} -\cos\dlphase\right) \nonumber\\ 
&\qquad\qquad\qquad\qquad\qquad\qquad\qquad\qquad\qquad\quad\qquad+6\pi\mu \ahead (1-\rho)c^2\left( \frac{3}{2} + \cos\dlphase \right)~~\bigg]~~
\bigg\},
\end{align}
and substituting into Eq.~\ref{deltaUHelixGenform} gives the result of Eqs.~\ref{deltaUgenformCalculated},\ref{QgenformCalculated}.

\end{document}